\documentclass[%
reprint,
superscriptaddress,
nofootinbib,
 amsmath,amssymb,
 aps,
 prx,
]{revtex4-2}
\usepackage{booktabs}
\usepackage{xcolor}
\usepackage{graphicx}
\usepackage{dcolumn}
\usepackage{bm}
\usepackage{hyperref}
\usepackage{chngcntr}
\usepackage{verbatim}
\usepackage{amsmath}
\usepackage{mathtools}
\usepackage{amssymb}
\usepackage{amsfonts}
\usepackage{soul}
\usepackage{multirow}
\usepackage{array}
\usepackage{makecell}

\newcommand{\edit}[1]{#1}

\begin{document}

\title{Fixation and extinction in  time-fluctuating spatially structured metapopulations}

\author{Matthew Asker}
\email{asker@evolbio.mpg.de}
\affiliation{%
 Department of Applied Mathematics, School of Mathematics, University of Leeds, Leeds LS2 9JT, United Kingdom
}%
\homepage{https://eedfp.com}
\author{Mohamed Swailem}
\affiliation{
 Department of Physics \& Center for Soft Matter and Biological Physics, MC 0435, Robeson Hall, 850 West Campus Drive, Virginia Tech, Blacksburg, Virginia 24061, USA
}
\author{Uwe C. T\"auber}
\affiliation{
 Department of Physics \& Center for Soft Matter and Biological Physics, MC 0435, Robeson Hall, 850 West Campus Drive, Virginia Tech, Blacksburg, Virginia 24061, USA
}
\author{Mauro Mobilia}
\email{M.Mobilia@leeds.ac.uk}
\affiliation{%
 Department of Applied Mathematics, School of Mathematics, University of Leeds, Leeds LS2 9JT, United Kingdom
}

\date{\today}

\begin{abstract}
Bacteria  evolve in volatile environments and complex spatial structures. Migration, fluctuations and environmental variability therefore have a significant impact on the evolution of microbial populations.
Here, we consider a class of spatially explicit metapopulation models arranged as regular (circulation) graphs where wild-type and mutant cells compete  in a {\it time-fluctuating}  environment in which demes (subpopulations) are connected by slow cell migration. 
The carrying capacity is the same at each deme and endlessly switches between two values associated to harsh and mild environmental conditions. 
It is known that environmental variability can lead to population bottlenecks, following which the population is prone to fluctuation-induced extinction. 
Here, we analyse how slow migration, spatial structure, and fluctuations affect the phenomena of fixation and extinction on  clique, cycle, and square lattice metapopulations. 
When  the carrying capacity remains large,  bottlenecks are weak and deme extinction can be ignored. 
The dynamics is thus captured by a coarse-grained description within which the probability and mean time of fixation are obtained analytically. 
This allows us to show that, in contrast to what happens in static environments, the mutant fixation probability depends on the rate of migration. 
We also show that the fixation probability and mean fixation time can exhibit a non-monotonic dependence on the switching rate. 
When the carrying capacity is small under harsh conditions, bottlenecks are strong, and the metapopulation evolution is shaped by the coupling of deme extinction and strain competition. 
This yields rich dynamical scenarios, among which we identify the best conditions to eradicate mutants without dooming the metapopulation to extinction. We offer an interpretation of these findings in the context of an idealised treatment strategy and discuss possible generalisations of our models.
\end{abstract}

\maketitle

\section{Introduction}
\label{sec:introduction}

Microbial populations live in volatile and time-varying environments embedded in complex spatial settings, across which the distribution of microbes fluctuates. For instance, many organisms live in  densely packed
aggregates on surface-attached biofilms~\cite{ISME2016}, numerous commensal bacteria are distributed throughout the gastrointestinal tract~\cite{engel2013,garud2019}, and patients' organs are spatial environments between which bacteria can migrate~\cite{she2024}.
Moreover, natural environments are not static, e.g.  temperature, pH, or available resources change over time. These abiotic variations,
not caused by the organisms themselves, are referred to as
environmental fluctuations and can have a significant influence on the evolution of natural populations. For example,  
the gut microbiome of a host is exposed to fluctuations of great amplitude on various timescales, and these affect the diversity of the microbiota~\cite{cignarella2018,smits2017}.
In small populations, demographic fluctuations are another important form of randomness resulting in  fixation - where one strain takes over the community - or extinction~\cite{Ewens,Kimura}. 
Since the  variations of population size and composition are often interdependent~\cite{Roughgarden,melbinger2010,cremer2011,cremer2012,melbinger2015,chuang2009,verdon2024habitat}, this can lead to the coupling of environmental and demographic fluctuations~\cite{wienandEvolutionFluctuatingPopulation2017,wienandEcoevolutionaryDynamicsPopulation2018,taitelbaumPopulationDynamicsChanging2020a,taitelbaum2023evolutionary,shibasakiExclusionFittestPredicts2021,hernandez-navarroCoupledEnvironmentalDemographic2023,askerCoexistenceCompetingMicrobial2023,west2020,hernandez-navarroEcoevolutionaryDynamicsCooperative2024}. Their interplay  is particularly significant in  microbial communities, 
where it can lead to population bottlenecks, where new colonies consisting of a few cells are prone to fluctuations~\cite{Wahl02,Rainey2003,patwaAdaptationRatesLytic2010,Brockhurst2007a,Brockhurst2007b}. Population bottlenecks and fluctuations
are particularly relevant for the evolution of antimicrobial resistance, when cells surviving 
antibiotics treatment may replicate leading to the spread of resistance~\cite{coatesAntibioticinducedPopulationFluctuations2018,mahrt2021bottleneck,
hernandez-navarroCoupledEnvironmentalDemographic2023,hernandez-navarroEcoevolutionaryDynamicsCooperative2024,LKUM2024}. 

How likely is a population to be taken over by a mutant or to go extinct? What is the typical time for these events to occur? 
These  are central  questions in evolution, and the answers depend on  the population's spatial structure as well as the environmental variations and fluctuations. In this context, it is important to understand the impact of spatial structure, migration, and  fluctuations on the spread of a mutant strain. 
A common approach to represent a spatially structured biological population is by dividing it into several demes - well-mixed subpopulations connected by cell migration - hence forming  a {\it metapopulation}~\cite{wrightEvolutionMendelianPopulations1931,kimuraSteppingStoneModel1964,Levins69,Hanski99,Lugo08,Butler09,Szczesny14,Peruzzo20,marrecUniversalModelSpatially2021,moawadEvolutionCooperationDemestructured2024}. 
The influence of the spatial arrangement of 
a population and stochastic fluctuations on mutants' fate has been studied in {\it static} environments both theoretically~\cite{wrightEvolutionMendelianPopulations1931,kimuraSteppingStoneModel1964,maruyamaFixationProbabilityMutant1970,barton93,whitlock97,whitlock03,liebermanEvolutionaryDynamicsGraphs2005,houch11,allen17,marrecUniversalModelSpatially2021,abbaraFrequentAsymmetricMigrations2023a,marrec2023,yagoobi21,tkadlec23,abbaraMutantFateSpatially2024} and experimentally~\cite{kryazhimskiy12,nahum15,chakrabortyExperimentalEvidenceThat2023,kreger23}. 
Maruyama notably showed that in a constant environment, 
when cell migration is symmetric and preserves  the overall mutant fraction, the fixation probability of a mutant is independent of the spatial structure and migration rate
~\cite{maruyamaFixationProbabilityMutant1970}. However, it has been shown that
random extinction and recolonisation can affect the mutant fixation probability on fully-connected (static) graphs, even when cell migration is symmetric~\cite{barton93}. In this case, deme extinction is immediately followed by recolonisation by a mixture of cells from other demes~\cite{Levins69,barton93,Hanski99}. 
Furthermore,
independent extinction and recolonisation by a single neighbour of demes   
has been studied on fully-connected (static) graphs~\cite{landeExtinctionTimesFinite1998}. 
Recently, the authors of Ref.~\cite{marrecUniversalModelSpatially2021} studied the  
the influence of slow migration on the fate of mutants on static graphs, and demonstrated that migration asymmetry can dramatically affect their fixation probability on certain spatial structures like the star graph.  
 However, the biologically relevant problem  of mutants evolving in time-varying spatially structured populations has been  
rather scarcely investigated, and the case of strains competing to colonise and fixate  demes prone to extinction  
remains under-explored.  

Here, we tackle these important issues by studying a class of time-fluctuating  microbial metapopulation models 
consisting of demes in which wild-type and mutant cells evolve in a time-varying environment represented by a  switching carrying capacity. We use coarse-grained descriptions of the dynamics to study the joint influence of environmental variability, demographic fluctuations, migration, and spatial structure on the evolution of  the metapopulation. 
 We obtain explicit results for 
cliques (island model~\cite{wrightEvolutionMendelianPopulations1931,kimuraSteppingStoneModel1964}, or fully-connected graph), cycles, and two-dimensional grids (with periodic boundaries).  
 In stark contrast with the evolution in static environments, we demonstrate that when bottlenecks are weak, the fixation probability on regular circulation graphs 
  depends on the migration rate.
 Moreover, we show that under the effect of environmental variability and fluctuations the  
 fixation probability and mean fixation time can exhibit a non-monotonic dependence on the switching rate. 
In the case of strong bottlenecks, arising when 
 the carrying capacity is  small under harsh conditions, the dynamics is characterised by deme extinction and strain competition coupled by environmental switching. This yields rich dynamical scenarios
among which we  identify the best  conditions 
 to eradicate mutants without risking metapopulation extinction.

 In the next section, we introduce the explicit spatial metapopulation model that we study and outline our methodology. In Sec.~\ref{sec:constant_env}, we present our results in the case of static environments.
 This paves the way to the detailed analysis  in time-fluctuating environments of Sec.~\ref{sec:changing_env}, with the weak and strong bottleneck regimes respectively studied in Secs.~\ref{sec:changing_env-mild} and \ref{sec:changing_env-harsh}.
 Sec.~\ref{sec:generality} is dedicated to a
  discussion of  our findings, assumptions and 
  possible generalisations.  
We present our conclusions  in Sec.~\ref{sec:conclusions}. Additional technical details 
 are given in a series of appendices.

\section{Model \& Methods}
\label{sec:model-and-methods_model}

\begin{figure*}
    \centering
    \includegraphics[width=\textwidth]{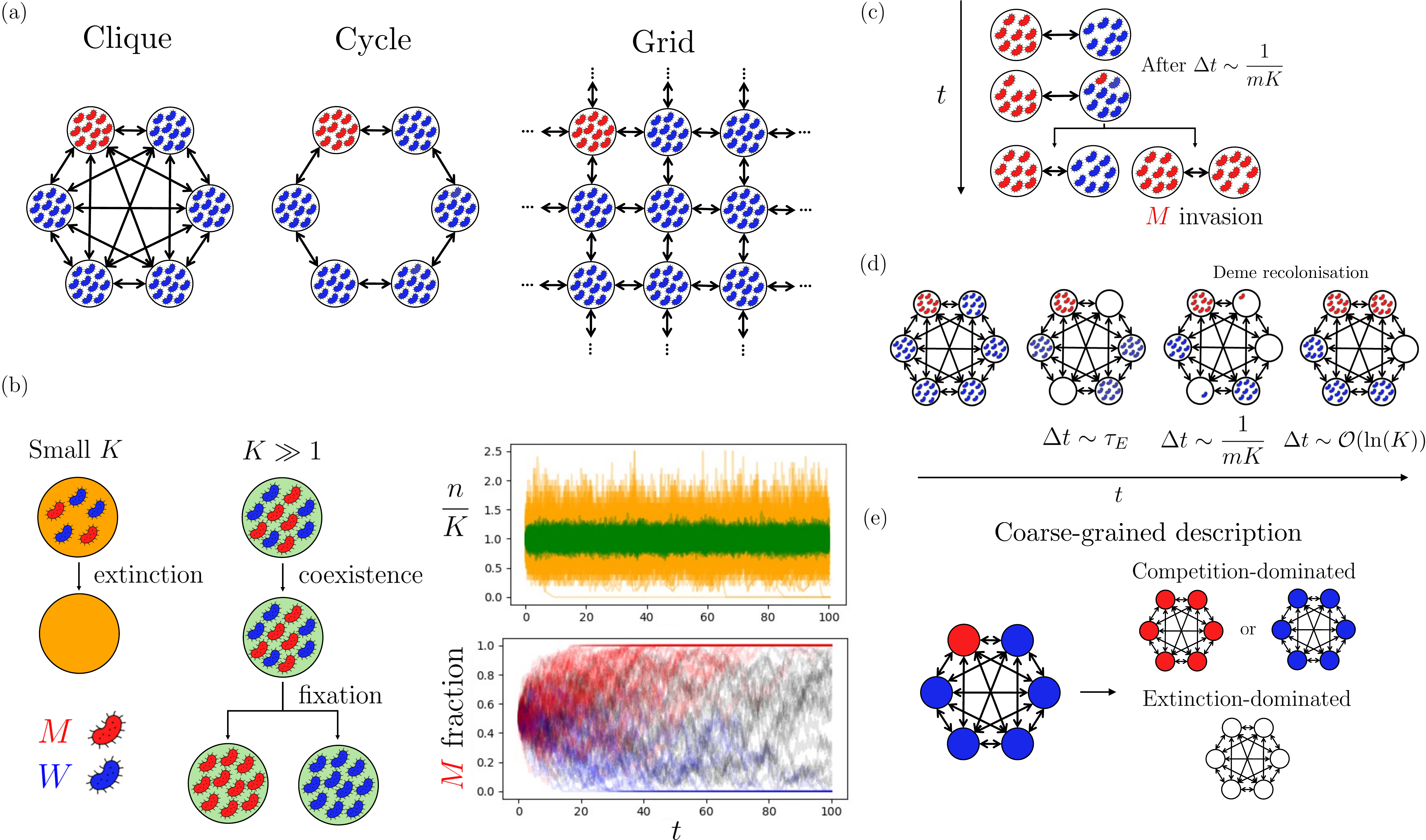}
    \caption{Metapopulation dynamics in a static environment. 
    (a) {\it Examples of metapopulation graphs}: a clique, cycle, and grid (from left to right). Neighbouring demes are connected by migration (double arrows). Initially, there is one mutant deme (red/light) and $\Omega -1$ wild-type demes  (blue / dark), and all demes have the same constant carrying capacity $K$. 
     (b) {\it Dynamics in a single deme.} Left: Wild-type $W$ cells (blue / dark) compete with mutants of type $M$ (red / light). When $K$ is small, the deme is prone to extinction. When $K$ is large, both types coexist prior to $W$ or $M$ fixation. Top right:  
     Realisations of the rescaled deme size $n/K$ vs. time $t$  for $K=5$ (orange/light) and $K=100$ (green/dark) illustrating how $n$ fluctuates about $K$. Bottom right:  Fraction of $M$ cells vs. $t$ in a deme with $K=100$. Deme extinction is not  observed. Transient coexistence of $W$ and $M$ is followed by the fixation of $W$ (blue traces) or $M$ (red traces). Here $s=0.01$.  (c) {\it Invasion of $W$ deme by an $M$ cell}:
     Any $M$ cell of a deme migrates to a neighbouring $W$ site with migration rate $m$ after a mean time $\Delta t =
     1/(mK)$, and then type $M$ either quickly fixates, producing a new $M$ deme (right), or does not fix leaving the pair of $M$ and $W$ demes unchanged (left).
     The same picture holds for the invasion of an $M$ deme by a $W$ cell; see text.
     (d) {\it Deme recolonisation} (here for the clique): Deme extinction occurs after a mean time $\tau_E$, and empty demes are then 
     recolonised by an invader from a neighbouring 
     surviving deme after  $\Delta t \sim
     1/(mK)$. A recolonised deme is rapidly taken over (in $\Delta t \sim \mathcal{O}(\ln(K))$). 
     (e) {\it Coarse-grained description of the metapopulation dynamics}: Each deme is always either fully $W$ (blue / dark) or $M$ (red / light) or empty (white). In this description, different scenarios arise, shown for the clique. Competition-dominated regime: all demes are occupied and there is always fixation of $W$ or $M$. Extinction-dominated regime: there are frequent deme extinctions and the metapopulation quickly goes extinct. 
     }
    \label{fig:cartoon-constant}
\end{figure*}

We consider a class of spatially explicit metapopulation models of $\Omega$ demes labelled by $x\in\{1,...,\Omega\}$, each of which at time $t$ consists of a well-mixed subpopulation of $n_W$ cells of wild-type $W$, and $n_M$ mutants of strain $M$. Each wild-type cell has a baseline fitness $f_W=1$ and all mutant cells have fitness $f_M=1+s$. We assume $0<s\ll 1$, giving  $M$  a small selective advantage over $W$.

The microbial metapopulation can be envisioned as a graph whose nodes $x\in\{1,\dots,\Omega\}$ are demes (also called
sites).
Each deme $x$ is  a well-mixed subpopulation
of size  $n(x)=\edit{\sum_{\alpha\in\{W,M\}} n_{\alpha}(x)}=n_W(x)+n_M(x)$ 
located at a node of the  
metapopulation graph. Thus, $N_{W/M}\equiv \sum_x n_{W/M}(x)$ is 
the number of $W/M$ individuals across the metapopulation, and
$N \equiv N_W + N_M$ is the total number of individuals
in the whole graph.
Here, we focus on  fully-connected graphs (as in the island model ~\cite{wrightEvolutionMendelianPopulations1931,kimuraSteppingStoneModel1964}), called cliques, and periodic one- and two-dimensional lattices called respectively 
cycles and grids; see Fig.~\ref{fig:cartoon-constant}(a). 
These are regular graphs, generally denoted by  ${\rm G}=\{{\rm clique, cycle, grid}\}$, 
of $\Omega$ demes connected by edges to their $q_{{\rm G}}$ nearest neighbours via cell migration 
at per capita rate $m$ (independently from division and death)~\cite{wrightEvolutionMendelianPopulations1931,kimuraSteppingStoneModel1964,maruyamaFixationProbabilityMutant1970,marrecUniversalModelSpatially2021,moawadEvolutionCooperationDemestructured2024,fruetSpatialStructureFacilitates2024a}; see Fig.~\ref{fig:cartoon-constant}(a-c). 
We study the eco-evolutionary dynamics of the metapopulation in the biologically relevant regime of slow migration (whereby intra-deme dynamics occur much faster than inter-deme dynamics; see below)~\cite{wrightISOLATIONDISTANCE1943,wrightPopulationStructureEvolution1949,marrecUniversalModelSpatially2021,moawadEvolutionCooperationDemestructured2024,keymerBacterialMetapopulationsNanofabricated2006}, and consider that initially  one deme is occupied entirely by mutants ($M$ deme), while the other $\Omega -1$ demes ($W$ demes) are all populated by $W$ cells; see Sec.~\ref{sec:generality}. 
All demes are assumed to have the same carrying capacity $K$ which 
 encodes environmental variability. In Sec.~\ref{sec:constant_env}, we assume that $K$ is constant, and in Sec.~\ref{sec:changing_env} we let the carrying capacity switch endlessly
 between two values representing mild and harsh  conditions~\cite{wienandEvolutionFluctuatingPopulation2017,wienandEcoevolutionaryDynamicsPopulation2018,taitelbaumPopulationDynamicsChanging2020a,west2020,shibasakiExclusionFittestPredicts2021,taitelbaum2023evolutionary,hernandez-navarroCoupledEnvironmentalDemographic2023,askerCoexistenceCompetingMicrobial2023,hernandez-navarroEcoevolutionaryDynamicsCooperative2024,LKUM2024}; see below. 
 
In close relation to the Moran process~\cite{Moran,Blythe07,traulsen2009stochastic,antal2006fixation} (see Appendix~\ref{appendix:slow_migration}), a reference model in mathematical biology~\cite{Ewens}, 
the intra-deme dynamics in a deme $x$  is thus represented by a multivariate birth-death process defined by the birth and death of a cell of type $\alpha\in\{W,M\}$ in that site according to the 
reactions~\cite{wienandEvolutionFluctuatingPopulation2017,wienandEcoevolutionaryDynamicsPopulation2018,askerCoexistenceCompetingMicrobial2023,askerCoexistenceCompetingMicrobial2023,LKUM2024}
\begin{align}
    n_{\alpha}\stackrel{T^+_{\alpha}}{\longrightarrow} n_{\alpha}+1 \quad \text{and} \quad n_{\alpha}\stackrel{T^-_{\alpha}}{\longrightarrow} n_{\alpha}-1,
\label{eq:BD}
\end{align}
occurring at the transition rates
\begin{equation}
\label{eq:intra_transition_rates}
        T^+_\alpha(x) = \frac{f_\alpha}{\overline{f}} n_\alpha \quad \text{and}\quad
        T^-_\alpha(x) = \frac{n}{K}n_\alpha,
\end{equation} %
where  $\overline{f}\equiv(n_W f_W + n_Mf_M)/n$ is the average fitness in deme $x$, and $K$ here denotes  the constant carrying capacity in a static environment and its time-switching version in a dynamical environment; see below. In this formulation, without loss of generality, selection operates
 on birth events. This can be generalised to include selection on deaths; see e.g.~\cite{melbinger2010,cremer2011}.
 
The inter-deme dynamics stems from the migration 
of one cell of type $\alpha\in \{W,M\}$
from the site $x$ 
 to one of its $q_{\rm G}$ neighbouring demes denoted by $y$ at a per-capita migration rate $m$.
 Here, for the sake of simplicity, we assume that the migration rate is the same  in all directions and for both types (symmetric migration); see Sec.~\ref{sec:generality} for a discussion of these assumptions. The inter-deme dynamics for all cells at deme $x$ with its neighbouring demes labelled $y$ is therefore  
 implemented 
according to the reaction  
 \begin{equation}
\label{Eq:Migreact}
\bigl[n_{\alpha}(x),n_{\alpha}(y)\bigr]\stackrel{T^{m,{\rm G}}_\alpha}{\longrightarrow} \bigl[n_{\alpha}(x)-1,n_{\alpha}(y)+1\bigr], 
 \end{equation}
 occurring at the migration transition rate
\begin{equation}
\label{eq:migration-transition}
        T^{m,{\rm G}}_\alpha(x) = \frac{m n_\alpha}{q_{\rm G}}.
\end{equation} 
\edit{On a given metapopulation, if the number of cells migrating into a deme equals the number of cells migrating out, and this is true for all demes, we say that the metapopulation is a circulation.} {Precisely, the condition here for a metapopulation to be a circulation is given by
\begin{equation}
    \label{eq:circulation_condition}
    \sum_{y~\text{n.n.}~x}\sum_{\alpha} T^{m,{\rm G}}_\alpha(x)=\sum_{y~\text{n.n.}~x}\sum_{\alpha} T^{m,{\rm G}}_\alpha(y)\quad\text{for all }x,
\end{equation}
where $y~\text{n.n.}~x$ denotes the sum over the $q_G$ neighbours $y$ of the deme $x$. 
In our case, this simplifies considerably to give $n(x)=\sum_{y~\text{n.n.}~x}n(y)/q_G$
for all $x$. Intuitively, this means that for our setup, if each deme has the same population size, then the metapopulation is a circulation: there is the same incoming and outgoing migration flow at each deme. This will indeed be the case under sufficiently large carrying capacity.}

Microbial communities generally live in time-varying conditions, and are often subject to  sudden and drastic environmental changes.
Here, environmental variability is encoded in the time-variation of the binary carrying capacity~\cite{wienandEvolutionFluctuatingPopulation2017,wienandEcoevolutionaryDynamicsPopulation2018,taitelbaumPopulationDynamicsChanging2020a,west2020,shibasakiExclusionFittestPredicts2021,taitelbaum2023evolutionary,hernandez-navarroCoupledEnvironmentalDemographic2023,askerCoexistenceCompetingMicrobial2023,hernandez-navarroEcoevolutionaryDynamicsCooperative2024,LKUM2024} 
\begin{equation}
 \label{eq:K(t)}
 K(t)=\frac{1}{2}\left[K_- + K_- +\xi(t) (K_+ - K_-)\right],
\end{equation}
driven by
a random telegraph process
$\xi(t)\in \{-1,1\}$.
 The coloured dichotomous Markov noise (DMN) $\xi(t)$  switches between $\pm 1$
according to $\xi\to -\xi$ at rate $\nu$ for the symmetric DMN (see Sec.~\ref{sec:generality} and Appendix~\ref{appendix:bias} for the generalisation to asymmetric switching)~\cite{bena2006,HL06,Ridolfi11}.
 The carrying capacity, equal across demes,  thus switches at a rate $\nu$ between a value $K=K_+$ in a mild environment (e.g. abundance of nutrients, lack of toxin) and $K=K_-<K_+$ under harsh environmental conditions (e.g. lack of nutrients, abundance of toxin) according to $K_-\overset{\nu}{\rightleftharpoons}K_+$, and
thus  represents
random  cycles of mild and harsh conditions (feast and famine cycles). The randomly time-switching $K(t)$ drives the size of each deme, and is hence  responsible for the coupling of demographic fluctuations with environmental variability, an effect particularly important when 
there are population bottlenecks~\cite{taitelbaumPopulationDynamicsChanging2020a,west2020,taitelbaum2023evolutionary,hernandez-navarroCoupledEnvironmentalDemographic2023,askerCoexistenceCompetingMicrobial2023,hernandez-navarroEcoevolutionaryDynamicsCooperative2024,LKUM2024,wienandEvolutionFluctuatingPopulation2017,wienandEcoevolutionaryDynamicsPopulation2018}; see below.
Here, the  DMN is at stationarity, i.e. it is initialised from its long-time distribution where instantaneous correlations become time-independent (while the autocovariance is a function of the time difference)  \cite{Gardiner,HL06}. This means that the average of the DMN
vanishes, $\langle \xi(t)\rangle=0$, and its autocovariance coincides with its autocorrelation reading $\langle \xi(t)\xi(t')\rangle=e^{-2\nu|t-t'|}$~\cite{bena2006,HL06,Ridolfi11}, where $\langle \cdot \rangle$ denotes the ensemble average and  $1/(2\nu)$ is the finite correlation time (with $t,t'\to \infty$). As a consequence,  the  fluctuating carrying capacity is always at stationarity, with a constant average $\langle K(t)\rangle=\langle K\rangle=(K_++ K_-)/2$ and an autocorrelation
$\langle K(t)K(t')\rangle=\langle K\rangle e^{-2\nu|t-t'|}$.
 In our simulations with symmetric random switching, the carrying capacity is initially drawn from its stationary distribution, with $K(0)=K_+$ or $K(0)=K_-$ each with probability $1/2$; see Sec.~\ref{sec:generality} and Appendix~\ref{appendix:bias}.

The full individual-based model  is therefore 
a continuous-time multivariate Markov process defined by the reactions and transition rates Eqs.~\eqref{eq:BD}-\eqref{eq:migration-transition} 
that satisfies the master equation Eq.~\eqref{eq:ME}
discussed in Appendix~\ref{appendix:ME}.
The microscopic intra- and inter-deme dynamics encoded in the 
master equation~\eqref{eq:ME} has been simulated using the Monte Carlo method described in Appendix~\ref{appendix:model-and-methods_simulation}. 
The eco-evolutionary dynamics of a single deme is outlined in Appendix~\ref{appendix:eco-evolutionary-dynamics-in-a-deme}. It is worth noting that 
$n , n_{W/M}$, $T_{W/M}^{\pm}$, and $T_{W/M}^{m,{\rm G}}$ are all  quantities that depend on the site $x$ and time $t$, and on  $\xi$
in a time-varying environment. 
However, for notational simplicity, we often drop the explicit dependence 
on some or all of the variables
$x,t$, and $\xi$. 
Below, we combine coarse-grained analytical approximations and
individual-based stochastic simulations to
study how the spatial structure, migration, and demographic 
fluctuations influence the fixation and extinction properties of the microbial metapopulation.
\section{Static environments}
\label{sec:constant_env}
We first consider a static environment where the carrying capacity $K$ of each deme is constant.
In this setting, the size $n$ of each deme rapidly reaches and fluctuates about $K$, with $n\approx K$ when $K\gg 1$; see  Fig.~\ref{fig:cartoon-constant}(b, top right). The expected number of migrants per unit time and deme is thus  $mK$. The occurrence of migration events, alongside
the competition between $M$ and $W$ to take over demes of the other type, increases with $K$. 
Cell migration and competition are however limited when $K$ is small: regardless of their type, demes of small size are  prone to extinction  in a mean time $\tau_E(K)$; see Fig.~\ref{fig:cartoon-constant}(b,d).
For independent demes of size $K$, the deme mean extinction time $\tau_E(K)$ can be obtained from a logistic birth-death process (see Appendix~\ref{appendix:extinction_time}) yielding
\begin{equation}
 \label{eq:tauE}
\tau_E(K)\approx \frac{e^{K}}{K},
 \end{equation}
when $K\gg 1$; see  Fig.~\ref{fig:constant-environment}(b,top). We thus refer to $K$ as ``small'' where
fixation of a deme (intra-deme dynamics) occurs on  slower timescale than its extinction. Similarly, $K$ is referred to as large (i.e. $K\gg1$) where fixation of a site occurs faster than deme extinction.
In our analysis, we
distinguish between different dynamical scenarios through
\begin{equation}
 \label{eq:psi}
\psi(m,K)\equiv mK\tau_E,
 \end{equation}
giving the average number of migration events during the typical deme extinction time. With Eq.~\eqref{eq:tauE}, we have 
$\psi(m,K)\approx me^K$ when $K\gg 1$. 
In the regime where $\psi\gg 1$, many migration
events occur before any deme extinction, and the dynamics is thus dominated by $M/W$ competition. 
When $\psi(m,K)< 1$, migration is ineffective and  
there is fast extinction of all demes. An intermediate regime where some demes are empty and others occupied by $W$ or $M$ arises when  $\psi(m,K)\gtrsim 1$. To rationalise this picture in the coarse-grained description of Ref.~\cite{landeExtinctionTimesFinite1998}, it is useful to track the number of occupied demes $j=0,1,\dots, \Omega$ (by either $W$ or $M$ cells). For cliques, as shown in Appendix~\ref{appendix:site_occupancy}, we find that the long-time fraction 
of occupied demes is
\begin{equation}
\label{eq:Omega}
\frac{j}{\Omega} \to
\frac{\Omega_\text{occ}(m,K)}{\Omega}\approx \begin{cases}
1 & \text{if $\psi \gg 1$}, \\
\frac{\psi-1}{\psi} & \text{if $\psi\gtrsim 1$}, \\
0 &  \text{if $\psi< 1$}.
\end{cases}
\end{equation}
The expression  of $\Omega_\text{occ}$ ignores spatial correlations and hence is not accurate for cycles and grids
if $\psi(m,K)$ is not much larger than 1. 
However, $\psi(m,K)$
allows us to efficiently distinguish between the
regimes dominated by $M/W$ competition ($\psi\gg1$) and deme extinction ($\psi<1$); see  Appendix~\ref{appendix:site_occupancy}. The crossover intermediate regime ($\psi \gtrsim 1$) is discussed in detail in Appendix~\ref{appendix:inter}.

Henceforth, we refer to ``invasion'' when a cell of type $M/W$ migrates to and fixates in a $W/M$ deme, and to ``recolonisation'' when a cell of either type migrates into an empty deme and takes it over; see Fig.~\ref{fig:cartoon-constant}(c,d). 
\edit{The competition-dominated dynamics is characterised by invasions, while the extinction-dominated regime consists of extinctions and some recolonisations prior to the final extinction of the metapopulation.}  
A summary of the timescales for the different processes and regimes considered in this work can  be found  in Tables \ref{tab:timescales} and \ref{tab:regimes}.

\subsection{Competition-dominated dynamics}
\label{sec:constant_env-large_K}
\begin{figure*}
    \centering
    \includegraphics[width=\textwidth]{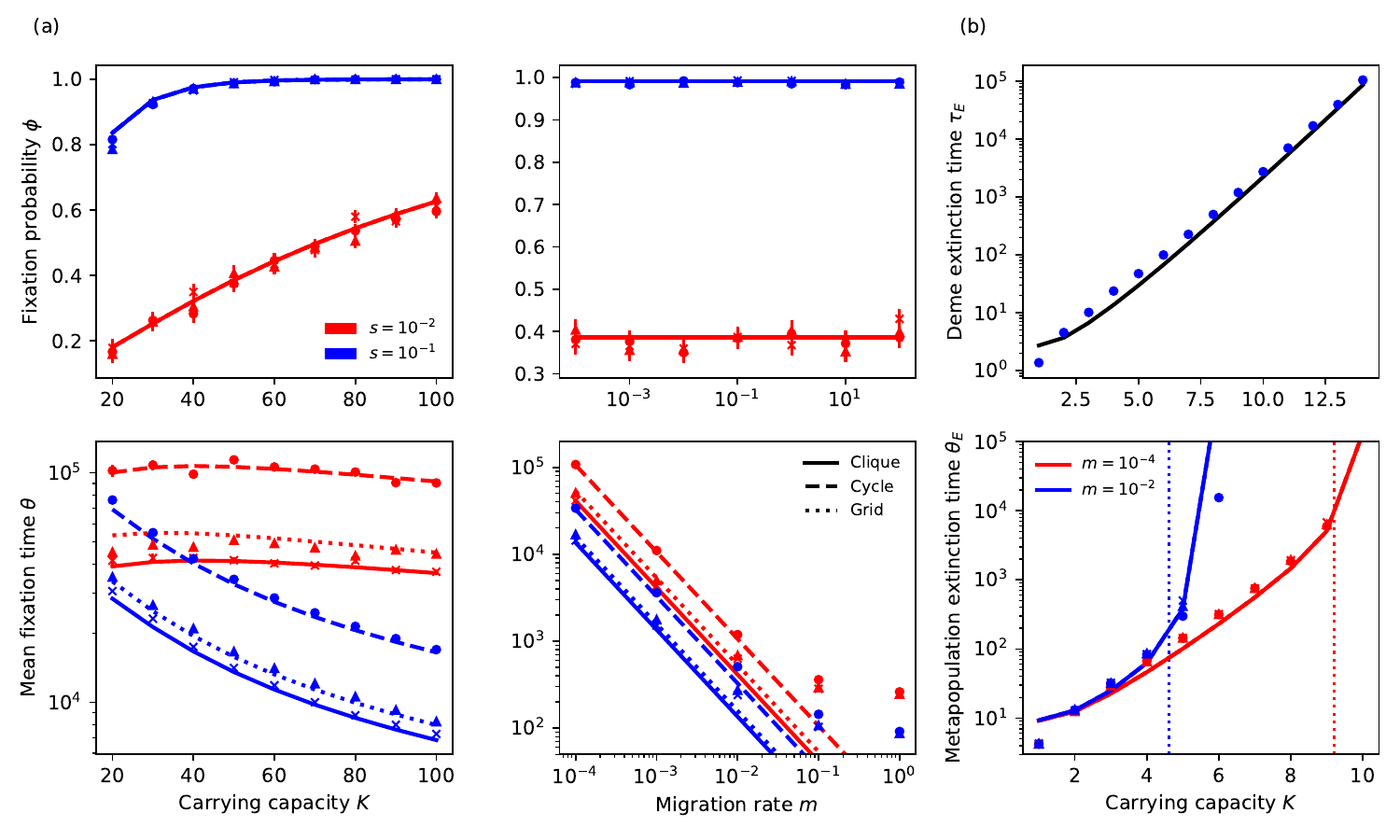}
    \caption{
    (a): \textit{Competition-dominated dynamics}, $\psi\gg 1$. (Top left) $M$ fixation probability  $\phi$ vs. constant carrying capacity $K$; (Bottom left) unconditional mean fixation time $\theta$ vs. $K$; (Top right)  $\phi$ vs. per capita migration rate $m$; (Bottom right) $\theta$ vs. $m$. Markers are simulation results and lines are predictions of Eq.~\eqref{eq:static-fixation} for $s=0.1$ (blue) and $s=0.01$ (red) on a clique (solid lines / crosses), cycle (dashed lines / circles), and grid (dotted lines / triangles). In (left), $m=10^{-4}, \Omega=16$, and in (right), $K=50, \Omega=16$.
    In (top), markers for the same $s$ are almost indistinguishable indicating independence of the spatial structure.
     (b): \textit{Extinction-dominated dynamics}, $\psi<1$. (Top) Mean extinction time of a single deme $\tau_E$ vs. $K$  ($m=0$). Circles are simulation data,  line shows the predictions of Eq.~\eqref{eq:tauE}.
     (Bottom) Metapopulation mean extinction time $\theta_E$ vs. $K$
     for  $\Omega=16$ and $m=10^{-2}$ (blue) and 
      $10^{-4}$ (red). Markers are simulation results and thick lines are predictions of Eq.~\eqref{eq:S9} for cliques (solid lines / crosses), cycles  (dashed lines / circles), and  grids (dotted lines / triangles). Thin dashed vertical lines are guides to the eye showing  $\psi=1$
      for   $m=10^{-2}$ (blue) and 
      $10^{-4}$ (red). Selection plays no role in this regime, so simulation data for (b) has been obtained with $s=0$.
    In  all panels, there is initially one $M$ deme and $\Omega -1$ demes occupied by $W$.
In panels (a,top) and (b,bottom), dashed lines overlap with solid lines and so are not visible. Error bars are plotted in each case but are typically too small to see.
    }
    \label{fig:constant-environment}
\end{figure*}

When $\psi\gg1$ with \edit{$mK< 1$ (slow migration; see Eq.~\eqref{eq:slowmigr})}, the carrying capacity is large enough for  many migrations to occur on the timescale $\tau_E$  of deme extinction, with intra-deme dynamics occurring faster than inter-deme dynamics. Since
every deme expects many incoming cells
in time $\tau_E$,  deme extinction  is unlikely and can be neglected. In this regime,  the dynamics is dominated by local $M/W$ competition: 
$W$ and $M$ cells compete in each deme to fixate the local subpopulation; see Fig.~\ref{fig:cartoon-constant}(b). 

As in Refs.~\cite{marrecUniversalModelSpatially2021,marrec2023,moawadEvolutionCooperationDemestructured2024}, the slow migration condition allows us to adopt a coarse-grained description treating each deme as a single entity of type $W$ or $M$. This is because in the regime of slow migration, 
the mean time for an $M$ or $W$ invader to fixate  a deme is much shorter compared to $1/(mK)$, the expected time between migrations (see Appendix~\ref{appendix:extinction_time}, Table~\ref{tab:timescales} and Eq.~\ref{eq:slowmigr}). When the dynamics is dominated by $W/M$ competition and \edit{$mK<1$}, at a coarse-grained  level, each deme can thus be regarded as being either entirely occupied by $W$ or $M$ individuals; see 
Fig.~\ref{fig:cartoon-constant}(c,e). 
In this regime, each sequential migration is an invasion attempt, with  a cell from an $M/W$ site trying to invade a neighbouring  $W/M$ deme; see Fig.~\ref{fig:cartoon-constant}(c). 
Here, an $M/W$ invasion is the fixation of a single $M/W$ mutant in a deme consisting of $K-1$ cells of type $W/M$.

In the realm of the coarse-grained description, the state of the metapopulation in this regime is denoted by $i$, where  $i=0,1,\dots,\Omega$ is the number of demes of type $M$ leaving $\Omega-i$ demes of type $W$. 
The probability $\rho_{M/W}$ of invasion by an $M/W$ migrant is here given by the probability that a single $M/W$ cell takes over a population of constant size $K$ in a Moran process~\cite{Moran,Ewens,Blythe07,traulsen2009stochastic,antal2006fixation} and, as shown  in Appendix~\ref{appendix:slow_migration}, reads 
\begin{equation}
\label{eq:rhoMW}
\begin{aligned}
    \rho_{M}(K)&=\frac{1}{1+s}\left[\frac{s}{1-(1+s)^{-K}}\right],\\
    \rho_{W}(K)&=\frac{1}{(1+s)^K}\left[\frac{s}{1-(1+s)^{-K}}\right].
\end{aligned}
\end{equation}
In each time unit,
a deme receives from and sends to its neighbours an average of $mK$ cells.
Importantly, only edges connecting $M$ and $W$ demes
can lead to invasions; see Fig.~\ref{fig:cartoon-constant}(c). 
These are ``active edges''
and their  number in state $i$
on graph ${\rm G}$ is denoted by $E_{{\rm G}}(i)$, where here we consider  ${\rm G}=\{{\rm clique, cycle, grid}\}$. Moreover,  migration from  a deme  can occur to any of the $q_{\rm G}$ neighbours of the deme, where 
$q_{\rm clique}=\Omega-1$, $q_{\rm cycle}=2$, $q_{\rm grid}=4$,
and $q_{\rm d-dim}=2d$ for a $d$-dimensional regular lattice. 
The number of active edges generally varies with the metapopulation state and  the spatial structure, and is difficult to determine. However, a clique being the fully-connected graph, the $i$ demes of type $M$ are connected to the $\Omega -i$ demes of type $W$, yielding $E_{{\rm clique}}(i)=i(\Omega-i)$. The $M$ demes of a clique form a single unbreakable cluster since all demes are connected. For a cycle, if
the initial state is $i=1$, the $M$ deme is initially  connected to exactly two $W$ demes. This property is conserved by the coarse-grained dynamics on a cycle, with an unbreakable cluster of $M$ demes always connected to a cluster of $W$ demes by two active edges until $W$ or $M$ fixes the metapopulation, yielding
$E_{{\rm cycle}}(i)=2~\text{~for  $i\neq 0,\Omega$}$; see  Fig.~\ref{fig:cartoon-constant}(a,c) and below.
The unbreakable nature of the $M$ cluster in these two cases (and the symmetric nature of the graphs), means there is only one possible metapopulation state for a given size of $M$ cluster. This allows us to obtain the above explicit expressions for $E_{\rm clique,cycle}(i)$.

The number of active edges in a grid is difficult 
to find because the cluster of demes is not unbreakable on a two-dimensional lattice, and there are many possible metapopulation states for a given number of $M$ demes. However, in Appendix~\ref{appendix:2d-heuristic}, we show that the
average number of active edges on a grid, starting from the metapopulation state $(i,\Omega-i)=(1,\Omega-1)$,
can be approximated by $2\sqrt{\pi i}$. We will therefore approximate 
$E_{{\rm grid}}(i)\approx 2\sqrt{\pi i}$ for  $i\neq 0,\Omega$. 
When $i= 0,\Omega$, one strain fixates the entire metapopulation, where all  demes are $M$ if
$i=\Omega$ and  all demes are $W$ when $i=0$, and hence $E_{{\rm G}}(0)=E_{{\rm G}}(\Omega)=0$.

In the coarse-grained description of the competition-dominated dynamics,
starting from a single $M$ deme ($i=1$), there are
$mKE_{\rm G}(i)/q_{\rm G}$ expected 
migration attempts per unit time to grow the number of $M$ demes by invading neighbouring $W$ demes. Since the probability of an $M$ invasion 
is $\rho_M$, given by Eq.~\eqref{eq:rhoMW}, the number of $M$ demes grows at a rate $mKE_{\rm G}(i)\rho_M/q_{\rm G}$.
Similarly a $W$ invader attempts to increase the number of  $W$ demes by invading $M$ demes, hence reducing the {number of $M$ demes}, at a rate $mKE_{\rm G}(i)\rho_W/q_{\rm G}$. 
In this representation, $M$ and $W$ invasions therefore
act at the interface of $M$ and $W$ demes by increasing or reducing the number $i$ of $M$ demes at respective rates~\cite{marrecUniversalModelSpatially2021,moawadEvolutionCooperationDemestructured2024,fruetSpatialStructureFacilitates2024a}, 
\begin{equation}
\label{eq:Moran_slow}
\begin{aligned}
    T_i^+(m,{\rm G},K)&=mK\frac{E_{\rm G}(i)}{q_{\rm G}}\rho_{M},\\
    T_i^-(m,{\rm G},K)&=mK\frac{E_{\rm G}(i)}{q_{\rm G}}\rho_{W}.
\end{aligned}
\end{equation}
These rates, together with
the fact that the timescale of intra-deme dynamics is $1/s$ (timescale of fixation of a single isolated deme~\cite{Moran,Ewens,Blythe07,traulsen2009stochastic,antal2006fixation}; see Table~\ref{tab:timescales})
can be used to define slow migration in the competition-dominated regime. To this end, we notice that the  growth of the cluster of $i$ mutant demes in time $1/s$ is $T^+_i(m, {\rm G}, K)/s=mK
E_{{\rm G}}(i)\rho_M/(q_{{\rm G}}s)$.\footnote{Here, we use the growth of the $M$ cluster to define slow migration as the mutant has a fitness advantage. Therefore, $\rho_M>\rho_W$ and $T^+_i(m,\mathrm{G},K)>T_i^-(m,\mathrm{G},K)$ in general.}
In the  coarse-grained description,
slow migration is the regime where invasion can be regarded as being instantaneous, with fixation of a successful $M$ invader
occurring before the next invasion. This requires that the average number of successful $M$ \edit{invasions} on the timescale of the intra-deme dynamics is less than one, i.e. $T^+_i(m, {\rm G}, K)/s< 1$. Therefore, the condition for
slow migration here is
\begin{equation}
\label{eq:slowmigr}
    m < \frac{sq_{{\rm G}}}{K\rho_{M}E_{{\rm G}}(i)}\leq\frac{s}{K\rho_{M}},
\end{equation}
where  we have used $q_{\rm G}/E_{\rm G}(i)\leq 1$. 
When $s\ll 1$ and $Ks\gg1$, we have $\rho_M\sim s$~\cite{Moran,Ewens,Blythe07,traulsen2009stochastic,antal2006fixation} and there is slow migration when \edit
{$mK \lesssim 1$, i.e. the expected number of migrants from a deme per unit time is less than one.}
 For typical values used here, e.g. $\Omega=16, K=100$ and $s=0.1$,
 we estimate that there is slow migration if
 $m\lesssim 10^{-2}$, which is in line with the values of $m\in [10^{-5},10^{-2}]$ used in our examples, and explains the deviations reported in Fig.~\ref{fig:constant-environment}(a,bottom right) for larger $m$.
The coarse-grained  competition-dominated dynamics  is thus a birth-death process 
for the number $i$ of $M$ demes, with absorbing boundaries at $i=\Omega$ ($M$ fixation) and  $i=0$ ($W$ fixation); see Appendix~\ref{appendix:circulation}. %
In this representation, the $M$ fixation probability
in a metapopulation of size $\Omega$, spatially structured as a graph ${\rm G}$, consisting initially of $i$ mutant demes is denoted
$\phi_i^{{\rm G}}$, and the  unconditional (i.e. regardless of whether $M$ or $W$ takes over~\cite{Ewens,antal2006fixation,traulsen2009stochastic}) mean fixation time (uMFT) denoted $\theta_i^{{\rm G}}$.
These quantities satisfy the first-step equations~\cite{allenIntroductionStochasticProcesses2010b,Ewens,antal2006fixation,traulsen2009stochastic}
\begin{equation}
 \label{eq:first-phi-theta}
\begin{aligned}
 (T_i^++T_i^-)\phi_{i}^{{\rm G}}&=T_i^+\phi_{i+1}^{{\rm G}}+T_i^-\phi_{i-1}^{{\rm G}},
 \\
 (T_i^++T_i^-)\theta_{i}^{{\rm G}}&=1+T_i^+\theta_{i+1}^{{\rm G}}+T_i^-\theta_{i-1}^{{\rm G}},
\end{aligned}
\end{equation}
for $i=1,\dots, \Omega-1$, with 
 boundary conditions $\phi_0^{{\rm G}}=1- \phi_{\Omega}^{{\rm G}}=0$
and $\theta_0^{{\rm G}}=\theta_{\Omega}^{{\rm G}}=0$.
Eqs.~\eqref{eq:first-phi-theta} can be solved exactly~\cite{antal2006fixation,traulsen2009stochastic} (see also Appendix~\ref{appendix:slow_migration}).
Here, we are chiefly interested in the fixation
of a single initial $M$ deme, $i=1$, 
and simply write $\phi^{{\rm G}}\equiv \phi_1^{{\rm G}}$
and $\theta^{{\rm G}}\equiv \theta_1^{{\rm G}}$,
finding
\begin{equation}
\label{eq:static-fixation}
\begin{aligned}
    \phi^{{\rm G}}(K)&=\phi(K)=\frac{1-\gamma}{1-\gamma^\Omega},\\
    \theta^{{\rm G}}(m,K)&=\frac{1-\gamma}{1-\gamma^\Omega}\sum_{k=1}^{\Omega-1}\sum_{n=1}^k \frac{\gamma^{k-n}}{T^+_n(m,{\rm G},K)},
\end{aligned}
\end{equation}
where $\gamma\equiv T_i^- /T_i^+ = \rho_{W}/\rho_{M}\approx \exp(-Ks)$ is a quantity independent of $m$ and ${\rm G}$. As noted in Refs.~\cite{marrecUniversalModelSpatially2021,abbaraFrequentAsymmetricMigrations2023a,
moawadEvolutionCooperationDemestructured2024}
the fixation probability $\phi^{{\rm G}}=\phi$ is therefore independent of the  migration rate and spatial structure. This remarkable result stems from the graphs considered here being circulations; see Eq.~\eqref{eq:circulation_condition}.
In {\it static environments}, 
a generalised circulation theorem ensures that the fixation probability is independent of  $m$ and ${\rm G}$ for circulation graphs~\cite{liebermanEvolutionaryDynamicsGraphs2005,marrecUniversalModelSpatially2021,moawadEvolutionCooperationDemestructured2024}, a feature displayed in the stochastic simulations of Fig.~\ref{fig:constant-environment}(a,top) 
for the full microscopic model. 
In excellent agreement with simulation data of Fig.~\ref{fig:constant-environment}(a,top), we find that the $M$ fixation probability increases almost exponentially with $Ks$ and approaches $1$ when $Ks\gg 1$, $\phi^{{\rm G}}\approx1$. This stems from 
the invasion of $W$ demes being increasingly likely (and the invasion of $M$ demes exponentially less likely)
when the 
 average  number of migrations ($mK$) increases along with $K$.
 When   $Ks\ll 1$, the
competition is effectively neutral, and in this case  $\phi^{{\rm G}}\approx 1/\Omega$. In good agreement with simulation results of Fig.~\ref{fig:constant-environment}(a,bottom),
 Eq.~\eqref{eq:static-fixation} predicts
 that the uMFT  decreases with the migration rate 
 $\theta^{{\rm G}}\sim 1/m$ and, for given parameters, the uMFT is shortest on cliques,
 while it is larger on cycles than on grids. Intuitively, for higher $m$ and more connected graphs, migrants spread faster leading to quicker invasion and fixation.

\subsection{Extinction-dominated dynamics}
\label{sec:constant_env-small_K}
In the extinction-dominated regime $\psi <1$
with  \edit{$mK< 1$}  (slow migration), we do not expect any deme invasions in a time $\tau_E$,
enabling us to adopt a suitable coarse-grained description. Deme invasions being negligible under slow migration,
the timescale of extinction dynamics is much shorter than that of $M/W$ competition, and site extinction dominates over deme invasion; see Fig.~\ref{fig:cartoon-constant}(e). The dynamics in this  extinction-dominated regime is hence governed by the random extinction of demes, regardless of their type. Following Ref.~\cite{landeExtinctionTimesFinite1998}, in the realm of the coarse-grained
description outlined in Appendix~\ref{appendix:extinction_time}, demes are regarded as being either 
 occupied (by either $W$ or $M$ cells) or empty. Deme
extinction occurs randomly while
empty demes  may be recolonised by  migrations from 
 occupied neighbouring demes, which occur rarely in this regime. Since spatial structure may only influence the dynamics via migrations, the coarse-grained dynamics
 is largely independent of the spatial structure in this regime where \edit{$mK<1$}, and can
be represented  by a birth-death process
 for the number of occupied demes, assuming that the site extinction and recolonisation occur instantaneously \cite{landeExtinctionTimesFinite1998}; see Appendix~\ref{appendix:extinction_time}. 
 In this coarse-grained representation, 
 when all demes are initially occupied ($n=K$), the metapopulation mean extinction time (mMET), denoted by $\theta_E$,
  is given by Eq.~\eqref{eq:S9} in Appendix~\ref{appendix:extinction_time}.
 When $\Omega \gg 1$ and $\psi<1$, the leading contribution to the mMET arises from the term $i=n$ in the innermost sum, yielding 
\begin{equation}
\label{eq:thetaE-approx}
\theta_E(K,\Omega)\approx\tau_E(\ln(\Omega)+\gamma_\text{EM}),
\end{equation}
where $\gamma_\text{EM}\approx 0.577...$ is the Euler-Mascheroni constant.
Eq.~\eqref{eq:thetaE-approx} gives a good approximation of the mMET as reported in Fig.~\ref{fig:extinction-vs-omega}.
As $\psi$ increases with $K$, at the upper-limit of the extinction-dominated limit (where $\psi$ approaches $1$),
 the mMET grows almost exponentially with $K$ and logarithmically with $\Omega$, $\theta_E\approx e^K\ln{(\Omega)}/K$. Eq.~\eqref{eq:S9} thus predicts a rapid growth 
of the  mMET when  $\psi\gtrsim 1$, as shown in Fig.~\ref{fig:constant-environment}(b,bottom)
where simulation results are found to be in good agreement with Eq.~\eqref{eq:S9}.

\edit{
When $\psi \gtrsim 1$, the competition- and extinction-dominated dynamics are separated by an intermediate crossover regime analysed  in detail in Appendix~\ref{appendix:inter}.}

\section{Time-fluctuating environments}
\label{sec:changing_env}
\begin{figure*}
    \centering
    \includegraphics[width=\textwidth]{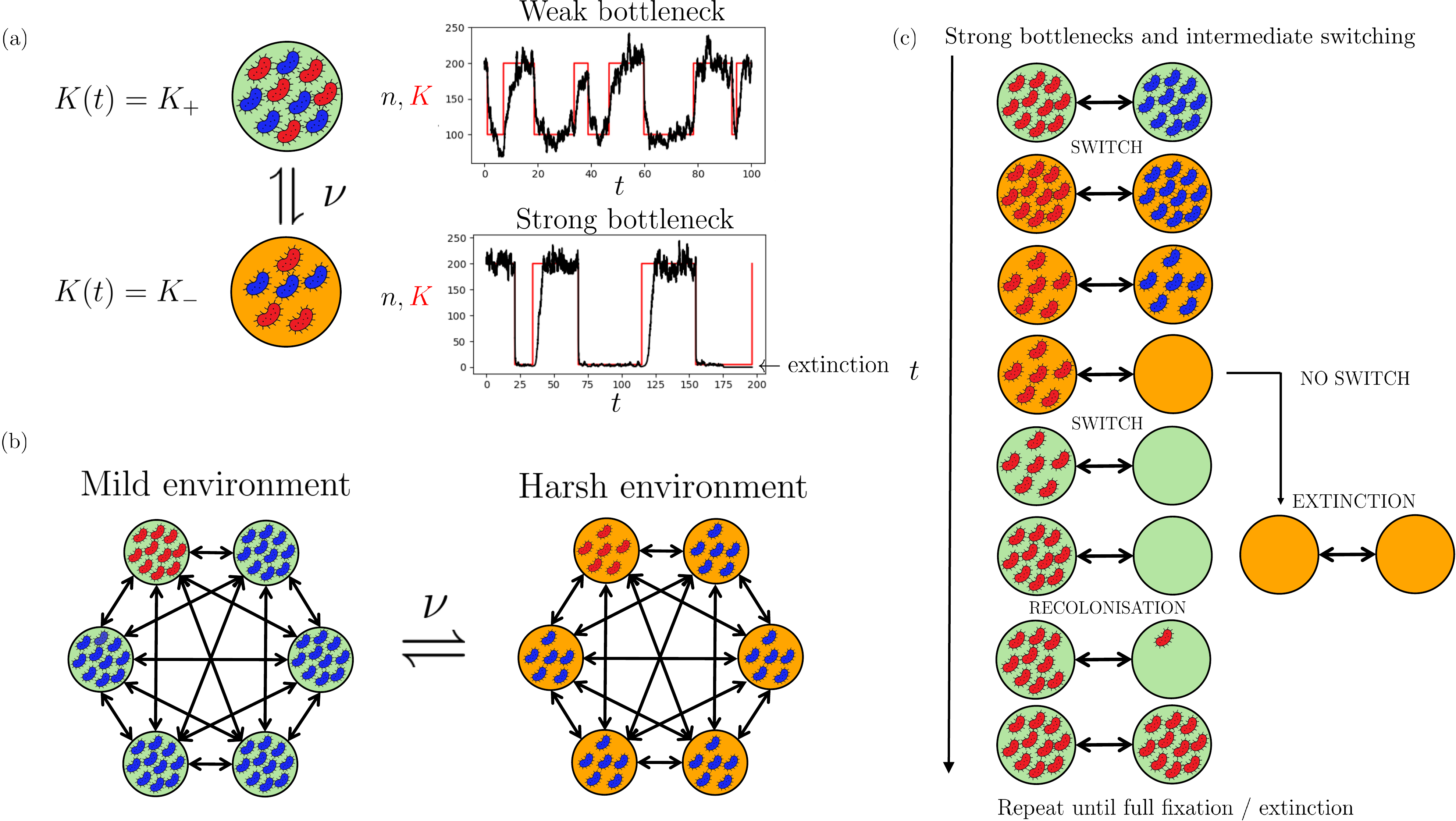}
    \caption{
    (a) Left: single deme in time-switching environment. The carrying capacity $K(t)$ encodes environmental variability by  switching between $K=K_+$ (mild environment, green / light) and $K(t)=K_-<K_+$ (harsh environment, orange / dark) at symmetric rate $\nu$ (see also Appendix~\ref{appendix:bias}). Communities are larger in the mild environment. When $K(t)$ switches at an intermediate rate $\nu\lesssim 1$, each deme experiences bottlenecks prior to deme fixation at an average frequency $\nu/2$; see text. Right: $n$ and $K$ vs. time in the intermediate switching regime where the size $n$ of a deme undergoes bottlenecks. Parameters are: 
    $K_+=200$, $\nu=0.05$ and $K_-=100$ (top) and $K_-=5$ (bottom).
    The bottlenecks are weak when $\psi(m,K_-)\gg1$ (top, right)
    where deme extinction is unlikely. When $\psi(m,K_-)<1$, there are strong bottlenecks and  each deme can go extinct in the harsh environment     (bottom, right). 
    (b) Clique metapopulation with $\Omega=6$ connected demes (double arrows). All demes have the same time-switching carrying capacity $K(t)$ encoding environmental variability across the metapopulation, with each deme in the same environmental state. (c) Example evolution across two nearest-neighbour demes in a switching environment subject to strong bottlenecks in the intermediate switching regime; see text. Starting in the mild environment where $K=K_+$, the carrying capacity switches to $K_-$ (harsh environment) after $t\sim 1/\nu$. Following the $K_+\to K_-$  switch, each deme size decreases and each subpopulation is subject to strong demographic fluctuations and hence prone to extinction. In the absence of recolonisation of empty demes, effective only in the mild state, all demes go extinct. If there is a switch back to the mild environment $K_-\to K_+$ prior to total extinction, empty demes can be rescued by migration and recolonised by incoming $W$ or $M$ cells from neighbouring demes. In the sketch, an empty deme is recolonised by a mutant
    in the mild environment and becomes an $M$ deme.
     The cycle continues until the entire metapopulation consists of only $W$ or $M$ demes, or metapopulation extinction.}
    \label{fig:cartoon-switching}
\end{figure*}
Microbial populations generally evolve in time-varying 
environments, and are often subject to conditions  changing suddenly and drastically, e.g.
experiencing cycles of harsh and mild environmental states~\cite{wittingMicrofluidicSystemCultivation2024,shibasakiExclusionFittestPredicts2021,rodriguez-verdugoRateEnvironmentalFluctuations2019,coatesAntibioticinducedPopulationFluctuations2018,hengge-aronis_survival_1993,morleyEffectsFreezethawStress1983,vasi_long-term_1994,wahlEvaluatingImpactPopulation2002,fux_survival_2005,Brockhurst2007a,acar_stochastic_2008,proft2009microbial,caporaso_moving_2011,himeoka_dynamics_2020,Tu20},  see Fig.~\ref{fig:cartoon-switching}. These variations cause fluctuations often associated with population {\it bottlenecks}, arising when the deme size is drastically reduced, e.g. due to nutrient scarcity or exposure to toxins
\cite{wahlEvaluatingImpactPopulation2002,Brockhurst2007a,Brockhurst2007b,patwaAdaptationRatesLytic2010,mahrt2021bottleneck}.
 Here, environmental variability is encoded in the 
time-fluctuating carrying capacity $K(t)$ of Eq.~\eqref{eq:K(t)}
driven by the DMN $\xi(t)\in \{-1,1\}$~\cite{bena2006,HL06,Ridolfi11,wienandEvolutionFluctuatingPopulation2017,wienandEcoevolutionaryDynamicsPopulation2018,taitelbaumPopulationDynamicsChanging2020a,west2020,shibasakiExclusionFittestPredicts2021,taitelbaum2023evolutionary,hernandez-navarroCoupledEnvironmentalDemographic2023,askerCoexistenceCompetingMicrobial2023,hernandez-navarroEcoevolutionaryDynamicsCooperative2024,LKUM2024}; see Sec.~\ref{sec:model-and-methods_model}. Since the  dynamics of the deme size $n$ occurs on a timescale of order $1$ (see Appendix~\ref{appendix:eco-evolutionary-dynamics-in-a-deme}), the variable  $n$  tracks $K(t)$ and experiences a  
bottleneck whenever the carrying capacity switches from $K_+$ to $K_-$ at a rate  $\nu \lesssim 1$;
see Fig.~\ref{fig:cartoon-switching}(a,right)~\cite{wienandEvolutionFluctuatingPopulation2017,wienandEcoevolutionaryDynamicsPopulation2018,taitelbaumPopulationDynamicsChanging2020a}.

In order to study the joint effect of
migration and  fluctuations on the metapopulation dynamics,
we assume $K_+\gg 1$ such that demographic fluctuations are weak in the mild environment.
In what follows, 
we distinguish between
weak bottlenecks, where $\psi(m,K_-)\gg 1$
and deme extinction is negligible, and 
strong bottlenecks, where $\psi(m,K_-) < 1$ and deme extinctions dominate. 
\subsection{Weak bottlenecks: $\psi(m,K_-)\gg  1$}
\label{sec:changing_env-mild}
\begin{figure*}
    \centering
    \includegraphics[width=\linewidth]{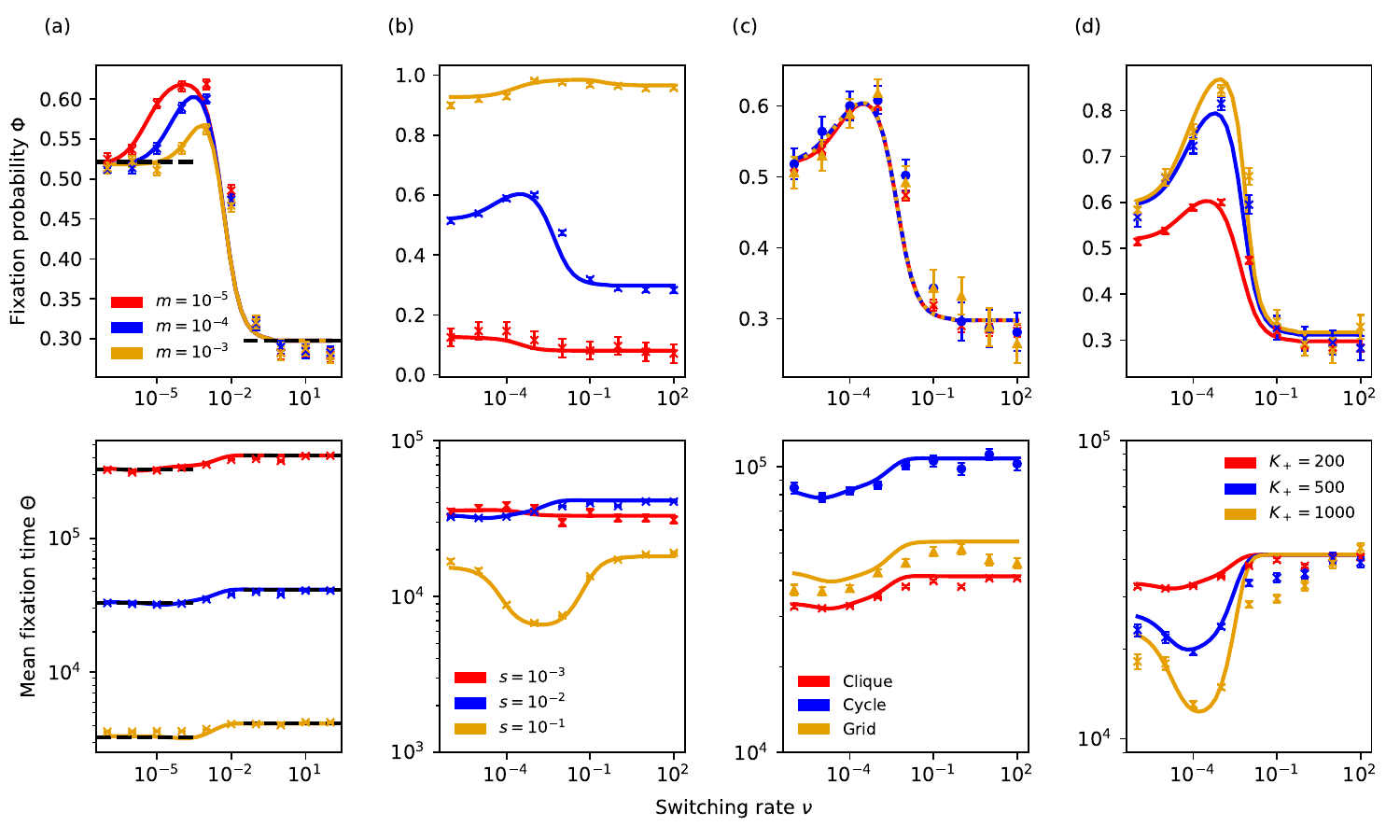}
    \caption{
    Fixation probability $\Phi^{\rm G}$ and mean fixation time $\Theta^{\rm G}$ against switching rate $\nu$ for various parameters. Each panel shows $\Phi^{\rm G}$ vs. $\nu$ (top) and  $\Theta^{\rm G}$ vs. $\nu$ (bottom).
    Markers show simulation results and lines are predictions of 
    Eq.~\eqref{eq:weak-bottleneck-solution}. 
    (a,b) $\Phi^{\rm clique}(\nu)$ and $\Theta^{\rm clique}(\nu)$  for a clique metapopulation and different values of $m$ in (a) and $s$ in (b). (a)
    $m=10^{-5}$ (red), $m=10^{-4}$ (blue), $m=10^{-3}$ (yellow), and $s=0.01$. (b) $s=10^{-3}$ (red), $s=10^{-2}$ (blue), $s=10^{-1}$ (yellow), and $m=10^{-4}$. 
    Dashed black lines are guides to the eye 
    showing $\Phi_{0,\infty}$ in (a,top) and $\Theta_{0,\infty}$ in (a,bottom); see text.
    Other parameters are $\Omega=16$, $K_+=200$, and $K_-=20$. 
    (c)
    $\Phi^{\rm G}(\nu)$ and $\Theta^{\rm G}(\nu)$ for clique (red,  crosses), cycle (blue, circles), and grid (yellow, triangles) metapopulations.  Other parameters are $\Omega=16$, $K_+=200$, $K_-=20$, $s=0.01$, $m=10^{-4}$.
    (d) $\Phi^{\rm clique}(\nu)$ and $\Theta^{\rm clique}(\nu)$  for a clique metapopulation with 
    $K_+=200$ (red), $K_+=500$ (blue), and $K_+=1000$ (yellow). Deviations occur for $\Theta$ with $K_+=1000$ since the slow-migration condition is not satisfied in the mild environment. Other parameters are $\Omega=16$, $K_-=20$, and $s=0.01$, $m=10^{-4}$.  In all examples, there is initially a single $M$ deme and $\Omega -1$ others of type $W$; see text.}
    \label{fig:weak-bottleneck}
\end{figure*}
When $K_+> K_-$ with $\psi(m,K_-)\gg  1$, each deme experiences a weak  bottleneck at an average frequency $\nu/2$
when $\nu\lesssim 1$~\cite{wienandEvolutionFluctuatingPopulation2017,wienandEcoevolutionaryDynamicsPopulation2018,taitelbaumPopulationDynamicsChanging2020a}; see Fig.~\ref{fig:cartoon-switching}(a,top). 
The condition $\psi(m,K_-)\approx me^{K_-}\gg  1$ ensures that deme extinction can be neglected, with  metapopulation dynamics dominated by
 $M/W$ competition. The metapopulation fate can thus 
 be captured by a two-state coarse-grained description
 similar to that of Sec.~\ref{sec:constant_env-large_K}. Since the  deme size in the environment $\xi$, and thus the number of migrating cells, varies with 
 $K(t)$,  it is  useful to introduce the long-time average deme size in environmental state $\xi=\pm 1$ under switching rate $\nu$ denoted by
 $\mathcal{N}_{\xi}(\nu)\equiv \mathcal{N}_{\pm}(\nu)$; see below.

We first discuss the metapopulation fate
 in  the limit of slow and fast environmental switching, and then return to the above case of weak bottlenecks with $\nu\lesssim 1$.
  
 When the environment varies very slowly, $\nu \ll 1$, 
 the carrying capacity remains at its initial value, i.e. $K(t)=K_+$ or $K(t)=K_-$ each with a probability $1/2$, until  invasions lead to the fixation of $W$ or $M$. In other words, the time between switches is longer than the mean fixation time in each environment such that $\nu\max(\theta^{\rm G}(m,K_+), \theta^{\rm G}(m,K_-))<1$, where $\theta^{\rm G}(m,K)$ is given by Eq.~\eqref{eq:static-fixation}. In the slow switching regime, the $M$ fixation probability and uMFT 
 on a metapopulation spatially arranged as a regular graph ${\rm G}$
 are respectively denoted by $\Phi_0^{\rm G}$ and $\Theta_0^{\rm G}$. The quantities are obtained
 by averaging their static counterparts~\eqref{eq:static-fixation} 
 over the stationary distribution of $K$, yielding for symmetric switching
 \begin{equation}
\label{eq:slow-fix}
\begin{aligned}
    &\Phi_0^{\rm G}(m,K_{\pm})=\Phi_{0}(K_{\pm})=\frac{1}{2}\left[\phi(K_+)+\phi(K_-)\right],\\
    &\Theta_0^{\rm G}(m,K_{\pm})=\frac{1}{2}\left[\theta^{{\rm G}}(m, K_+)+\theta^{{\rm G}}(m,K_-)\right].
\end{aligned}
\end{equation}
When the environment varies very quickly, $\nu \gg 1$, 
 the DMN self averages before 
 invasion-mediated fixation occurs,  
 and the 
 carrying capacity of each deme rapidly reaches the effective value
 \begin{equation}
  \label{eq:curlyK}
  {\cal K}\equiv \frac{2K_+K_-}{K_++K_-},
 \end{equation}
 the harmonic mean of $K_+$ and $K_-$, with ${\cal K}\approx 2K_-$ if $K_-\gg 1$ and $\mathcal{N}_{\pm}(\infty)\to {\cal K}$ when ${\cal K}\gg 1$~\cite{wienandEvolutionFluctuatingPopulation2017,wienandEcoevolutionaryDynamicsPopulation2018,taitelbaumPopulationDynamicsChanging2020a,askerCoexistenceCompetingMicrobial2023}; see Appendix~\ref{appendix:eco-evolutionary-dynamics-in-a-deme}. In this fast switching regime, the $M$ fixation probability and uMFT on a metapopulation spatially arranged as a regular graph ${\rm G}$,  respectively denoted by $\Phi_\infty^{\rm G}$ and $\Theta_\infty^{\rm G}$, are obtained 
by replacing $K$ with ${\cal K}$ in Eq.~\eqref{eq:static-fixation}, yielding
 \begin{equation}
\label{eq:fast-fix}
\begin{aligned}
    &\Phi_\infty^{\rm G}(m,{\cal K})=\Phi_\infty({\cal K})=\phi({\cal K}),\\
    &\Theta_\infty^{\rm G}(m,K_{\pm})=\theta^{{\rm G}}(m, {\cal K}).
\end{aligned}
\end{equation}
From these expressions and  Eq.~\eqref{eq:static-fixation}, we notice 
 the fixation probability 
 in the regime of slow and fast switching is independent
 of the 
migration rate and spatial 
structure: $\Phi_0^{\rm G}=\Phi_0$ and $\Phi_\infty^{\rm G}=\Phi_\infty$. However, the metapopulation uMFT 
depends explicitly 
on the migration rate $m$ and  ${\rm G}$, with $\Theta_0^{\rm G} \sim 1/m$ and  $\Theta_\infty^{\rm G}\sim 1/m$.

Under intermediate switching rate, when   $\nu\lesssim 1$,
the coupling of demographic and environmental fluctuations plays a key role, while cell migration depends on the deme size that in turn varies with the environmental state. Here, the metapopulation dynamics cannot be directly related to its static counterpart. The average deme size
$\mathcal{N}_{\xi}(\nu)$ depends non-trivially on  $\nu$ and  $\xi$, and generally needs  to be computed by sampling long-time simulations.
However, analytical progress can be made by  approximating the   distribution of the size $n$ of an isolated deme in the environmental state $\xi$
by the joint probability density
$p_{\xi}(n;\nu)$ 
of the  piecewise deterministic Markov process (PDMP), where $n$ and $\xi$ are variables, and $\nu$ a parameter, obtained by 
ignoring demographic fluctuations~\cite{davisPiecewiseDeterministicMarkovProcesses1984,hufton2016,wienandEvolutionFluctuatingPopulation2017,wienandEcoevolutionaryDynamicsPopulation2018} (see Appendix~\ref{appendix:eco-evolutionary-dynamics-in-a-deme}): 
\begin{equation}
\label{eq:N_PDMPs}
\begin{aligned}
p_{\xi}(n;\nu)=\begin{cases}
                \frac{\mathcal{Z}_+}{n^2} \left(\frac{K_+ - n}{n}\right)^{\nu-1} \left(\frac{n-K_-}{n}\right)^{\nu} \quad \text{if $\xi=+1$,}\\
                \frac{\mathcal{Z}_-}{n^2} \left(\frac{K_+ - n}{n}\right)^{\nu} \left(\frac{n-K_-}{n}\right)^{\nu-1}
                \quad \text{if $\xi=-1$}.
               \end{cases}
\end{aligned}
\end{equation}
The density $p_{\xi}(n;\nu)$ has support $[K_-,K_+]$, and the normalisation
constants $\mathcal{Z}_\pm$ ensure  $\int_{K_-}^{K_+} p_{\xi}(n;\nu)~\text{d}n=1$. 
The $M/W$ competition characterising the intermediate switching regime
dynamics can be described by the coarse-grained representation
of Sec.~\ref{sec:constant_env-large_K} generalised to a time-fluctuating environment following  Refs.~\cite{wienandEvolutionFluctuatingPopulation2017,wienandEcoevolutionaryDynamicsPopulation2018,askerCoexistenceCompetingMicrobial2023}.
Here, we analyse  the  influence of $\nu$ and $m$ on the
$M$ fixation probability,  $\Phi_i^{\rm G}(\nu,m)$, and uMFT, $\Theta_i^{\rm G}(\nu,m)$,   in a metapopulation consisting of $i$ mutants demes and $\Omega -i$ $W$-demes spatially arranged as a regular graph ${\rm G}$. To this end, 
we consider a birth-death process
for the number $i=0,\dots,\Omega$ of $M$ demes. As in  Sec.~\ref{sec:constant_env-large_K},
we assume that there is initially a single $M$ deme ($i=1$). The 
 effective rates, denoted by $\mathcal{T}^{\pm}_{i,\xi}$, for the increase or decrease by one
 of the number $i$ of $M$ demes in the environmental state $\xi$
 depend on the expected number of  migrating
 cells, which in turn depends on the deme size that is now a time-fluctuating quantity driven by Eq.~\eqref{eq:K(t)}.
 In a time-varying environment, the expected number of migrants from a deme, $mn$, is approximated by $m\mathcal{N}_{\xi}(\nu)$, where the 
 the long-time mean deme size in the environmental state $\xi$ is
 obtained from   the PDMP density according to
\begin{equation}
\label{eq:average_Ns}
    \mathcal{N}_{\xi}(\nu)= \int_{K_-}^{K_+}n p_{\xi}(n; \nu/s)~\text{d}n,
\end{equation}
 where, as in Refs.~\cite{wienandEvolutionFluctuatingPopulation2017,wienandEcoevolutionaryDynamicsPopulation2018,taitelbaumPopulationDynamicsChanging2020a,taitelbaum2023evolutionary,west2020},  the switching rate has been rescaled, $\nu \to \nu/s$,
by the timescale of the deme fixation dynamics (see Appendix~\ref{appendix:eco-evolutionary-dynamics-in-a-deme}) where there are an average of ${\cal O}(\nu/s)$  switches on the  deme fixation timescale~\cite{wienandEcoevolutionaryDynamicsPopulation2018,taitelbaumPopulationDynamicsChanging2020a}. The (marginal) average deme size regardless of $\xi$ is given by 
$\mathcal{N}(\nu)=\frac{1}{2}\sum_{\xi}\mathcal{N}_{\xi}(\nu)$ and known to be a decreasing function of $\nu$~\cite{wienandEvolutionFluctuatingPopulation2017,wienandEcoevolutionaryDynamicsPopulation2018}. As in static environments (see Eq.~\eqref{eq:Moran_slow}) the transition rates 
$ \mathcal{T}^{\pm}_i$  depend 
 on the spatial structure, via  $E_{{\rm G}}(i)/q_{{\rm G}}$, 
 and on the probability $\rho_{M/W,\xi}(\nu)$
 that
an $M/W$ migrant invades a $W/M$ deme in the environment $\xi$.
Putting everything together, this yields the effective transition rates
\begin{equation}
\label{eq:effectT}
    \begin{aligned}
    \mathcal{T}^{+}_{i,\xi}(\nu,m,{\rm G})=m\mathcal{N}_{\xi}(\nu) \frac{E_{{\rm G}}(i)}{q_{\rm G}}\rho_{M,\xi}(\nu),\\
    \mathcal{T}^{-}_{i,\xi}(\nu,m,{\rm G})=m\mathcal{N}_{\xi}(\nu) \frac{E_{{\rm G}}(i)}{q_{\rm G}}\rho_{W,\xi}(\nu),
\end{aligned}
\end{equation}
 where, by analogy with Eq.~\eqref{eq:rhoMW}, we have introduced
 \begin{equation}
 \label{eq:rhoMWweak}
\begin{aligned}
   \rho_{M,\xi}(\nu)&\equiv\frac{s}{1+s}\frac{1}{1-(1+s)^{-\mathcal{N}_{\xi}(\nu)}},\\
    \rho_{W,\xi}(\nu)&\equiv \frac{s}{(1+s)^{\mathcal{N}_{\xi}(\nu)}}\frac{1}{1-(1+s)^{-\mathcal{N}_{\xi}(\nu)}}.
\end{aligned}
\end{equation}
With Eq.~\eqref{eq:effectT}, by  dropping all explicit dependence except on $i$ and $\xi$, we obtain 
 the $M$ fixation probability starting from the environmental state $\xi$ with $i$ mutant demes on a graph $\rm G$, denoted by
 $\Phi_{i,\xi}^{{\rm G}}(\nu,m)$, as the solution of
 the $\nu$-dependent first-step analysis equation
\begin{equation}
\label{eq:firststepphi}
\left[\mathcal{T}^{+}_{i,\xi} + \mathcal{T}^{-}_{i,\xi}+\nu\right]\Phi_{i,\xi}^{{\rm G}}
=
\mathcal{T}^{+}_{i,\xi} \Phi_{i+1,\xi}^{{\rm G}}+
\mathcal{T}^{-}_{i,\xi}\Phi_{i-1,\xi}^{{\rm G}}
+\nu\Phi_{i,-\xi}^{{\rm G}},
\end{equation}
subject to the boundary conditions $\Phi_{0,\xi}^{{\rm G}}= 0$ and $\Phi_{\Omega,\xi}^{{\rm G}}= 1$. 
The metapopulation uMFT starting from the same initial conditions, denoted by $\Theta_{i,\xi}^{{\rm G}}$, similarly 
satisfies
\begin{equation}
\label{eq:firststeptau}
\left[\mathcal{T}^{+}_{i,\xi} + \mathcal{T}^{-}_{i,\xi}+\nu\right]\Theta_{i,\xi}^{{\rm G}}= 1+
\mathcal{T}^{+}_{i,\xi}\Theta_{i+1,\xi}^{{\rm G}}+
\mathcal{T}^{-}_{i,\xi}\Theta_{i-1,\xi}^{{\rm G}} +\nu\Theta_{i,-\xi}^{{\rm G}},
\end{equation}
with boundary conditions $\Theta_{0,\xi}^{{\rm G}} = \Theta_{\Omega,\xi}^{{\rm G}}= 0$. 
Eqs.~\eqref{eq:firststepphi} and \eqref{eq:firststeptau} 
generalise Eqs.~\eqref{eq:first-phi-theta}  to a time-switching environment, with the last terms on the RHS accounting for environmental switching, and coupling  $\Phi_{i,\xi}^{{\rm G}} $ to  $\Phi_{i,-\xi}^{{\rm G}}$ and $\Theta^{{\rm G}}_{i,\xi}$ to  $\Theta^{{\rm G}}_{i,-\xi}$.  Eqs.~\eqref{eq:firststepphi} and~\eqref{eq:firststeptau} can be solved numerically using standard methods. The  $M$ fixation probability $\Phi_i^{{\rm G}}(\nu)$ and uMFT $\Theta_ i^{{\rm G}}(\nu)$ regardless of $\xi$ are obtained by
averaging  over the stationary distribution of $\xi$, yielding
\begin{equation}
\begin{aligned}
\Phi_i^{\rm G}(\nu,m)&= \frac{1}{2}\sum_{\xi}\Phi_{i,\xi}^{{\rm G}}(\nu,m), \\
\Theta_ i^{\rm G}(\nu,m)&=\frac{1}{2}\sum_{\xi}\Theta_{i,\xi}^{{\rm G}}(\nu,m),
\end{aligned}
\label{eq:marginalisedsolution}
\end{equation}
 where we have reinstated the explicit dependence on $\nu$ and $m$.
As we specifically consider the initial condition of a single 
$M$ deme, we  set $i=1$ in Eq.~\eqref{eq:marginalisedsolution} and simplify the notation by writing 
\begin{equation}
\label{eq:weak-bottleneck-solution}
    \Phi^{\rm G}(\nu,m) = \Phi_1^{\rm G}(\nu,m) \text{ and } \Theta^{\rm G}(\nu,m)=\Theta_1^{\rm G}(\nu,m).
\end{equation}
Eq.~\eqref{eq:weak-bottleneck-solution} are the expressions of the $M$ fixation probability and metapopulation  uMFT in the realm of the combined coarse-grained  and PDMP description. This approach is valid under slow migration rate (\edit{$mK<1$}) and weak selection strength ($s\ll 1$) for the assumption  
$n\approx \mathcal{N}_{\xi}(\nu)$ to hold  at each  invasion; see Appendix~\ref{appendix:eco-evolutionary-dynamics-in-a-deme}. 
In Fig.~\ref{fig:weak-bottleneck}, the comparison  of the predictions of Eq.~\eqref{eq:weak-bottleneck-solution} 
with the simulation results of the full model on the regular graphs ${\rm G}=\{{\rm clique, cycle, grid}\}$ 
shows that Eq.~\eqref{eq:weak-bottleneck-solution}  captures well the 
dependence of  $\Phi^{\rm G}$ and $\Theta^{\rm G}$
on $\nu$, $m$, $s$ and $K_+$. In particular, Eq.~\eqref{eq:weak-bottleneck-solution}
reproduces on all ${\rm G}$ the non-monotonic $\nu$-dependence of  $\Phi^{\rm G}$ and $\Theta^{\rm G}$ (when it exhibits this feature),  as well as their behaviour when $\nu \to 0,\infty$ given by Eqs.~\eqref{eq:slow-fix} and \eqref{eq:fast-fix}. 

A striking feature of  $\Phi^{\rm G}$ and $\Theta^{\rm G}$ is their 
 dependence on spatial migration. In Fig.~\ref{fig:weak-bottleneck}(a,top), we indeed find that simulation data for $\Phi^{\rm clique}(\nu,m)$
vary noticeably with $m$ in the range $\nu\in [10^{-4},10^{-1}]$. These deviations, of up  to $20\%$,  exceed the error bars and are reasonably well captured by  Eq.~\eqref{eq:weak-bottleneck-solution}.
In Fig.~\ref{fig:weak-bottleneck}(c,top), we notice that both simulation results and  predictions of  Eq.~\eqref{eq:weak-bottleneck-solution} for $\Phi^{\rm G}(\nu,m)$ differ  slightly for each graph ${\rm G}$, whereas Fig.~\ref{fig:weak-bottleneck}(c,bottom) shows that $\Theta^{\rm G}$ clearly depends on the spatial structure. The explicit dependence of the fixation probability on  migration and spatial structure is in stark contrast 
with the result Eq.~\eqref{eq:static-fixation} obtained in static environments,  and is therefore a signature of the eco-evolutionary dynamics in time-fluctuating environments. As shown in Appendix~\ref{appendix:circulation}, the correspondence demonstrated in Ref.~\cite{marrecUniversalModelSpatially2021} between $\Phi^{\rm G}$ and the  fixation probability
of a  random walk for the number $i=0,1,\dots, \Omega$ of mutant demes with
hopping probabilities independent of $m$
and  ${\rm G}$, and 
absorbing states $0, \Omega$,  breaks down in    time-varying environments.
This leads to the dependence  of $\Phi^{\rm G}$ and $\Theta^{\rm G}$ on $m$ and ${\rm G}$ in time-switching environments. 

Another distinctive feature of  $\Phi^{\rm G}$ and $\Theta^{\rm G}$ is their 
  non-monotonic $\nu$-dependence  when the other parameters ($s$, $m$, $K_{\pm}$, $\Omega$) are kept fixed. In particular, 
Fig.~\ref{fig:weak-bottleneck} shows that  
 $\Phi^{\rm G}$ may exhibit a sharp peak
 in the regime of intermediate  $\nu\in [10^{-4},10^{-1}]$ that is well 
 captured by Eq.~\eqref{eq:weak-bottleneck-solution}.
  These results are in marked contrast with their counterparts  in a single deme, which vary monotonically with $\nu$~\cite{wienandEvolutionFluctuatingPopulation2017,wienandEcoevolutionaryDynamicsPopulation2018,taitelbaum2023evolutionary}. 
 The non-monotonic $\nu$-dependence of $\Phi^{\rm G}$ and $\Theta^{\rm G}$ is therefore an inherent effect of spatial migration. Intuitively, this behaviour stems, on the one hand, from  more $M$ invasions
occurring when the deme size is as close as possible to $n\approx K_+$; see Eq.~\eqref{eq:effectT}. On the other hand, the average deme size is a decreasing function of $\nu$~\cite{wienandEvolutionFluctuatingPopulation2017,wienandEcoevolutionaryDynamicsPopulation2018,taitelbaumPopulationDynamicsChanging2020a}; see Appendix~\ref{appendix:eco-evolutionary-dynamics-in-a-deme}. 
Therefore, optimising the probability of $M$ fixation requires two considerations:
the environment should avoid remaining stuck in the harsh environment for too long, which can happen with a probability close to $1/2$ when $\nu\ll 1$; and the environment should not switch too frequently (i.e. $\nu \gg 1$), as this reduces the effective deme size ($n\approx {\cal K}$), which can be significantly smaller than $K_+$.
Hence, the best conditions for the fixation of $M$ are for a range of $\nu$ in  the intermediate regime. Since the uMFT is longer in the harsh than in the mild environment (where there are fewer migration events; see Fig.~\ref{fig:constant-environment}(a,bottom left)), a similar reasoning leads to a minimum mean fixation time for $\nu$ in the intermediate regime; see Fig.~\ref{fig:weak-bottleneck}(b,d). 

\subsection{Strong bottlenecks: $\psi(m,K_-)<1$}
\label{sec:changing_env-harsh}
\begin{figure*}
    \centering
    \includegraphics[width=\textwidth]{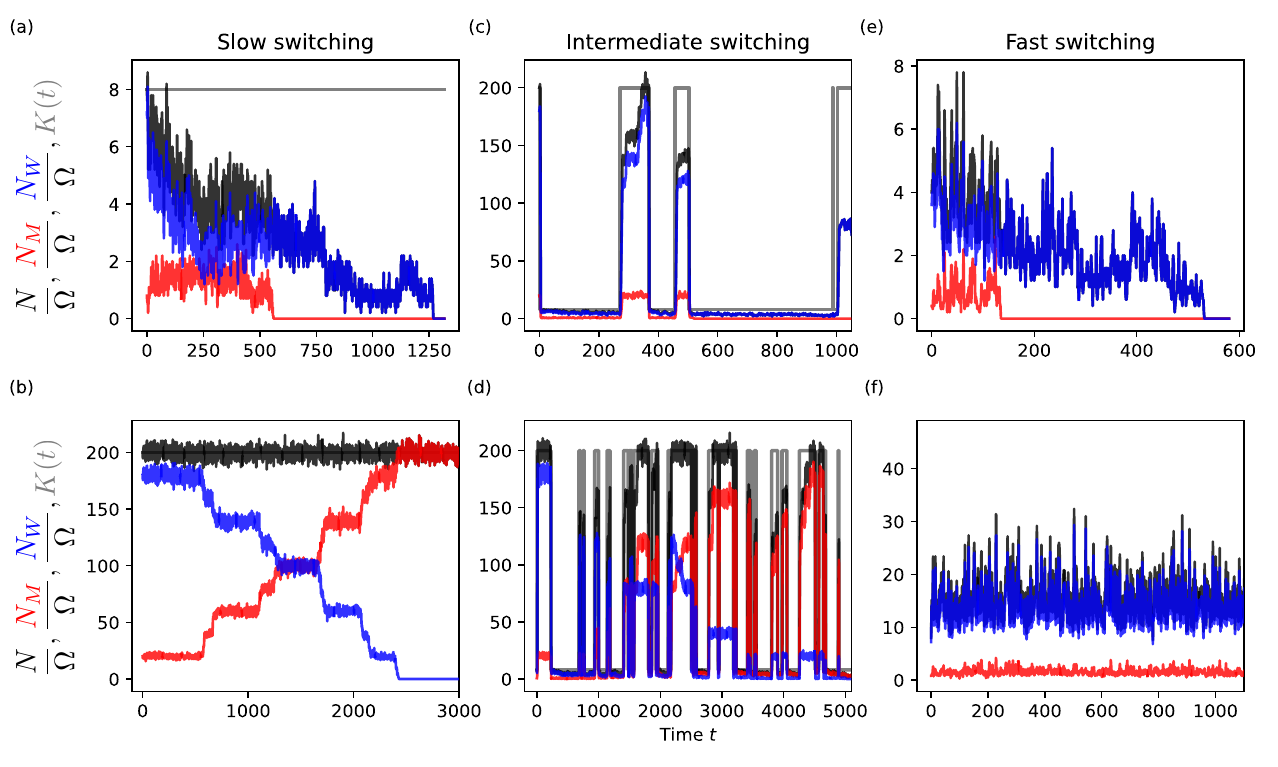}
    \caption{
    Typical single realisations of $N/\Omega$ (black), $N_{M}/\Omega$ (red), $N_{W}/\Omega$ (blue), and $K(t)$ (grey) against time for different values of $K_-$ and $\nu$.
    (a,b): Here, $\nu=10^{-4}$ and $K_-=8$. In (a), $K=K_-$ at $t=0$ and $M$ and then $W$ quickly go extinct. In (b),  $K=K_+$ at $t=0$ and $M$ fixes the population while $W$ goes extinct. (c,d): Here,
    $\nu=10^{-2}$ and $K_-=8$. In (c), mutants survive the first few bottlenecks but their abundance is low leading to the fixation of $W$ and removal of $M$ after four bottlenecks ($t\gtrsim 1000$).
    In (d), mutants survive the first bottlenecks and spread in the mild state where they recolonise and invade demes. They are eventually able to fix the population. (e,f): Here,
    $\nu=10$, and $K_-=4$ in (e) and $K_-=10$ in (f). $K(t)$ switches very frequently and is not shown for clarity. In (e), the deme size is $n\approx 2K_-=8$ and the dynamics is dominated by deme extinction leading to the rapid extinction of the metapopulation. In (e), the deme size is $n\approx 2K_-=20$ and there is $M/W$ competition that leads to 
    fixation of $M$ and extinction of $W$
    after a typical time $t\sim \theta^{\rm clique}(2K_-)\gtrsim 10^4$ (not shown). Similar results are obtained on other regular graphs ${\rm G}$; see text.
    Other parameters are $\Omega=10$, $s=0.1$, $m=10^{-4}$, and $K_+=200$. In all panels, initially there is a single $M$ deme and $\Omega -1$ demes occupied by $W$.  
    }
    \label{fig:clique-strong-bottleneck-trajectories}
\end{figure*}
\begin{figure*}
    \centering
    \includegraphics[width=\textwidth]{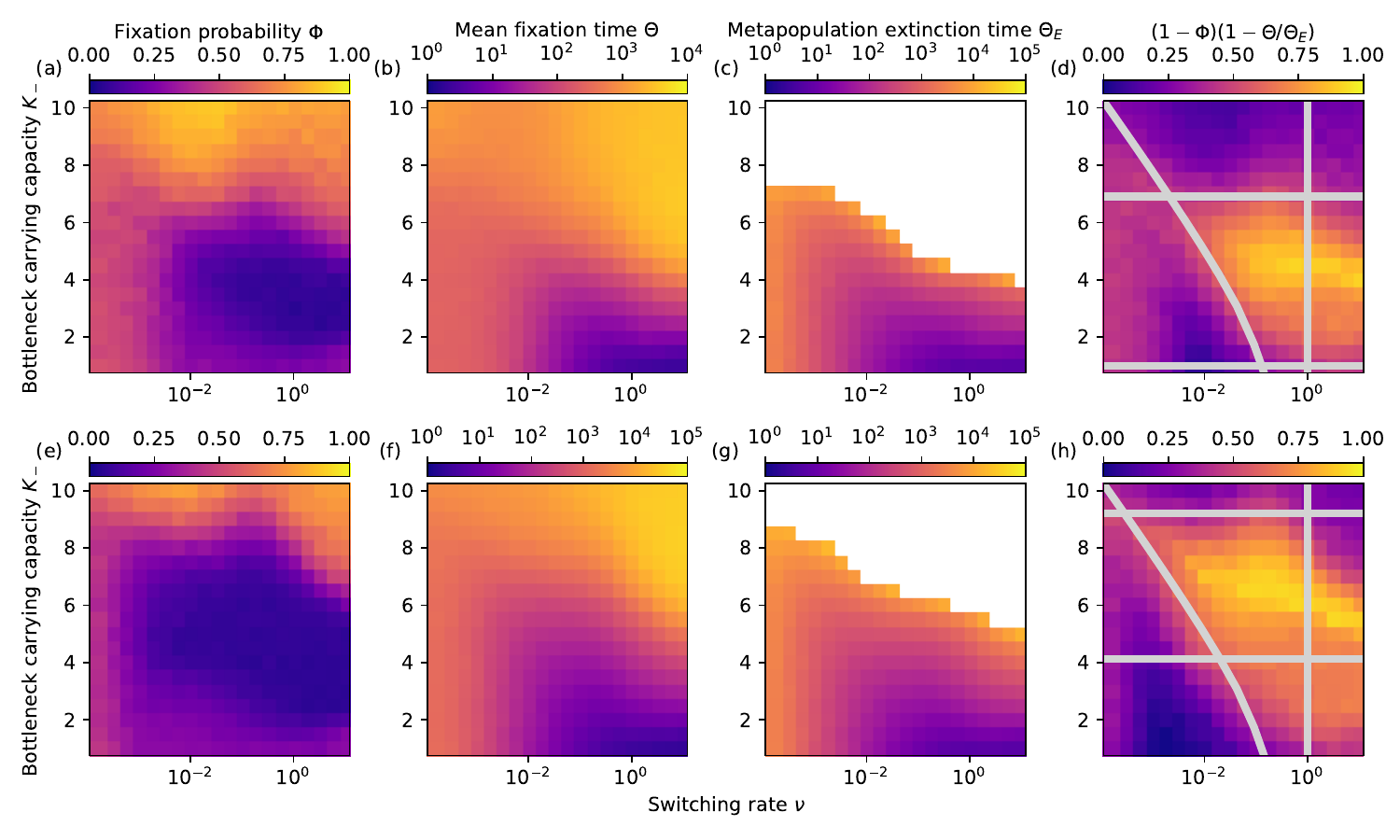}
    \caption{Near-optimal condition for the idealised treatment strategy. $(\nu,K_-)$ heatmaps of 
    $\Phi$, $\Theta$, $\Theta_E$ and $(1-\Phi)(1-\Theta/\Theta_E)$ 
   for a  clique metapopulation. The migration rate is $m=10^{-3}$ in (a-d) and $m=10^{-4}$ in (e-h). 
   White space in panels (c) and (g) indicates where at least one realisation for those parameters did not reach extinction by $t=10^5$, i.e. $\Theta_E\sim10^5$ or larger. 
   Grey lines in panels (d) and (h) are show the near-optimal conditions for the idealised treatment strategy, given by Eq.~\eqref{eq:near-opt}:  $\psi(m,K_-)<1$ below the top horizontal line,   $mK_+\theta_E>1$ above the bottom horizontal line, and $\nu\theta_E>1$
   above  the curved line, while the vertical line indicates where $\nu<1$. Here, $\theta_E$ is obtained from Eq.~\eqref{eq:thetaE-approx}.
   The near-optimal treatment conditions is the yellowish cloud at the centre of the area enclosed by these lines.   Similar results are obtained on other regular graphs ${\rm G}$; see text and Fig.~\ref{fig:other-strong-bottleneck}.
   Other parameters $\Omega=16$, $s=0.1$, and $K_+=200$. In all panels, initially there is a single 
   $M$ deme and $\Omega -1$ demes occupied by $W$.
}
    \label{fig:clique-strong-bottleneck}
\end{figure*}
When $\psi(m,K_-)< 1$, each deme can undergo strong bottlenecks; see Fig.~\ref{fig:cartoon-switching}(a) and below.
In the harsh environment $\xi=-1$ ($K=K_-$), the entire metapopulation 
experiences extinction,
in an observable time $\theta_E(K_-,\Omega)$, denoted $\theta_E$ here for conciseness. However, in the mild state $\xi=+1$ ($K=K_+$), deme extinction can be neglected and each site can be regarded as being occupied by either $W$ or $M$ cells. The dynamics in the harsh state is thus dominated by extinction, whereas the $M/W$ competition characterising the mild state 
is aptly captured by the  two-state coarse-grained
description of Section~\ref{sec:constant_env-large_K}. Environmental switching thus couples regimes that are dominated in turn by
deme extinction  and  $M/W$ competition, yielding complex dynamical scenarios  whose analysis is difficult. However, 
we can gain valuable insight  by considering first the limits $\nu\ll1$, $\nu\gg 1$, and then the case of intermediate switching where $\nu\lesssim1$.

 When the environment varies very slowly, $\nu\ll 1$, 
 the $K(t)$ remains at its initial value for long periods, that is $K=K_\pm$ if $\xi(0)=\pm1$ each with a probability $1/2$. On the one hand, if initially $\xi=-1$ (harsh environment), $K=K_-$
 and each deme is prone to extinction after a mean time $\tau_E(K_-)$,
which eventually leads to the  collapse of the metapopulation after a mMET $\theta_E\approx e^{K_-}\ln{(\Omega)}/K_-$ when $\Omega \gg 1$ and $K_-\gg 1$; see Eqs.~\eqref{eq:tauE} and \eqref{eq:thetaE-approx} and Fig.~\ref{fig:clique-strong-bottleneck-trajectories}(a). 
 On the other hand, if  initially $\xi=+1$ (mild condition), $n\approx K_+$ and there is $M/W$ competition characterised by the 
 fixation of $M$ with a probability $\phi(K_+)$ approaching 1 when $K_+s\gg 1$; see Eq.~\eqref{eq:static-fixation} and Fig.~\ref{fig:clique-strong-bottleneck-trajectories}(b). As a result, when $\nu \ll1$ and $K_+s\gg 1$, there are  two equally likely outcomes illustrated in  Fig.~\ref{fig:clique-strong-bottleneck-trajectories}: either the extinction of the metapopulation in a mean time $\theta_E$ as in Fig.~\ref{fig:clique-strong-bottleneck-trajectories}(a), or the fixation of $M$ after a mean time $\theta^{\rm G}(K_+)$ as in Fig.~\ref{fig:clique-strong-bottleneck-trajectories}(b).
 
 In frequently varying environments, when $\nu\gg 1$, 
 the size of each deme readily settles about the effective carrying capacity  Eq.~\eqref{eq:curlyK}, with $n\approx {\cal K}$ (when ${\cal K}\gg 1$) after $t\sim 1$~\cite{wienandEvolutionFluctuatingPopulation2017,wienandEcoevolutionaryDynamicsPopulation2018}. Since ${\cal K}\approx 2K_-$ when $K_+\gg K_-$, if $\psi(m,2K_-)<1$  the dynamics is characterised by the extinction of individual demes 
 and then of the entire metapopulation after  mean times  $\tau_E(2K_-)$ and  $\theta_E(\Omega,2K_-)$; see Fig.~\ref{fig:clique-strong-bottleneck-trajectories}(e). However, if $\psi(m,2K_-)\gg 1$ and  $2K_-s\gg 1$, 
 the dynamics is characterised by  $M/W$ competition with $M$ most likely to fix the metapopulation after a mean time $\theta^{\rm G}(2K_-)$, as illustrated by  Fig.~\ref{fig:clique-strong-bottleneck-trajectories}(f).

In slowly and rapidly changing environments, regardless of the spatial structure, the metapopulation subject to strong bottlenecks is therefore always at risk of either  complete extinction or
of being taken over by mutants. 
 
In the intermediate switching regime, $\nu \lesssim 1$ with $\psi(m, K_-)<1$, each deme is subject to strong bottlenecks~\cite{wienandEvolutionFluctuatingPopulation2017,wienandEcoevolutionaryDynamicsPopulation2018,taitelbaumPopulationDynamicsChanging2020a}; see Appendix \ref{appendix:model-details} {(and first paragraph of Sec.~\ref{sec:changing_env})}. In this regime the entire metapopulation therefore experiences strong bottlenecks and can avoid extinction for extended periods of time while either strain can prevail. 
In the harsh environmental state ($K=K_-$),
 the dynamics is always dominated by deme extinction. In the mild state ($K=K_+\gg K_-$), there is recolonisation of empty demes that rapidly become either $W$ or $M$ demes, followed by invasions and $M/W$ competition.  
In order to ensure that the collapse of the metapopulation is unlikely to be observed,  the mean time spent in either environmental state needs to be shorter than the metapopulation mean extinction time in the harsh environment, i.e. $1/\nu < \theta_E$. 
Moreover, when $\nu\sim 1/\tau_E(K_-)>1/\theta_E$, 
 numerous demes go extinct in the harsh environment before  switching to the mild state. 
 Hence, when $\nu\lesssim 1$ and $\nu\theta_E>1$, 
 the metapopulation is unlikely to go extinct
 and transiently  consists of a mixture of empty demes and $W/M$ demes before either $W$ or $M$ eventually takes over. In this regime, mutants are likely to be removed from the metapopulation 
 when there is a small fraction  of them in the 
 harsh environment; see  Fig.~\ref{fig:clique-strong-bottleneck-trajectories}(c). 
 At each strong bottleneck, 
 $M$ demes have a finite probability  to
 go extinct before switching to the mild environment, where surviving mutants can invade $W$ demes and recolonise empty demes.  
 In a scenario illustrated by Fig.~\ref{fig:clique-strong-bottleneck-trajectories}(c), there are periods of duration $\sim 1/\nu$ during which
 the number of mutants remains low and prone to extinction
when $K=K_-$, followed by periods in $K_+$ where the number of $M$ demes increases (due to $M/W$ competition). Each bottleneck can thus be regarded as an
 attempt to remove $M$ demes, whereas each switch $K_-\to K_+$ can be envisioned as a  rescue of mutants. This cycle  repeats itself until $M$ demes are entirely removed  after enough bottlenecks. The metapopulation thus consists of a fluctuating number of $W$ demes and empty demes.
 This scenario is the most likely to occur  
 when the initial fraction  of $M$ demes is small.
 Another possible outcome, illustrated by Fig.~\ref{fig:clique-strong-bottleneck-trajectories}(d), occurs when mutants surviving the harsh conditions invade and are successful in recolonising many  demes
 in the mild environment. Mutants can thus significantly increase the number of $M$ demes, exceeding that of  $W$ demes. In this case,  bottlenecks can be seen as an attempt to remove  $W$ and $M$ demes, and the most likely outcome is the removal of $W$ demes. In this scenario, the metapopulation eventually consists of a fluctuating number of empty demes and 
 mutant demes, as illustrated by Fig.~\ref{fig:clique-strong-bottleneck-trajectories}(d). 
 The results of Fig.~\ref{fig:clique-strong-bottleneck-trajectories} have been obtained for cliques, but the same qualitative behaviour is expected for any regular graphs ${\rm G}$, with the spatial structure affecting the  
 the long-term fraction of occupied demes and therefore the probability of removal of each bottleneck. However, phenomena 
 operated by extinction are mostly independent of ${\rm G}$ and $m$, as illustrated by Fig.~\ref{fig:other-strong-bottleneck}.
 
 {\it A hypothetical  idealised treatment strategy:} 
  In this intermediate switching regime, the metapopulation is likely to avoid extinction  in the harsh environment  if $\nu\theta_E\gtrsim 1$. Moreover, when
   $mK_+\theta_E\gtrsim1$, then enough demes are recolonised in the mild environment to ensure that the metapopulation will not readily go extinct after a bottleneck. Hence, the metapopulation is likely to avoid extinction
   when $\nu\theta_E>1$ and $mK_+\theta_E>1$. In this scenario, either $W$ or $M$ can be entirely removed, with respective probabilities $\Phi$ and $1-\Phi$, after a mean time $\Theta$, while metapopulation extinction occurs after a mean time $\Theta_E\gg \Theta$. As  an application, we consider
 a hypothetical  idealised 
 treatment 
 to efficiently remove unwanted mutants by controlling the environmental conditions via the parameters $K_-$ and $\nu$.
 In this context, $M$ cells are interpreted as unwanted mutants that have a selective advantage over the $W$ cells we would like the population to consist of, here represented by the metapopulation consisting initially of $\Omega-1$ demes of type $W$ and a single $M$ deme. 
 In a healthy host, cells replicate in a controlled, self-regulating manner. Mutations may lead to the loss of  self-regulation of cell replication, and such mutant cells replicate rapidly. This is the case of cancerous cells, that appear as rare mutants before possibly proliferating. Thus, initial conditions like those considered here are relevant, even for smaller systems. If allowed to proliferate, cancer cells will outcompete the slower-growing healthy cells, leading to a growing tumour. Therefore, in this motivating context, we wish to remove these  aggressive cells while not eliminating healthy ones.
 The idealised treatment strategy consists of finding the set of near-optimal environmental conditions 
 to remove $M$ cells
 and minimise  the risk of extinction of the entire metapopulation. 
 This corresponds to determining the range of $K_-$
 and $\nu$ for which $\Phi$ and $\Theta/\Theta_E$ are minimal.
 According to the above discussion,
the near-optimal conditions for this idealised treatment strategy 
on a regular graph ${\rm G}$
are
  \begin{equation}
 \label{eq:near-opt}
 \psi(m,K_-)<1, \quad \nu\lesssim 1, \quad 
\nu\theta_E\gtrsim 1, 
\quad mK_+\theta_E\gtrsim1.
 \end{equation}
  Under these conditions, illustrated in Fig.~\ref{fig:clique-strong-bottleneck}, which depend on $m$ but not on the spatial structure ${\rm G}$, 
  environmental variability  generates a succession of strong bottlenecks 
 at a frequency ensuring that the 
 mutant type is the most likely to go extinct in a mean time that is much shorter than the metapopulation mean extinction time. While determining analytically $\Phi$ and $\Theta/\Theta_E$ satisfying Eq.~\eqref{eq:near-opt} is challenging, this can 
 be done efficiently numerically as illustrated by the heatmaps of Fig.~\ref{fig:clique-strong-bottleneck}, and be summarised 
 by maximising the composite quantity $(1-\Phi)\left(1-\Theta/\Theta_E\right)$, as shown in Fig.~\ref{fig:clique-strong-bottleneck}(d,h).
 In the examples of  Fig.~\ref{fig:clique-strong-bottleneck}, we find that the near-optimal treatment conditions are $10^{-2}\lesssim\nu\lesssim 1$
 and for $K_-$ that changes with $m$: $K_-\in [2,7]$ for $m=10^{-3}$ and $K_-\in [4,9]$ for $m=10^{-4}$. 
 The idealised treatment strategy therefore consists of
 letting the metapopulation evolve under the near optimal conditions Eq.~\eqref{eq:near-opt}, under which it undergoes a series of strong bottlenecks whose expected outcome is the removal of mutants. 
  Once all mutants are removed, as in Fig.~\ref{fig:clique-strong-bottleneck-trajectories}(c), the final course of the treatment consists of keeping the metapopulation in the mild environment (with $K=K_+$), where $W$ cells would spread and finally take over all the demes. In the example of Fig.~\ref{fig:clique-strong-bottleneck-trajectories}(c), this would be achieved by setting $K=K_+$ after $t\gtrsim 1000$. This idealised treatment strategy, illustrated for clique metapopulations in Fig.~\ref{fig:clique-strong-bottleneck}, qualitatively holds on regular graphs ${\rm G}$, with small influence of the spatial structure on the shape of the heatmap when $m$ is kept fixed, as seen by comparing Figs.~\ref{fig:clique-strong-bottleneck}(e-h) and \ref{fig:other-strong-bottleneck}. 
\section{Discussion, generalisations and robustness}
\label{sec:generality}

Here, we discuss our main results by critically reviewing our assumptions and outline possible generalisations.
We have studied the eco-evolutionary dynamics of a metapopulation consisting of $\Omega$ identical demes with the same carrying capacity $K$, containing wild-type $W$ and mutant $M$ cells, that are connected by slow migration  and arranged according to regular circulation graphs. While our approach holds for any regular graph, we have specifically considered the 
 examples  of  cliques (island model),  cycles, and square  grids (with periodic boundaries) which are all {\it circulation} graphs, i.e. the rate of in-flow and out-flow 
migration is the same at each deme.
This has allowed us to consider the impact of various graph structures on the metapopulation dynamics.
We have analysed the metapopulation dynamics  in a static environment where $K$ is constant, and in a time-varying environment where $K$ switches endlessly between $K_+$ and $K_-<K_+$ at a rate $\nu$; see Eq.~\eqref{eq:K(t)}. 
In static environments, the deme size fluctuates about $K$ and the metapopulation dynamics is characterised by  either
$M/W$ competition (when $\psi\gg 1$), or by deme extinction (when $\psi <1$). We have used suitable coarse-grained descriptions to analytically characterise the fate of the population in those regimes; see Fig.~\ref{fig:constant-environment}. When, as here, the metapopulation is spatially arranged on circulation graphs, the circulation theorem~\cite{marrecUniversalModelSpatially2021,liebermanEvolutionaryDynamicsGraphs2005} guarantees that the fixation probability
in the competition-dominated regime is independent of the migration rate and the spatial structure. 
We have also devised a coarse-grained three-state description of the dynamical equilibrium in the crossover regime (where $\psi \gtrsim 1$)
where in the long run there is a mixture of occupied  demes of type $W$ or $M$ and empty demes; see Appendix~\ref{appendix:inter}.
In time-fluctuating environments, when $K$ switches neither too quickly nor too slowly, each deme is subject to bottlenecks that can be weak when $K_-$ is large enough to ensure $ \psi(m,K_-)\gg1$. 
Deme extinction can be neglected in the
 weak bottleneck regime, and we have combined a coarse-grained description with a  PDMP approximation to characterise the $W/M$ competition in time-varying environments in the absence of deme extinction. This has allowed us to  show that weak bottlenecks 
lead to a {\it non-monotonic} dependence of the mutant fixation probability $\Phi^{{\rm G}}$ and mean fixation time  $\Theta^{{\rm G}}$ on the switching rate $\nu$,
with an explicit dependence  on the migration rate, whereas 
the spatial structure has an unnoticeable effect on $\Phi^{{\rm G}}$,  regardless of spatial correlations, 
but influences $\Theta^{{\rm G}}$. 
When demes are subject to strong bottlenecks, metapopulation extinction is a likely outcome under slow and fast switching ($\nu\ll1$  and $\nu\gg1$), whereas the 
overall extinction can be avoided for long periods
 under intermediate switching, when $W/M$ competition and deme extinction dynamics are coupled.
As a hypothetical application, we have considered an  idealised treatment strategy
for the rapid removal of the mutant 
conditioned on minimising the risk of overall extinction.
 
The coarse-grained descriptions adopted in the static and dynamic environments track the dynamics of the number of $M$ demes, which, in the case of the clique and cycle, is a single unbreakable cluster of $M$ demes ($M$-cluster). This requires starting from such a cluster, where here we assumed the natural initial condition of a single $M$ deme. Under these considerations, the number of active edges  connecting $W$ and $M$ demes 
in cliques and cycles is known exactly, making these graphs  particularly amenable to detailed analysis. It is also possible to capture the number of active edges exactly for other starting configurations of these graphs (e.g. two or more neighbouring $M$ demes) provided that the unbreakable structure of the  $M$-cluster is preserved at all times $t\geq 0$. For the sake of concreteness and simplicity, we have focussed on a class of regular circulation graphs.  In two dimensions,  
spatial correlations between demes  are more complex, and the
coarse-grained description of 
the $M/W$ competition dynamics on a grid has required approximations of the 
the number of active edges; see Appendix~\ref{appendix:2d-heuristic}.
A similar approximation is expected to hold on hypercubic lattices (with periodic boundaries). 
These considerations on the role of the initial condition and spatial structure do not matter when the metapopulation dynamics is dominated by the extinction of demes since these occur randomly. As a consequence, the ``idealised treatment strategy'' based on the dynamic coupling of competition and deme extinction to remove a targeted strain is expected to hold on more complex structures, including generic regular graphs. 

In this work, we have focussed on the biologically relevant regime of slow migration, which is well known to increase population fragmentation and hence influences its evolution and diversity~\cite{wrightISOLATIONDISTANCE1943,slatkin1981}. Here, the assumption of slow migration is crucial for the coarse-grained description of the metapopulation dynamics, and the values considered in our examples,  $m\approx 10^{-5}-10^{-2}$, are comparable with those used in microfluidic experimental setups~\cite{keymerBacterialMetapopulationsNanofabricated2006}. For $m\gg1$, the behaviour of a single well-mixed deme is recovered; see Ref.~\cite{wienandEvolutionFluctuatingPopulation2017}. For intermediate $m$, the dynamics is characterised by coarsening, i.e. the slow  growth of domain sizes over time \cite{KSBbook,UCTbook}. For the sake of simplicity and without loss of generality, we have assumed that migration occurs without any directional preference and with the same rate for $M$ and $W$. These assumptions can  be relaxed and the coarse-grained description be readily generalised to the case of directional and type-specific migration~\cite{marrec2023}, yielding the same qualitative behaviour discussed here for circulation graphs.   We note however that asymmetric directional migration  significantly affects the evolutionary dynamics on
non-circulation graphs, like the star~\cite{marrecUniversalModelSpatially2021,moawadEvolutionCooperationDemestructured2024,abbaraFrequentAsymmetricMigrations2023a,abbaraMutantFateSpatially2024}. It would be interesting to study the evolution on these non-circulation graphs in  time-varying environments in the case of symmetric and directional migration.
For computational tractability, we have chiefly considered metapopulations consisting of $16$ demes of size ranging  between $1$ and $200$, which are much smaller systems than most realistic microbial communities.  However, with
microfluidic devices and single-cell techniques, it is possible to perform spatially structured experiments with  $10$ to $100$ cells 
per microhabitat patch, which are conditions close to those used here~\cite{keymerBacterialMetapopulationsNanofabricated2006,totlani2020scalable,hsu2019microbial}.
 Moreover, {\it in-vivo} host-associated metapopulations are often fragmented into
a limited number of relatively small demes, e.g. $\Omega \approx 25$ and $K\approx 1000$  in mouse lymph nodes~\cite{Ganchua2020,VandenBroeck2006,fruetSpatialStructureFacilitates2024a}. 

Here, we have conveniently represented environmental variability by the   random  switching of the  carrying capacity $K=\{K_+,K_-\}$ driven by a symmetric dichotomous Markov  noise (DMN), with the extension to asymmetric DMN outlined in Appendix~\ref{appendix:bias}.
DMN is commonly used to model evolutionary processes because it is 
simple to simulate and analyse, and closely reproduces the binary conditions  used in many laboratory-controlled experiments. These are 
typically carried out
in periodically changing environments
~\cite{acar_stochastic_2008,Lambert2014,abdul2021}. It has however been shown that  letting $K$ vary periodically
between $K_+$ and $K_-$ with a period $1/(2\nu)$ leads to essentially the same dynamics~\cite{taitelbaumPopulationDynamicsChanging2020a}. Moreover, 
the relationship
between
DMN  used here and other common forms of environmental
noise has been extensively studied~\cite{taitelbaumPopulationDynamicsChanging2020a,taitelbaum2023evolutionary,HL06,Ridolfi11}, showing that DMN
is a convenient and non-limiting choice to model environmental variability. 

\section{Conclusions}
\label{sec:conclusions}
Cells evolve in spatially structured settings, where the competition with those nearby is stronger than those further afield, subject to never-ending environmental changes. Spatial structure and environmental variability  impact the eco-evolutionary dynamics of microbial populations significantly, but their joint influence is scarcely considered. Mutations frequently arise in cell communities and some may increase cell proliferation while having deleterious effects to a host organism, e.g. leading to cancerous cells or causing resistance to antibiotics~\cite{coatesAntibioticinducedPopulationFluctuations2018,mahrt2021bottleneck,hernandez-navarroCoupledEnvironmentalDemographic2023,LKUM2024}. Motivated by these issues, and inspired by recent advances in microfluidics allowing experiments to track dynamics at the single-cell level~\cite{keymerBacterialMetapopulationsNanofabricated2006,totlani2020scalable,hsu2019microbial}, we have investigated
the classic example of a rare mutant having a selective advantage over wild-type resident cells
occupying a spatially structured population in time-fluctuating environments.
 Here, we have considered a class of metapopulation models spatially arranged as regular (circulation) graphs
where cells of  wild and mutant types compete in a {\it time-fluctuating}  environment. The metapopulation consists  of demes (subpopulations) with the same carrying capacity, connected to each other by slow cell migration.
 We represent environmental variability by letting the 
carrying capacity
 endlessly switch between two values associated  to harsh and mild  conditions. 
 In this framework, we have studied how migration, spatial structure,  and fluctuations influence the probability and time for the mutant or wild-type to take over the 
 metapopulation, and under which conditions extinction of demes and the entire metapopulation occurs.
 This allows us to identify when environmental variability coupled to demographic fluctuations can be utilised to remove the mutant.
  
We have first considered the case where demes fluctuate about
a constant carrying capacity in static environments. We have thus characterised analytically and using stochastic simulations a regime dominated by the competition between the mutants and wild-type cells, another where there is deme extinction, as well as a crossover regime combining local competition and extinction.  
 In time-varying environments, various qualitatively different dynamical scenarios arise and  environmental fluctuations can significantly influence
the evolution of metapopulations. When the rate of switching is neither too slow nor too fast, demes experience  bottlenecks
and the population is  prone to fluctuations or extinction.
When the fluctuating carrying capacity remains large and bottlenecks are weak, deme extinction is negligible. The dynamics is thus dominated by the competition between wild-type cells and mutants to invade and take over demes, and eventually the population, which we characterise by devising a suitable  coarse-grained description of the individual-based model when migration is slow.  This allows us to determine the fixation probability and mean fixation time by combining analytical and computational tools, and to show that these quantities can vary non-monotonically with the environmental switching rate. We find that in the regime of weak bottlenecks,  
the mutant fixation probability on regular circulation graphs depends on the migration rate, which is in stark contrast with what happens in static environments, while the spatial structure has no noticeable influence. 
 When the carrying capacity is small under harsh conditions, bottlenecks are strong and  there is a dynamical coupling of 
 strain competition in the mild environmental state and 
 deme extinction in the harsh environment. This yields rich dynamical scenarios  among which we 
 identify a mechanism, expected to hold on any regular graph, 
 driven by environmental variability and fluctuations to efficiently eradicate one strain. As a hypothetical application, we have thus proposed an idealised treatment strategy to remove the mutant, assumed to be unwanted and favoured by selection. We have  shown that, 
 when each deme is subject to strong bottlenecks
 at intermediate switching rates, the mutant can be efficiently removed by demographic fluctuations arising  in the harsh environment  without exposing the entire population to a risk of rapid extinction. 
 We have thus determined the near-optimal conditions on the
 switching rate and bottleneck strength for 
 this idealised treatment strategy and found that these are qualitatively the same on other graphs.

In summary, our analysis sheds further light on the 
influence of the spatial structure, migration, and fluctuations on
the spread of a mutant strain in time-fluctuating environments. 
We have identified and characterised various dynamical scenarios, displaying a complex dependence on the switching and migration rates. 
We  have also shown how environmental variability and fluctuations can be utilised to achieve desired
evolutionary outcomes like the efficient removal of a
pathogenic mutant. 
While we have made a number of simplifying assumptions,  allowing us to make analytical progress, 
many of these   can  be relaxed without affecting the results or the methodology. Our approach
holds for arbitrary regular  graphs and
can be generalised to more complex spatial settings.
We therefore believe that the model studied  here has numerous potential applications. For instance, it 
mirrors the \textit{in vitro} evolution of a mutant across an array of microfluidic devices, where cells migrate between ``microhabitat patches'' either via microchannels or pipette, with  bottlenecks implemented via a strict control of the nutrient level in each device.\\

{\noindent{\bf Data availability statement:}
 The data and codes  that support the findings of this study are openly available at the following URL/DOI: \href{https://doi.org/10.5518/1660}{10.5518/1660} \cite{codes-data}.}

\begin{acknowledgments}
We would like to thank K. Distefano, L. Hern\'andez-Navarro, J. Jim\'enez, S. Mu\~noz-Montero, M. Pleimling, and A. M. Rucklidge for fruitful discussions.
M. M. gratefully acknowledges funding from the U.K. Engineering and Physical Sciences Research Council (EPSRC) under the Grant No. EP/V014439/1 for the project ‘DMS-EPSRC Eco-Evolutionary Dynamics of Fluctuating Populations’. The support of a Ph.D. scholarship to M. A. by the EPSRC Grant No. EP/T517860/1 is also thankfully acknowledged.  
M. S. and U. C. T.’s contribution to this research was supported by the U.S. National Science Foundation, Division of Mathematical Sciences under Award No. NSF DMS-2128587. 
This work was undertaken on ARC4, part of the High Performance Computing facilities at the University of Leeds, UK.
\end{acknowledgments}

\subsection*{Author contributions}\label{sec:authinfo.subsec:authcont}
{\bf Matthew Asker}: Conceptualisation (supporting), Formal analysis, Data curation, Investigation, Methodology, Software, Validation, Visualisation, Writing -- original draft, Writing -- review \& editing. 
{\bf Mohamed Swailem}:   Conceptualisation (supporting), Formal analysis, Data curation, Investigation,  Software, Validation,  Writing -- review \& editing. 
{\bf Uwe C. T\"auber}: 
Funding acquisition, Supervision,  Resources,  Writing -- review \& editing. 
{\bf Mauro Mobilia}: Conceptualisation, 
Formal analysis,  Methodology, Validation, 
Funding acquisition, Supervision,  Resources,   Writing -- original draft, Writing -- review \& editing.  

\begin{table*}
\centering
\caption{Typical timescales in metapopulation dynamics}
\label{tab:timescales}
\begin{tabular}{lll}
\toprule
\textbf{Process} & \textbf{Timescale for} & \textbf{Expression} \\
\midrule
Deme size growth & \makecell[l]{Time for deme size to reach\\ equilibrium} & $\edit{\sim}\ln K \edit{= \mathcal{O}(1)}$ \\
Deme fixation & \makecell[l]{Time for fixation\\ in single deme} & $\edit{\sim}1/s$ \cite{antal2006fixation,traulsen2009stochastic,Ewens} \\
Migration & \makecell[l]{Expected time between\\ migration events} & $\frac{1}{mK}$ \\
Deme extinction & \makecell[l]{Mean extinction time\\ of isolated deme} & $\tau_E(K) \approx \frac{e^K}{K}$ \\
Metapopulation extinction & \makecell[l]{Extinction time\\ for full metapopulation} & $\theta_E \approx \tau_E \ln \Omega$ \\
Metapopulation fixation & \makecell[l]{Unconditional mean\\ fixation time} & Unwieldy expression \\
Environmental switching & \makecell[l]{Correlation time of the\\ dichotomous Markov noise} & $\frac{1}{2\nu}$ \\
\bottomrule
\end{tabular}
\end{table*}

\begin{table*}
\centering
\caption{Dynamical regimes and their characteristics}
\label{tab:regimes}
\begin{tabular}{llll}
\toprule
\textbf{Regime} & \textbf{Condition} & \textbf{Characteristics} & \textbf{Dependence} \\
\midrule
\multicolumn{4}{c}{\textit{Static environment}} \\
\midrule
Competition-dominated & $\psi \gg 1$ & 
\makecell[l]{Mutant and wild-type species compete for demes \\ until fixation of the metapopulation.} & 
\makecell[l]{$\phi$ independent of $m$ and $\rm{G}$} \\[2ex]

Extinction-dominated & $\psi < 1$ & 
\makecell[l]{Rapid metapopulation extinction} & 
\makecell[l]{$\theta_E$ independent of $\mathrm{G}$} \\[1ex]

Intermediate & $\psi \gtrsim 1$ & 
\makecell[l]{Competition between demes with stochastic deme\\ extinction and recolonisation. Fraction of occupied\\ demes: $\frac{\Omega_{\text{occ}}}{\Omega} \approx 1 - \frac{1}{\psi}$} & 
\makecell[l]{Weak $\mathrm{G}$ dependence of $\phi_{\text{int}}$} \\[2ex]
\midrule
\multicolumn{4}{c}{\textit{Time-varying environment}} \\
\midrule
Weak bottlenecks & $\psi(m, K_-) \gg 1$ & 
\makecell[l]{Non-monotonic $\nu$-dependence of \\ $\Phi^{\rm G}$, $\Theta^{\rm G}$} & 
\makecell[l]{$\Phi^{\rm G}$ depends on $m$} \\[1ex]

Strong bottlenecks & $\psi(m, K_-) < 1$ & 
\makecell[l]{Extinction likely at $\nu$ extremes. \\ Mutant removal optimal at intermediate $\nu$} & 
\makecell[l]{Independent of $\mathrm{G}$} \\[2ex]
\bottomrule
\end{tabular}
\end{table*}

\appendix
\section{Further details on the model}
\label{appendix:model-details}
In this section, we provide further details on the 
model by discussing the master equation encoding its individual-based dynamics, and give further details of the size distribution of a single deme. \\

\subsection{Master equation}
\label{appendix:ME}
As discussed in Sec.~\ref{sec:model-and-methods_model}, 
the individual-based model 
is a continuous-time multivariate Markov process defined by the reaction and transition rates Eq.~\eqref{eq:BD}-\eqref{eq:migration-transition}. The intra and inter-deme 
dynamics is encoded in a master equation for the probability $P(\{n_W,n_M\}, \xi,t)$ that at time $t$ the metapopulation is in the environmental state $\xi$ and
configuration $\{n_W,n_M\}\equiv(\dots,n_W(x), n_M(x), \dots )$, where
$n_{W/M}(x)$ denotes the number of 
cells of type $W/M$ in deme $x=1,\dots,\Omega$.
The master equation (ME) for the metapopulation dynamics subject to environmental switching   on a 
regular
graph ${\rm G}=\{{\rm clique, cycle, grid}\}$ with degrees (or number of nearest neighbours) given by
$q_{\text{clique}}=\Omega-1$, $q_{\text{cycle}} =2$, or $q_{\text{grid}}=4$ 
reads
\begin{widetext}
\begin{equation}
    \label{eq:ME}
    \begin{aligned}
        \frac{\partial P(\{n_W,n_M\},\xi,t)}{\partial t}&=\sum_{x=1}^{\Omega}\sum_{\alpha}\big[\left(\mathbb{E}_\alpha^-(x)-1\right)T_\alpha^-(x)P(\{n_W,n_M\},\xi,t) +\left(\mathbb{E}_\alpha^+(x)-1\right)T_\alpha^+(x)P(\{n_W,n_M\},\xi,t)\big]\\     
        &+\frac{1}{2}\sum_{x=1}^{\Omega}\sum_{\text{ $y$ {\rm n.n.}  $x$ }}\Big[
        \left( \mathbb{E}_W^+(y)\mathbb{E}_W^-(x)-1\right)T^{m,{\rm G}}_W(x)+ \left( \mathbb{E}_W^+(x)\mathbb{E}_W^-(y)-1\right)T^{m,{\rm G}}_W(y)
        \Big]P(\{n_W,n_M\},\xi,t)\\
        &+\frac{1}{2}\sum_{x=1}^{\Omega}\sum_{\text{ $y$ {\rm n.n.}  $x$ }}\Big[
        \left( \mathbb{E}_M^+(y)\mathbb{E}_M^-(x)-1\right)T^{m,{\rm G}}_M(x)+ \left( \mathbb{E}_M^+(x)\mathbb{E}_M^-(y)-1\right)T^{m,{\rm G}}_M(y)
        \Big]P(\{n_W,n_M\},\xi,t)\\
        &+\nu\left[P(\{n_W,n_M\},-\xi,t)-P(\{n_W,n_M\},\xi,t)\right],
    \end{aligned}
\end{equation}
\end{widetext}
where $y$ {\rm n.n.}  $x$  denotes the sum over the $q_{\rm G}$ neighbours $y$ of the deme $x$ and $P(\dots)=0$
whenever any of $T_{\alpha}^{\pm}$ or $T^{m,{\rm G}}_{\alpha}$ is negative. The shift operators $\mathbb{E}_\alpha^\pm(x)$ act by raising or decreasing by one 
the number of cells of type $\alpha$ in deme $x$. For example,  $\mathbb{E}_{W}^{\pm}(x) [n_{W}(x) P(\dots, n_W(x), n_M(x) \dots, \xi,t)]=(n_{W}(x)\pm 1) P(\cdots, n_W(x)\pm 1, n_M(x) \cdots, \xi,t)$ and $\mathbb{E}_{M}^{\pm}(x) [n_{W}(x) P(\cdots, n_W(x), n_M(x) \cdots, \xi,t)]=n_{W}(x) P(\cdots, n_W(x), n_M(x)\pm 1 \cdots, \xi,t)$. The first line on the right-hand-side (RHS) of  Eq.~\eqref{eq:ME} encodes the intra-deme birth-death  dynamics, the second and third lines represent the  
inter-deme dynamics via inward and outward migration, and the last line 
 accounts for symmetric random environmental switching. Here, the ME has specifically been formulated in the presence of environmental switching, but its static-environment counterpart is readily obtained from Eq.~\eqref{eq:ME}: it  suffices to set $\nu=0$ and to replace $K(t)$ by a constant carrying capacity $K$, yielding 
 the ME  for  $P(\{n_W,n_M\},t)$ that is the probability  
 to find the metapopulation in a given state $\{n_W,n_M\}$ at time $t$ (with no environmental dependence).
Moreover, by setting $\Omega=1$ and $m=0$ in Eq.~\eqref{eq:ME}, the second and third lines on the RHS cancel, we obtain the ME encoding the intra-deme dynamics of a single isolated deme~\cite{wienandEvolutionFluctuatingPopulation2017,wienandEcoevolutionaryDynamicsPopulation2018}.
 
 While the ME Eq.~\eqref{eq:ME} holds for any regular 
 graphs ${\rm G}$,
in  our examples we consider specifically the regular  circulation graphs ${\rm G}=\{{\rm clique, cycle, grid}\}$. 
 The space-dependent individual-based dynamics encoded in the
ME Eq.~\eqref{eq:ME} has been simulated using the Monte Carlo method described in Appendix~\ref{appendix:model-and-methods_simulation}. It is worth noting  that demographic fluctuations eventually lead to the extinction of the entire
metapopulation, in all regimes. However, this phenomenon occurs after a time growing dramatically with the system size, and it can generally not be observed in sufficiently large metapopulations; see Fig.~\ref{fig:constant-environment}(b,bottom right).

\subsection{Eco-evolutionary dynamics of a single deme}
\label{appendix:eco-evolutionary-dynamics-in-a-deme}
Since the metapopulation consists of a graph of connected demes, all with the same carrying capacity,
we can gain significant insight into its dynamics by looking into its building block. In this section, we therefore  
analyse the eco-evolutionary dynamics of a single isolated deme (when $m=0$).

For an isolated deme, there is only intra-deme dynamics according to the birth-death process defined by  Eqs.~\eqref{eq:BD} and \eqref{eq:intra_transition_rates}. The ME for the dynamics of a single isolated deme is thus given by setting $\Omega=1$ and $m=0$ in
Eq.~\eqref{eq:ME}; see Appendix~\ref{appendix:ME}. 
The corresponding  intra-deme dynamics in a time-varying environment can be simulated  using 
exact methods like the Gillespie algorithm~\cite{Gillespie76}, as in Refs~\cite{wienandEvolutionFluctuatingPopulation2017,wienandEcoevolutionaryDynamicsPopulation2018,taitelbaumPopulationDynamicsChanging2020a,west2020,shibasakiExclusionFittestPredicts2021,hernandez-navarroCoupledEnvironmentalDemographic2023,askerCoexistenceCompetingMicrobial2023,hernandez-navarroEcoevolutionaryDynamicsCooperative2024}. 
It is instructive to ignore all forms of fluctuations and 
consider the mean-field approximation of an isolated deme 
dynamics subject to a constant carrying capacity $K\gg 1$. 
Following Refs.~\cite{wienandEvolutionFluctuatingPopulation2017,wienandEcoevolutionaryDynamicsPopulation2018}, with the transition rates Eq.~\eqref{eq:intra_transition_rates}, the mean-field eco-evolutionary dynamics of a single isolated deme is 
characterised by rate equations for the size $n$ of the deme and the fraction $x=n_M/n$ of mutants in the deme, which read
\begin{equation}
 \label{eq:MFdeme}
 \begin{aligned}
 \dot{n}&=\sum_{\alpha} T_{\alpha}^+-\sum_{\alpha} T_{\alpha}^-=n\left(1-\frac{n}{K}\right),\\
  \dot{x}&=\frac{T_{M}^+- T_{M}^-}{n}-x\frac{\dot{n}}{n}=\frac{sx(1-x)}{1+sx},
 \end{aligned}
\end{equation}
where the dot indicates the time derivative and  we have used $f_M=1+s$ and $f_W=1$. The prefactors of these decoupled rate equations predict the relaxation of the deme size towards the constant carrying capacity, with $n\to K$ on a timescale $t\sim 1$, and the  growth of the fraction of mutants, with $x\to 1$  on a timescale $t\sim 1/s$. In the total absence of fluctuations, when $0<s\ll1$ (small selective advantage to $M$),  in
the mean-field picture, 
the deme size quickly approaches the carrying capacity and
there is a timescale separation between $n$ and $x$, respectively the fast and slow variables, where $n$ evolves on a timescale of order $1$ and $x$ on a timescale $\sim 1/s \gg 1$; see Table~\ref{tab:timescales}. Thus, $W$ cells are thus slowly wiped out by mutants that take over the deme on a timescale $t\sim 1/s \gg 1$~\cite{wienandEvolutionFluctuatingPopulation2017,wienandEcoevolutionaryDynamicsPopulation2018}. It is worth noting that with the effective transition rates Eq.~\eqref{eq:effectT} we have assumed that  invasions always occur after deme size and composition relaxation. This means that  
Eqs.~\eqref{eq:marginalisedsolution} and
\eqref{eq:weak-bottleneck-solution} assume a timescale separation 
between $n$ and $x$ in addition to the slow migration assumption.

It is also relevant to consider the intra-deme dynamics of a single isolated site subject to a finite constant carrying capacity $K$. As explained in Appendix~\ref{appendix:slow_migration}, the intra-deme dynamics can be well approximated by a Moran process for a deme of constant size $n=K$~\cite{Moran,Ewens,Blythe07,traulsen2009stochastic,wienandEvolutionFluctuatingPopulation2017,wienandEcoevolutionaryDynamicsPopulation2018}, and characterised by the fixation probability and mean fixation time given by Eq.~\eqref{eq:Moran_fixation}. The probability $\rho_{M/W}(K)$  that a single $M/W$ cell takes over a $W/M$ deme of size $K$
is given by Eq.~\eqref{eq:rhoMW}.

When demographic fluctuations can be neglected and randomness only arises from environmental variability via Eq.~\eqref{eq:K(t)}, the intra-deme dynamics of an isolated deme is well-captured by the 
 piecewise deterministic Markov process (PDMP) for $n$, obtained by 
ignoring demographic fluctuations~\cite{davisPiecewiseDeterministicMarkovProcesses1984,wienandEvolutionFluctuatingPopulation2017,wienandEcoevolutionaryDynamicsPopulation2018}. In the realm of the PDMP approximation, the deme size thus satisfies
a deterministic logistic equation in each environmental state, 
subject to a carrying capacity that switches when the environment changes ($K=K_{\pm}$
when $\xi=\pm 1$), yielding
  the $n$-PDMP~\cite{wienandEvolutionFluctuatingPopulation2017,wienandEcoevolutionaryDynamicsPopulation2018,taitelbaumPopulationDynamicsChanging2020a,askerCoexistenceCompetingMicrobial2023,LKUM2024,hernandez-navarroCoupledEnvironmentalDemographic2023}
\begin{equation}
\label{PDMP_eq}
    \dot{n}=
    \begin{cases}
    n\left(1-\frac{n}{K_-}\right) & \text{if } \xi=-1, \\
    n\left(1-\frac{n}{K_+}\right)  & \text{if } \xi=+1, 
    \end{cases}
\end{equation}
that is decoupled from the mean-field equation for $x$
that is  as in Eq.~\eqref{eq:MFdeme}. 
The stationary joint probability density of this $n$-PDMP is given by Eq.~\eqref{eq:N_PDMPs}~\cite{wienandEcoevolutionaryDynamicsPopulation2018,taitelbaumPopulationDynamicsChanging2020a}, while the 
marginal probability density is
\begin{equation}
\label{eq:N_PDMP_marg}
p(n;\nu)=\frac{1}{2}\sum_{\xi}p_{\xi}(n;\nu)=\frac{\mathcal{Z}}{n^2} \left[\left(\frac{K_+ }{n}- 1\right) \left(1-\frac{K_-}{n}\right)\right]^{\nu-1},
\end{equation}
where $\mathcal{Z}$ is the normalisation constant and 
$n\in [K_-,K_+]$. Despite ignoring the effect of demographic noise, $p(n;\nu)$
aptly captures many properties of the quasi-stationary distribution of the size of an isolated single deme~\cite{wienandEcoevolutionaryDynamicsPopulation2018,taitelbaumPopulationDynamicsChanging2020a}. For instance, the long-time average deme 
size is accurately approximated by ${\cal N}(\nu)=\int_{K_-}^{K_+} n p(n;\nu) {\rm d}n$,
and is a decreasing function of $\nu$~\cite{wienandEvolutionFluctuatingPopulation2017,wienandEcoevolutionaryDynamicsPopulation2018}.

While  $p_{\xi}(n;\nu)$ and $p(n;\nu)$ 
give a PDMP description of the quasi-stationary distribution of  the size
of an isolated deme ($m=0$),  the $n$-PDMP stationary densities given by Eqs.~\eqref{eq:N_PDMPs} and \eqref{eq:N_PDMP_marg} are still a valid approximation of the long-time size distribution of $n$ in the presence of  migration between  many connected demes, as considered here. In fact, as shown below in Fig.~\ref{fig:deme-size-dist}, the influence of migration on the distribution of the deme size
is essentially unnoticeable, where the predictions of  Eqs.~\eqref{eq:N_PDMPs}  for a single isolated deme (no migration) are compared with the deme size distribution obtained from by sample averaging across all demes of the metapopulation in the presence of migration. Its main features are therefore well captured by Eq.~\eqref{eq:N_PDMPs}
and Eq.~\eqref{eq:N_PDMP_marg}.
As illustrated by Fig.~\ref{fig:deme-size-dist}, the density $p(n;\nu)$ correctly predicts that the deme size distribution is bimodal when $\nu<1$ and unimodal when $\nu>1$, and that it 
is peaked at $n\approx K_{\pm }$ when $\nu\ll1$ (slow switching) and   centred around $n\approx {\cal K}$  when $\nu \gg 1$ (fast switching),  see Eq.~\eqref{eq:curlyK}. The joint and marginal PDMP probability densities Eq.~\eqref{eq:N_PDMPs} and Eq.~\eqref{eq:N_PDMP_marg} provide valuable insight into the deme size distribution  when these are subject to weak bottlenecks and their extinction  can be neglected~\cite{wienandEvolutionFluctuatingPopulation2017,wienandEcoevolutionaryDynamicsPopulation2018,taitelbaumPopulationDynamicsChanging2020a,askerCoexistenceCompetingMicrobial2023,LKUM2024,hernandez-navarroCoupledEnvironmentalDemographic2023,taitelbaum2023evolutionary}.

\begin{figure*}
\phantomsection
    \centering
    \includegraphics[width=\linewidth]{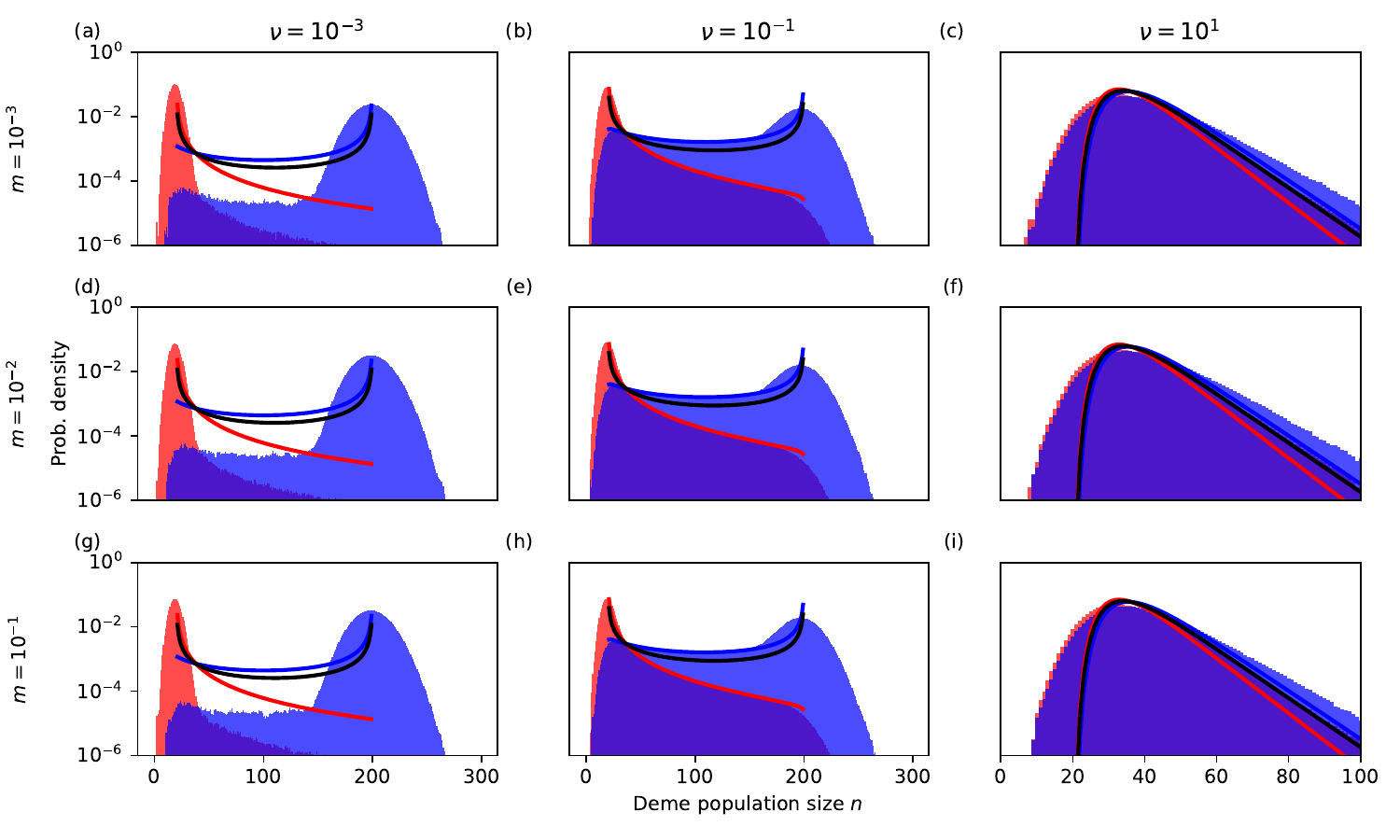}
    \caption{Quasi-stationary probability density  of deme population size ($n$-QPSD)  
    distribution on clique metapopulations for various parameters and its $n$-PDMP approximation given by Eqs.~\eqref{eq:N_PDMPs} and \eqref{eq:N_PDMP_marg}. Red and blue bars show data for the $n$-QPSD conditioned  on $K(t)=K_-$
    and $K(t)=K_+$, respectively. Red, blue, and black solid lines are the  $n$-PDMP stationary densities 
    $p_-(n;\nu)$, $p_+(n;\nu)$, and $p(n; \nu)$, respectively, given by Eqs.~\eqref{eq:N_PDMPs} and \eqref{eq:N_PDMP_marg}. Panels (a-c) are for $m=10^{-3}$, (d-f)  for $m=10^{-2}$, and (g-i) for $m=10^{-1}$. We have $\nu=10^{-3}$ in (a,d,g),  $\nu=10^{-1}$ in (b,e,h), and  $\nu=10^2$ in (c,f,i). Other parameters are $\Omega=16$, $K_+=200$, and $K_-=20$. All represent a single realisation tracked until $t=10^5$.}
    \label{fig:deme-size-dist}
\end{figure*}

In the main text, we have used the PDMP approximation to describe the size distribution of a single isolated deme subject to symmetric random switching of its carrying capacity Eq.~\eqref{eq:K(t)}. Here, we show that spatial migration has no noticeable influence on the size distribution of a single 
deme of a metapopulation structured as a clique (though the same holds for other spatial structures).
To this end, in Fig.~\ref{fig:deme-size-dist}, we compare the size distribution of a single 
deme in a metapopulation structured as a clique
in the presence of a migration per capita rate $m\in\{10^{-3},10^{-2},10^{-1}\}$ (obtained 
from stochastic simulations) with
the  predictions of  Eq.~\eqref{eq:N_PDMPs}.  These results illustrate that migration has no noticeable effect on the deme size distribution that can approximated by PDMP density
Eq.~\eqref{eq:N_PDMPs} (or Eq.~\eqref{eq:N_PDMP_marg}) in the absence and presence of migration.

As a consequence, the deme size distribution of any metapopulation considered here is well approximated by the joint and marginal PDMP densities $p_\pm(n;\nu)$ and $p(n;\nu)$, given by Eq.~\eqref{eq:N_PDMPs} and Eq.~\eqref{eq:N_PDMP_marg}.

Intuitively, this can be understood by noticing that the spatial structures considered here are circulation graphs, yielding the same inward and outward migration flow at each deme, and each deme has the same carrying capacity. As a consequence, the average number of cells per deme is expected to be independent of migration.
The latter remains well captured by  $p_\pm(n;\nu)$ and $p(n;\nu)$ regardless of migration rate, as seen in Fig.~\ref{fig:deme-size-dist}.

\section{Deme invasion}
\label{appendix:slow_migration}
In this section, we analyse the  process of invasion of a single deme subject to a constant carrying capacity $K$  when $\psi(m,K)\gg 1$; see Sec.~\ref{sec:constant_env-large_K}. In this competition-dominated regime, the extinction of demes can be neglected and their size rapidly fluctuates about $K$; see Fig.~\ref{fig:cartoon-constant}(a). In this scenario, we can assume that the deme size is constant $n(x)= K$, and describe the  deme dynamics by tracking the number of 
mutants $n_M$ and wild-type cells $n_W$ in the deme $x$. The deme composition 
$( n_M, n_W)= (n_M, K-n_M)$
thus changes according to the Moran process~\cite{Moran,Ewens,Blythe07,traulsen2009stochastic}
\begin{align}
    (n_M, n_W) &\stackrel{{\cal T}^{+}_{\text{Mo}}}{\longrightarrow} (n_M+1, n_W-1), \quad 
    \nonumber\\ (n_M, n_W)&\stackrel{{\cal T}^{-}_{\text{Mo}}}{\longrightarrow}   (n_M-1, n_W+1),
\label{Eq:BDReact}
\end{align}
where the transition rates are defined in terms of $T^{\pm}_{M/W}$,
given by Eq.~\eqref{eq:intra_transition_rates}, according to
~\cite{wienandEvolutionFluctuatingPopulation2017,wienandEcoevolutionaryDynamicsPopulation2018,askerCoexistenceCompetingMicrobial2023,hernandez-navarroCoupledEnvironmentalDemographic2023} 
\begin{align}
 \label{eq:Mo}
 {\cal T}^{+}_{\text{Mo}}(n_M)&=\frac{T^+_{M} T^-_{W}}{K} = \frac{f_{M}}{\overline{f}} \frac{n_M n_W}{K}=\frac{f_{M}}{\overline{f}} n_M\left(1-\frac{n_M}{K}\right),\nonumber\\
 {\cal T}^{-}_{\text{Mo}}(n_M)&=\frac{T^-_{M}T^+_{W}}{K} =\frac{f_{W}}{\overline{f}} \frac{n_M n_W}{K}=\frac{f_{W}}{\overline{f}} n_M\left(1-\frac{n_M}{K}\right).
\end{align}
These transition rates correspond to the  
the effective rates of increase and decrease in  the number of $M$ in a deme
of size $K$. This Moran process conserves the deme size by accompanying each birth
of an $M/W$ by the simultaneous death of a $W/M$, and is characterised by the absorbing states $(n_M,n_W)=(K,0)$ ($M$ deme) and $(n_M,n_W)=(0,K)$ ($W$ deme).
The $M$ fixation probability $\phi_\text{Mo}$ and unconditional mean fixation time (uMFT) $ \theta_\text{Mo}$ for this Moran process
are 
classical results,
and when there are initially $i$ cells of type $M$, they read~\cite{Ewens,antal2006fixation,Blythe07,traulsen2009stochastic}
\begin{equation}
\label{eq:Moran_fixation}
\hspace{-3mm}
\begin{aligned}
    \phi_\text{Mo}(i) &= \frac{1-\gamma_\text{Mo}^i}{1-\gamma_\text{Mo}^K},\\
    \theta_\text{Mo}(i) &= \phi_\text{Mo}(i) \sum_{n=i}^{K-1}\sum_{l=1}^n \frac{\gamma_\text{Mo}^{n-l}}{{\cal T}_\text{Mo}^+(l)} \\&- (1-\phi_\text{Mo}(i)) \sum_{n=1}^{i-1}\sum_{l=1}^n \frac{\gamma_\text{Mo}^{n-l}}{{\cal T}_\text{Mo}^+(l)}, 
\end{aligned}
\end{equation}
where $\gamma_\text{Mo}\equiv {\cal T}_\text{Mo}^-/{\cal T}_\text{Mo}^+=f_W/f_M=1/(1+s)$. The fixation probability of a single mutant ($i=1$) 
and of a single $W$ cell ($i=K-1$) 
are particularly relevant for our purposes, and explicitly read
\begin{equation}
\label{eq:Moran_rhos}
\begin{aligned}
   \rho_{M}&\equiv \phi_\text{Mo}(1)= \frac{s}{1+s}\left[\frac{1}{1-(1+s)^{-K}}\right],\\
    \rho_{W}&\equiv 1-\phi_\text{Mo}(K-1)=\frac{s}{(1+s)^K}\left[\frac{1}{1-(1+s)^{-K}}\right]. 
\end{aligned}
\end{equation}

\section{Deme and metapopulation mean extinction  times}
\label{appendix:extinction_time}

\begin{figure}
    \centering
    \includegraphics[width=\linewidth]{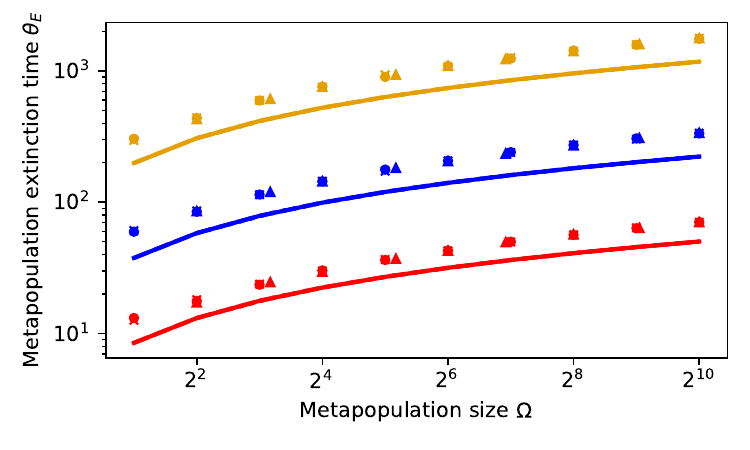}
    \caption{$\theta_E$ vs. metapopulation size $\Omega$ for $K=7$ (yellow), $K=5$ (blue), $K=3$ (red) and $m=10^{-4}$. 
       Markers are simulation results and lines are predictions of Eq.~\eqref{eq:S9} for cliques (solid lines / crosses), cycles  (dashed lines / circles), and  grids (dotted lines / triangles). For the grid, the values of $\Omega$ are chosen to maintain a square lattice, and thus occasionally differ slightly from the metapopulation size used in the cycle and clique.
       Markers of the same colour are almost indistinguishable. Deviations occur due to the approximation of $\tau_E(K)$ in Eq.~\eqref{eq:dMET_approx}. Selection plays no role in this regime, so results have been obtained with $s=0$. }
    \label{fig:extinction-vs-omega}
\end{figure}

In this section, we discuss the process of extinction of a single deme that has a  constant carrying capacity $K$ when $\psi(m,K)\ll 1$; see Sec.~\ref{sec:constant_env-small_K}.
In this extinction-dominated regime, we can assume that the deme size rapidly fluctuates about $K$, and extinction occurs  from a deme of constant size $n(x)=K$ prior to any invasion. Without loss of generality (see below), we hence assume that
extinction occurs from entirely occupied demes,
with the metapopulation consisting only of  $W$ and $M$ demes, all of size $K$. In this representation, the dynamics of a $W/M$ deme when $\psi\ll 1$ is given by the birth-death process of Sec.~\ref{sec:model-and-methods_model} with transition rates $T_{W/M}^{+}=n_{W/M}$ and $T_{W/M}^{-}=n_{W/M}^2/K$, subject to an absorbing boundary at $n_{W/M}=0$; see Eq.~\eqref{eq:intra_transition_rates}. Clearly therefore,
the deme dynamics is independent of its type, and
 the deme  mean extinction  time (dMET)
is the average time to reach $n_{W/M}=0$ and is 
the same for $W$ and $M$ demes (dMET is independent of $s$). A classical calculation (see, e.g.  
Sec.~6.7 in Ref.~\cite{allenIntroductionStochasticProcesses2010b}) for  an initially fully occupied deme of size $K$ yields
\begin{equation}
\label{eq:exact-deme-ext-time}
    \tau_E(K) = \sum_{n=0}^{K-1}\left(\frac{n!}{K^n}\sum_{i=n+1}^\infty\frac{1}{i}\frac{K^i}{i!}\right).
\end{equation}
The leading contribution to this expression arises from the term $n=0$: $\tau_E(K) \simeq \sum_{i=1}^\infty K^i/(i\cdot i!)$. 
 This expression, corresponding to the dMET of a deme initialised with a single cell (of either type), is a good approximation of Eq.~\eqref{eq:exact-deme-ext-time} which
indicates that the dMET is 
independent of selection and initial condition (for the leading order of $\tau_E$).
We can further simplify the leading contribution to the dMET by writing
\begin{equation*}
\begin{aligned}
\tau_E(K)& \simeq \sum_{n=1}^\infty \frac{K^n}{n!}\int_0^1 t^{n-1} \text{d}t,\\
&=\int_0^1 \frac{1}{t} \sum_{n=1}^\infty \frac{(Kt)^n}{n!}\text{d}t,\\
&=\int_0^K \frac{e^u - 1}{u}\text{d}u,
\end{aligned}
\end{equation*}
where we have used $u=Kt$. The main contribution to the last integral stem from the upper bound, yielding 
\begin{equation}
\label{eq:dMET_approx}
    \tau_E(K) \simeq \frac{e^K}{K}.
\end{equation}
The dMET hence increases almost exponentially with $K$, is independent of the deme  type, and its initial state.

The metapopulation mean extinction time (mMET) 
in the regime $\psi(m,K)\ll 1$ 
can be obtained analytically within the realm of the above coarse-grained description, in the spirit of the approach of Ref.~\cite{landeExtinctionTimesFinite1998} for cliques. The metapopulation thus consists 
initially of entirely occupied demes ($i$ mutant demes and $\Omega-i$
type $W$ demes). Since deme extinction here occurs prior to any invasion, we describe the metapopulation dynamics in terms of the number $j=0,1,\dots,\Omega$ of entirely occupied demes.
Residents (of either $W$ or $M$ type) of these filled demes
can recolonise a neighbouring empty site at a rate $B(j)$; see Fig.~\ref{fig:cartoon-constant}(d). In addition, each occupied deme goes extinct at a rate $D(j)$. This coarse-grained description of the
metapopulation dynamics is therefore a birth-death process with an absorbing state $j=0$ corresponding to the eventual extinction of the
metapopulation, and a reflecting boundary at $j=\Omega$ (all demes are occupied). In this picture, proceeding as above~\cite{allenIntroductionStochasticProcesses2010b}, the mMET reads
\begin{equation}
\label{eq:thetaE_1}
    \theta_E(K,\Omega)=\sum_{n=1}^{\Omega-1}\left[\left(\prod_{m=1}^{n-1}\frac{B(m)}{D(m)} \right) \sum_{j=n}^\Omega \frac{\prod_{l=1}^{j}\frac{B(l)}{D(l)}}{D(j)}\right].
\end{equation}
In the vein of Ref.~\cite{landeExtinctionTimesFinite1998},
the recolonisation-birth rate of occupied demes is 
$B(j)=mK j(1-j/\Omega)$, corresponding to a logistic growth with a rate proportional to  
the expected number of migrations from an occupied deme $mK$. Here, the extinction rate is  $D(j)=j/\tau_E$ and is inversely proportional to the mean local extinction time, that is the dMET. 
With 
$\psi=mK\tau_E$, using $\prod_{l=n}^{j-1}(1-\frac{l}{\Omega})=\frac{1}{\Omega^{j-n}}\frac{(\Omega-n)!}{(\Omega-j)!}$, Eq.~\eqref{eq:thetaE_1} can be rewritten as
\begin{equation}
\label{eq:S9}
    \theta_E(K,\Omega)=\tau_E(K)\sum_{n=1}^{\Omega} \sum_{j=n}^\Omega  \frac{1}{j}\left(\frac{\psi}{\Omega}\right)^{j-n}\frac{(\Omega-n)!}{(\Omega-j)!}.
\end{equation}
In the extinction-dominated regime $\psi \ll 1$, the main contribution to the inner sum stems from $j=n$, and the leading contribution to the mMET is therefore
\begin{equation}
\begin{aligned}
    \theta_E(K,\Omega)&\approx\tau_E(K) \sum_{n=1}^\Omega \frac{1}{n}  =\tau_E(K) H_\Omega,
\end{aligned}
\end{equation}
where $H_\Omega$ is the $\Omega$-th harmonic number. Asymptotically, we have $H_\Omega \simeq \ln(\Omega) + \gamma_\text{EM} + \mathcal{O}(\Omega^{-1})$ where $\gamma_\text{EM}\approx 0.577...$ is the Euler-Mascheroni constant.
This expression is independent of selection and, to leading order, generally does not depend on the initial state of the metapopulation. 
In the limit of a large metapopulation, $\Omega \gg 1$, the metapopulation mean extinction time in the regime $\psi(m,K)\ll1$, is asymptotically given by the simple expression Eq.~\eqref{eq:thetaE-approx}:
$\theta_E(K,\Omega)\simeq \tau_E(\ln(\Omega)+\gamma_\text{EM})$. When $\Omega \gg 1$ and $K\gg 1$, we simply have $\theta_E(K,\Omega)\simeq e^K\ln(\Omega)/K$. 

The result $\theta_E(K,\Omega)$ 
has been explicitly derived for cliques (island model), but is found to provide good qualitative insight into the extinction dynamics for cycles and grids; see Fig.~\ref{fig:constant-environment}.

\section{Stationary deme occupancy}
\label{appendix:site_occupancy}
\begin{figure}
    \centering
    \includegraphics[width=\linewidth]{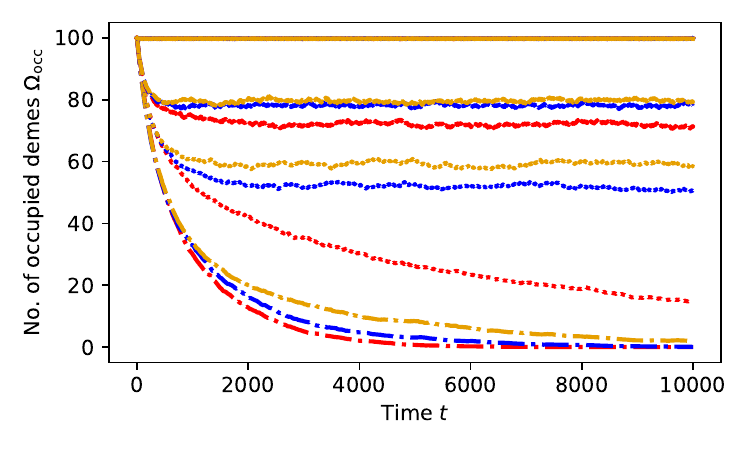}
    \caption{{\it Metapopulation occupancy}: $\Omega_\text{occ}$ vs. $t$ for cliques (yellow), cycles  (red), and  grids (blue), with $\Omega=100$ and $K=8$. 
       Simulation results averaged on 100 realisations for the stationary number  $\Omega_\text{occ}$ of occupied demes for $\psi=100$ (solid lines),  
        $\psi=5$ (dashed lines),  $\psi=2.5$ (dotted lines), and  $\psi<1$ (dash-dotted lines). Eq.~\eqref{eq:Omega} predicts  $\Omega_\text{occ}=100, 80,60,0$ for $\psi= 100, 5, 2.5$ and $\psi<1$, respectively.}
    \label{fig:metapopulation-occupancy}
\end{figure}

In the realm of the 
coarse-grained description of the extinction-dominated regime discussed in the previous section, we can use 
 the rates $B(j)=mK j(1-j/\Omega)$ and $D(j)=j/\tau_E$ to estimate the number of occupied demes $\Omega_\text{occ}$ in the metapopulation when the environment is static (constant carrying capacity $K$).
 At mean-field level, we can write the following balance equation \cite{Levins69}
\begin{equation}
\begin{aligned}
    \label{eq:mean-field_extinction} 
      \frac{{\rm d}}{{\rm d}t}\Omega_\text{occ}
    &=B(\Omega_\text{occ})-D(\Omega_\text{occ}),\\
    &=mK\Omega_\text{occ}\left(1-\frac{1}{\psi}-\frac{\Omega_\text{occ}}{\Omega}\right).
\end{aligned}
\end{equation}
The equilibria of this equation are
 $\Omega_\text{occ}=0$  and $\Omega_\text{occ}=\Omega\frac{\psi-1}{\psi}$ when $\psi>1$. The equilibrium $\Omega_\text{occ}=0$ is asymptotically stable when $\psi<1$ and unstable otherwise. This means that all demes go extinct, and there is extinction of the entire metapopulation when $\psi<1$. 
 When $\psi>1$, the equilibrium $\Omega_\text{occ}=\Omega\frac{\psi-1}{\psi}$ is asymptotically stable.
This   corresponds to a 
  fraction $1-1/\psi$ of the demes being entirely occupied, and there is fraction $1/\psi$ of empty demes. In the limit where   $\psi \gg 1$, we have $\Omega_\text{occ}\to \Omega$ and all demes and hence the metapopulation are fully occupied. Putting everything together, we obtain Eq.~\eqref{eq:Omega}.
  
 This mean-field derivation of  $\Omega_{\rm occ}$
  is accurate for large clique metapopulations but, as it 
   ignores spatial correlations, it is a crude approximation for cycles and grids; see Fig.~\ref{fig:metapopulation-occupancy}.
   In  particular, $\Omega_{\rm occ}$ overestimates the number of occupied demes in cycles when $\psi$ is not much larger than 1.
    However, $\psi= mK\tau_E$
allows us to distinguish between different
regimes  and provides a sound estimate of the number of occupied demes in the intermediate regime when $\psi(m,K)\approx me^K$ is sufficiently bigger than $1$.
 
\section{Average number of active edges on the square grid and the influence of the spatial structure}
\label{appendix:2d-heuristic}
In principle, the coarse-grained description of the $M/W$ competition holds  for any regular circulation graph. However, this approach requires the number of active edges for a given number of mutant demes to be known, which, except for cycles (one dimension) and cliques, is a difficult task due 
to complex spatial correlations between demes. 
Here, we consider the case of the square grid (with periodic boundaries), and illustrate how to estimate the average number of active edges when the metapopulation consists of only one single fully-occupied $M$ deme and all other demes are occupied by $W$ cells.
\begin{figure*}
\phantomsection
    \centering
    \includegraphics[width=\textwidth]{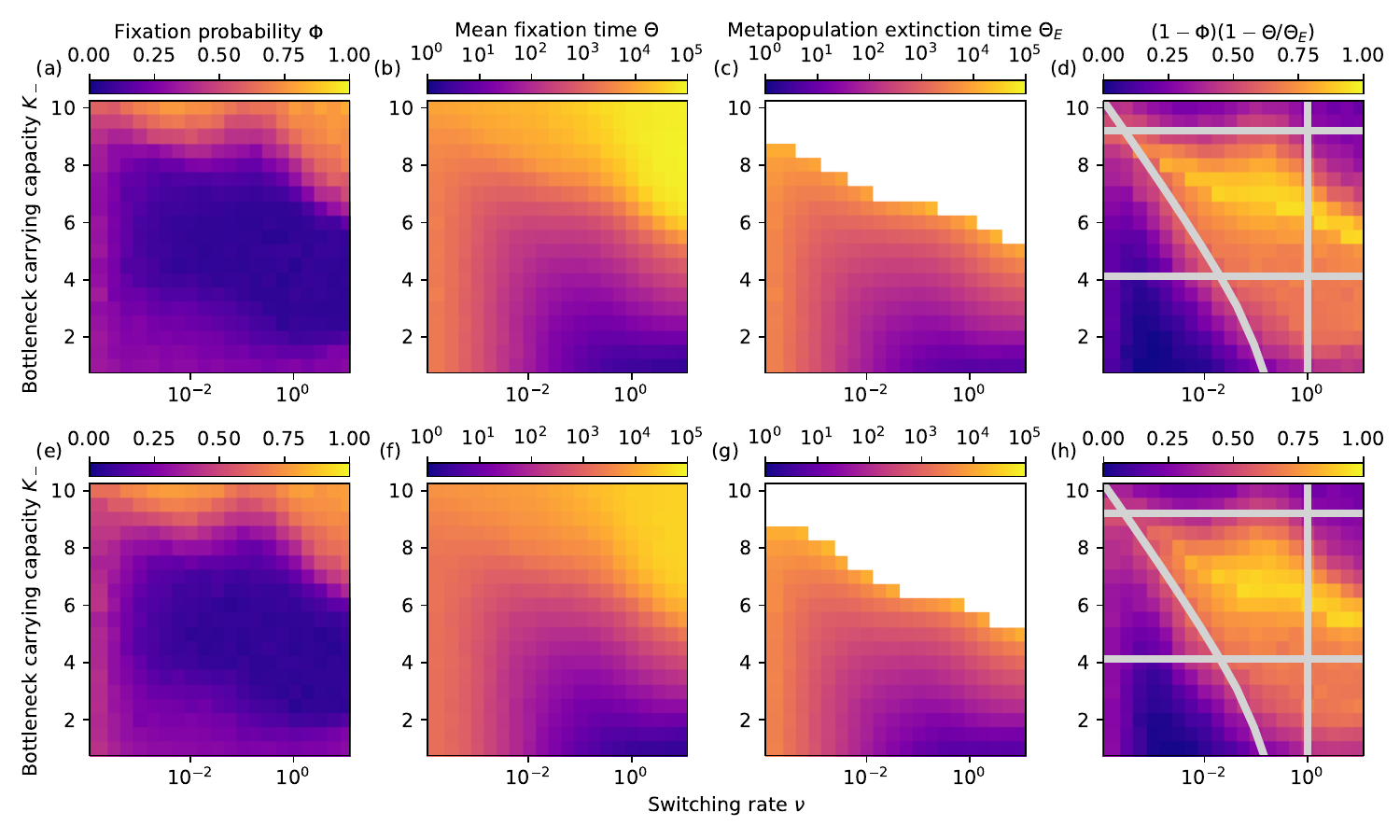}
    \caption{Near-optimal condition for the idealised treatment strategy on a cycle and grid metapopulation. $(\nu,K_-)$ heatmaps of 
    $\Phi$, $\Theta$, $\Theta_E$ and $(1-\Phi)(1-\Theta/\Theta_E)$ 
   for the cycle (a-d) and grid (e-h) metapopulations; see Appendix~\ref{appendix:2d-heuristic}. 
   White space in panels (c) and (g) indicates the parameter region where at least one realisation for those parameters did not reach extinction by $t=10^5$. Grey lines in panels (d) and (h) show the near-optimal conditions for the idealised treatment strategy:  $\psi(m,K_-)<1$ below the top horizontal line,   $mK_+\theta_E>1$ above the bottom horizontal line, and $\nu\theta_E>1$
   above  the curved line, while the vertical line indicates where $\nu<1$ and $\theta_E$ from Eq.~\eqref{eq:thetaE-approx}.
   The near-optimal treatment conditions is the yellowish cloud at the centre of the area enclosed by these lines.
   Other parameters are $\Omega=16$, $m=10^{-4}$, $s=0.1$, and $K_+=200$. In all panels, initially there is a single 
   $M$ deme and $\Omega -1$ demes occupied by $W$.}
    \label{fig:other-strong-bottleneck}
\end{figure*}
In the case of a large metapopulation, $\Omega \gg 1$,  
with unit spacing between neighbours, we assume that the mutant spreads outwards from the initial $M$ deme approximately forming an $M$-cluster with a circular front. 
If this circular $M$-cluster has a radius $r$, it has an area $\pi r^2$ containing a number $i$ of $M$ demes. The boundary of the circular $M$-cluster is of length $2\pi r$. Assuming that this length is equal to the number of $M$ demes on the boundary, we find that $r=\sqrt{i/\pi}$ and  there are $2\sqrt{i\pi}$ boundary demes given $i$ demes of type $M$. We therefore estimate that the average number 
active edges for a grid is 
$E_{\rm grid}(i)\approx2\sqrt{\pi i}$.
We have notably used this approximation in the 
transitions rates  Eq.~\eqref{eq:Moran_slow} and Eq.~\eqref{eq:effectT} for the 
 coarse-grained description of $M/W$ competition 
 in static and time-varying environments in the regime of weak bottlenecks.
 
  In Figs.~\ref{fig:constant-environment}(a,top) and \ref{fig:weak-bottleneck}(c), we have found that the spatial structure has a barely noticeable influence on the fixation probability \edit{under both constant carrying capacity and in the regime of weak bottlenecks. The mean fixation time also appears unchanged when the carrying capacity is constant.} Fig.~\ref{fig:other-strong-bottleneck} shows the heatmaps on a cycle  and a grid  metapopulation  for the 
 ``idealised treatment strategy'' proposed in Sec.~\ref{sec:changing_env-harsh}, which are almost identical. This is in accord with 
 Eqs.~\eqref{eq:near-opt}  predicting  that the same migration rate yields  the same near optimal conditions for the heatmaps of  metapopulation on any regular graph, here a cycle and a grid. Simulation results confirm spatial structure is only responsible for minor quantitative changes in the region of the heatmaps corresponding to the near-optimal ``treatment conditions''. This stems from the removal scenario characterising the idealised treatment strategy being due to deme extinction which is mostly independent of ${\rm G}$ and $m$.
\section{Intermediate dynamics in static environments}
\label{appendix:inter}

\begin{figure}
    \centering
    \includegraphics[width=\linewidth]{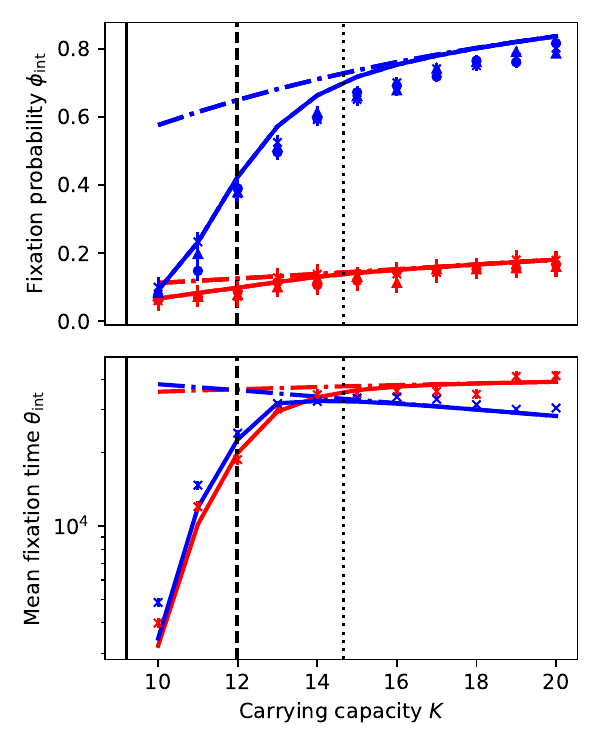}
    \caption{$\phi_{\rm int}$ vs. $K$
       for $s=0.1$ (blue) and $s=0.01$ (red) on clique (solid lines / crosses), cycle  (dashed lines / circles), and a grid (dotted lines / triangles). The different symbols and lines are almost indistinguishable. $\theta_{\rm int}$ vs. $K$ for the same parameters for a clique metapopulation.
       The vertical solid and dashed line indicate where the number of occupied demes $\Omega_\text{occ}\in [1,\Omega-1]$; see text. The dotted line indicates where $\psi=K\Omega$, i.e. every individual migrates in the time required for an independent deme extinction.
       Markers show simulation results, solid lines are predictions of Eq.~\eqref{eq:fixation_intermediate}, and dash-dot lines are predictions of Eq.~\eqref{eq:static-fixation}.
       Other parameters are $m=10^{-4}$ and $\Omega=16$.}
    \label{fig:intermediate-fixation}
\end{figure}

When $\psi \gtrsim 1$ with $mK<1$ (slow migration) in static environments, migration and deme extinction occur on the timescale $\tau_E$. In the long run, deme recolonisations and extinctions balance each other,
yielding a dynamical equilibrium consisting of $\Omega_{\rm occ}=\Omega(1-1/\psi)$ 
occupied demes and $\Omega-\Omega_{\rm occ}$ empty demes; see Fig.~\ref{fig:metapopulation-occupancy}. 
In this regime, the  three-state 
coarse-grained description of the dynamical equilibrium consists of a random mixture of 
empty demes, and occupied $W$ and $M$ demes; see Fig.~\ref{fig:cartoon-constant}(d).
After a mean time $\theta_{\rm int}^{\rm G}$
the metapopulation reaches the dynamical equilibrium consisting of a fraction $1-\Omega_{\rm occ}/\Omega=1/\psi$
empty demes, and the remaining demes are all of either type $M$ or $W$ with probability $\phi^G_{\rm int}$ and  
$1-\phi_{\rm int}^{\rm G}$, respectively. The dynamical equilibrium is thus defined by the quantities $\psi$, given by Eq.~\eqref{eq:psi}, and $\phi_{\rm int}^{\rm G}$ which is the  probability that mutants $M$ take over the $\Omega_{\rm occ}=\Omega(1-1/\psi)$ occupied demes, {where the unconditional mean fixation time is given by $\theta_{\rm int}^{\rm G}$.} In this section, we complete the results given in the main text by considering metapopulation intermediate dynamics on a regular graph ${\rm G}$, obtaining the explicit results  $\phi_{\rm int}^{\rm G}$ and $\theta_{\rm int}^{\rm G}$ for ${\rm G}=\{{\rm clique, cycle, grid}\}$ reported in Fig.~\ref{fig:intermediate-fixation} when the metapopulation initially consists of a single $M$ deme and $\Omega -1$ demes occupied by $W$. This region is concretely defined for values of $\psi$ such that $1<\psi<K\Omega$ for $\Omega \gg1$, where the lower bound ensures $\Omega_{\text{occ}}=1$ from Eq.~\eqref{eq:Omega} and the upper bound is obtained by considering the limiting case where every individual is expected to migrate on the timescale of deme extinction. Using that $\psi\approx me^K$ for $K\gg1$, we obtain the condition $\ln(1/m)\lesssim K\lesssim \ln(K \Omega/m)$. \edit{These bounds are illustrated by the vertical lines in Fig.~\ref{fig:intermediate-fixation}}.

The intermediate regime is characterised by $M/W$ competition and deme extinction. Therefore, in addition to invasions, a $W$ deme may become an $M$ deme through  extinction followed by a recolonisation, i.e. $W\rightarrow\emptyset\rightarrow M$, where $\emptyset$ indicates an extinct deme. Similarly, an $M$-deme can be changed into a $W$-deme via $M\rightarrow\emptyset\rightarrow W$. We assume that there is initially a single $M$ deme in the metapopulation (and $\Omega -1$ demes of type $W$).
With a probability $p_\text{surv}$ (see below), the 
initial $M$ deme survives the short transient as the metapopulation reaches the dynamical equilibrium
that consists of the $M$ deme, the remaining $W$ demes, and extinct demes. The number of $M$ demes $i=0,1,\dots,\Omega_{\rm occ}$ grows and shrinks through invasions and extinction-recolonisation events. 
We assume that immediately $\Omega_E=\Omega/\psi$ demes go extinct, so that 
the metapopulation quickly reaches its equilibrium occupancy $\Omega_\text{occ}=\Omega\left(1-1/\psi\right)$.
In this dynamical equilibrium, a $W$ deme can become an $M$ deme via   $W \rightarrow\emptyset$ ($W$ deme extinction) at rate $r_{{\rm ext},W}$ followed by $\emptyset \rightarrow M$ (recolonisation by $M$) at rate $r_{{\rm rec},M}^{\rm G}$. The overall extinction-recolonisation reaction $W\rightarrow\emptyset\rightarrow M$ thus occurs at  rate $1/(1/r_{{\rm ext},W} + 1/r_{{\rm rec},M}^{\rm G})$. Here, the rate of $W$ deme extinction is  
$r_{{\rm ext},W} =(\Omega_{\text{occ}} -i)/\tau_E$ and $\tau_E$ is given by Eq.~\eqref{eq:tauE}.  We proceed similarly for the extinction of an $M$ deme and its recolonisation into a $W$ site according to $M\rightarrow\emptyset\rightarrow W$. Taking also into account the rate  of invasion (see Eq.~\eqref{eq:Moran_slow}) the number of $M$ demes $i$ on 
a regular graph ${\rm G}$ varies 
according to the transition rates 
\begin{equation}
\label{eq:general-intermediate-K-rates}
\begin{aligned}
    \widetilde{T}_i^+(m, {\rm G},K)&=mK\frac{\widetilde{E}_{\rm G}(i)}{q_{\rm G}}\left[\rho_M + \frac{1}{\psi-1}\frac{q_{\rm G}}{\Omega}\frac{i (\Omega_\text{occ}-i)}{\widetilde{E}_{\rm G}(i)}\right],\\
    \widetilde{T}_i^-(m, {\rm G},K)&=mK\frac{\widetilde{E}_{\rm G}(i)}{q_{\rm G}}\left[\rho_W + \frac{1}{\psi-1}\frac{q_{\rm G}}{\Omega}\frac{i (\Omega_\text{occ}-i)}{\widetilde{E}_{\rm G}(i)}\right].
\end{aligned}
\end{equation}
It is important to note that the $\widetilde{E}_{\rm G}(i)$ in these expressions explicitly represent the number of active edges between $M$ and $W$ on the metapopulation given $i$ mutant demes, taking into account the presence of extinct demes. Therefore,  their expressions for a given graph structure $\rm G$ generally differ from those considered in the competition-dominated regime, denoted $E_{\rm G}(i)$.
With these rates, we can solve the following first-step analysis equations for the probability $\phi_{{\rm int},i}^{\rm G}$
that the dynamical equilibrium comprising initially $i$ demes of type $M$ consists of
occupied $M$
demes and extinct demes after a mean time $\theta_{{\rm int},i}^{\rm G}$
\begin{equation}
 \label{eq:first-phi-theta-int}
\begin{aligned}
 (\widetilde{T}_i^++\widetilde{T}_i^-)\phi_{{\rm int},i}^{{\rm G}}&=\widetilde{T}_i^+\phi_{{\rm int}, i+1}^{{\rm G}}+\widetilde{T}_i^-\phi_{{\rm int},i-1}^{{\rm G}},
 \\
 (\widetilde{T}_i^++\widetilde{T}_i^-)\theta_{{\rm int},i}^{{\rm G}}&=1+\widetilde{T}_i^+\theta_{{\rm int},i+1}^{{\rm G}}+\widetilde{T}_i^-\theta_{{\rm int},i-1}^{{\rm G}}.
\end{aligned}
\end{equation}
These equations are subject to 
the boundary conditions  $\phi_{{\rm int},0}^{\rm G}=0, \phi_{{\rm int},\Omega_{\rm occ}}^{\rm G}=1$
and $\theta_{{\rm int},0}^{\rm G}=
\theta_{{\rm int},\Omega_{\rm occ}}^{\rm G}=0$. We thus have $\phi_{\rm int}^{\rm G}\equiv p_\text{surv}\phi_{{\rm int},1}^{\rm G}$ and $\theta_{\rm int}^{\rm G}\equiv p_\text{surv}\theta_{{\rm int},1}^{\rm G}+(1-p_\text{surv})\tau_{E}$. The factor $p_\text{surv}=\frac{\Omega_{\text{occ}}}{\Omega}=1-1/\psi$
is the probability that the initial $M$ deme reaches the dynamical equilibrium (after what is assumed to be a short transient),   while the contribution $(1-p_\text{surv})\tau_{E}$ to $\theta_{\rm int}^{\rm G}$
accounts for the probability that the initial $M$ deme goes extinct in a mean time $\tau_E$ (given by Eq.~\eqref{eq:tauE}) before reaching the equilibrium.
The final expressions of $\phi_{\rm int}^{\rm G}$ and $\theta_{\rm int}^{\rm G}$
 thus read
\begin{equation}
\label{eq:fixation_intermediate}
    \begin{aligned}
    \phi^{\rm G}_\text{int}&=p_\text{surv}\frac{1}{1+\sum_{k=1}^{\Omega_\text{occ}-1} \prod_{m=1}^k \widetilde{\gamma}(m)},\\
    \theta^{\rm G}_\text{int}&=
    \phi^{\rm G}_\text{int}\sum_{k=1}^{\Omega_\text{occ}-1}\sum_{n=1}^k \frac{\prod_{m=n+1}^{k}\widetilde{\gamma}(m)}{\widetilde{T}^+_n} + (1-p_\text{surv})\tau_E,
\end{aligned}
\end{equation}
where
\begin{equation}
\label{eq:gammatilde}
    \widetilde{\gamma}_{\rm G}(i)\equiv \frac{\rho_W + \frac{1}{\psi-1}\frac{q_{\rm G}}{\Omega}\frac{i (\Omega_\text{occ}-i)}{\widetilde{E}_{\rm G}(i)}}{\rho_M + \frac{1}{\psi-1}\frac{q_{\rm G}}{\Omega}\frac{i (\Omega_\text{occ}-i)}{\widetilde{E}_{\rm G}(i)}},
\end{equation} 
and the upper limit of the first sum in $\phi^{\rm G}_\text{int}$ and $\theta^{\rm G}_\text{int}$ is rounded to the nearest integer. We 
find that $\phi^{\rm G}_\text{int}$ depends on the migration rate $m$, carrying capacity $K$, and the spatial structure $\rm G$ via $\widetilde{\gamma}_{\rm G}$ and $\Omega_\text{occ}$. In the case of the clique
discussed in the main text, the expression of Eq.~\eqref{eq:gammatilde}  simplifies to
\begin{equation}
    \widetilde{\gamma}_\text{clique}(i)\equiv\widetilde{\gamma}_\text{clique}=\frac{\rho_W + \frac{1}{\psi-1}}{\rho_M + \frac{1}{\psi-1}}.
\end{equation}
We notice that for all graphs ${\rm G}$,
the expressions of Eq.~\eqref{eq:fixation_intermediate} coincide with those of Eq.~\eqref{eq:static-fixation}  of the competition-dominated regime, with  $\widetilde{\gamma}_G(i)\stackrel{\psi\gg1}{\longrightarrow}\gamma=\rho_W/\rho_M$. In Fig.~\ref{fig:intermediate-fixation}, we find that the predictions of  Eq.~\eqref{eq:fixation_intermediate} are in good agreement with simulation results for all spatial structures ${\rm G}$. Moreover, we observe that the spatial dependence of
$\phi^{\rm G}_\text{int}$ and $\theta^{\rm G}_\text{int}$
is barely noticeable.

\section{Fixation probability in time-switching environments under weak bottlenecks}
\label{appendix:circulation}

In this section, we discuss in further detail the dependence of the fixation probability $\Phi^{\rm G}(\nu,m)$
on the migration rate $m$ and spatial structure ${\rm G}$ of the metapopulation in time-switching environments under weak bottlenecks.

In static environments, where $K$ is constant, a generalisation of the circulation theorem 
guarantees that the fixation probability is independent of the migration rate and spatial structure of the metapopulation arranged on a circulation graph; see Eq.~\eqref{eq:static-fixation}. This results from a correspondence between the fixation probability and the number of $M$ demes performing a biased random walk on $\{0,\dots,\Omega\}$
with a bias that is independent of $m$ and ${\rm G}$~\cite{marrecUniversalModelSpatially2021}.

In time-switching environments under weak bottlenecks (deme extinction is negligible)
the correspondence is
between the fixation $\Phi^{\rm G}(\nu,m)$ and the random walk  (with absorbing boundaries)
on $\{0,\dots,\Omega\}\times \{-1,1\}$
for the 
number of fully mutant demes in the environmental state $\xi=\pm 1$. As a consequence, 
$\Phi^{\rm G}(\nu,m)$ is the probability of absorption in the state 
$\Omega$. In this setting, defining the state of the random walk by $(i,\xi)$, where $i=0,1,\dots,\Omega$, the
random walk moves to the right ($i\to i+1$)
with a probability $r(i,\xi)$, to the left ($i\to i-1$)
with a probability $\ell(i,\xi)$, or switches environment ($\xi\to -\xi$) with probability $\epsilon(i,\xi)$, where 
\begin{equation}
\label{eq:RW}
\begin{aligned}
    r(i,\xi)&=\frac{m\mathcal{N}_{\xi}(\nu) \frac{E_{\rm G}(i)}{q_{\rm G}}\rho_{M,\xi}(\nu)}{m\mathcal{N}_{\xi}(\nu) \frac{E_{\rm G}(i)}{q_{\rm G}}\rho_{M,\xi}(\nu) + m\mathcal{N}_{\xi}(\nu) \frac{E_{\rm G}(i)}{q_{\rm G}}\rho_{W,\xi}(\nu) + \nu},\\
    \ell(i,\xi)&=\frac{m\mathcal{N}_{\xi}(\nu)\frac{E_{\rm G}(i)}{q_{\rm G}}\rho_{W,\xi}(\nu)}{m\mathcal{N}_{\xi}(\nu) \frac{E_{\rm G}(i)}{q_{\rm G}}\rho_{M,\xi}(\nu) + m\mathcal{N}_{\xi}(\nu) \frac{E_{\rm G}(i)}{q_{\rm G}}\rho_{W,\xi}(\nu) + \nu},\\
    \epsilon(i,\xi)&=\frac{\nu}{m\mathcal{N}_{\xi}(\nu) \frac{E_{\rm G}(i)}{q_{\rm G}}\rho_{M,\xi}(\nu) + m\mathcal{N}_{\xi}(\nu) \frac{E_{\rm G}(i)}{q_{\rm G}}\rho_{W,\xi}(\nu) + \nu}.
\end{aligned}
\end{equation}
 $\Phi^{\rm G}(\nu,m)$ thus coincides with the probability that the random walk defined by Eq.~\eqref{eq:RW} gets absorbed in the state $i=\Omega$. Unlike the case of the competition-dominated regime under the static environment, the fixation probability here typically depends on all parameters, including the migration rate.
 For the fixation probability to remain unchanged under parameter changes requires strict conditions. 
 This can be seen by assuming that for a parameter set $S_1$ the probabilities $r_1(i,\xi)$, $\ell_1(i,\xi)$, and $\epsilon_1(i,\xi)$ define the random walk corresponding to the fixation probability $\Phi_{i,\xi}^{\rm G}$. We therefore have
 \begin{equation}
     \Phi_{i,\xi}^{\rm G}=r_1(i,\xi)\Phi_{i+1,\xi}^{\rm G}+\ell_1(i,\xi)\Phi_{i-1,\xi}^{\rm G}+\epsilon_1(i,\xi)\Phi_{i,-\xi}^{\rm G}.
     \label{eq:1}
 \end{equation}
 We can also assume that under another set of parameters, say $S_2$, $\Phi_{i,\xi}^{\rm G}$ remains unchanged for all $i$ and $\xi$
 with a corresponding random walk defined by the probabilities $r_2(i,\xi)$, $\ell_2(i,\xi)$, and $\epsilon_2(i,\xi)$, such that 
 \begin{equation}
   \Phi_{i,\xi}^{\rm G}=r_2(i,\xi)\Phi_{i+1,\xi}^{\rm G}+\ell_2(i,\xi)\Phi_{i-1,\xi}^{\rm G}+\epsilon_2(i,\xi)\Phi_{i,-\xi}^{\rm G}.
   \label{eq:2}
 \end{equation}
 Subtracting the second equation from the first, and using conservation of probability, we find that
 \begin{equation}
 \label{eq:appendix-intermediate-step}
 \begin{aligned}
     &(r_1(i,\xi)-r_2(i,\xi))(\Phi_{i+1,\xi}^{\rm G}-\Phi_{i,-\xi}^{\rm G})\\
     &+(\ell_1(i,\xi)-\ell_2(i,\xi))(\Phi_{i-1,\xi}^{\rm G}-\Phi_{i,-\xi}^{\rm G})=0.
     \end{aligned}
 \end{equation}
 The cases of $\Phi_{i+1,\xi}^{\rm G}-\Phi_{i,-\xi}^{\rm G}=0$ and $\Phi_{i-1,\xi}^{\rm G}-\Phi_{i,-\xi}^{\rm G}=0$ imply that the fixation probability of all transient states is identical, and therefore we neglect this unphysical case. Defining $\Delta r(i,\xi)=r_1(i,\xi)-r_2(i,\xi)$ and $\Delta \ell (i,\xi)=\ell_1(i,\xi)-\ell_2(i,\xi)$, 
 Eq.~\eqref{eq:appendix-intermediate-step} yields
 \begin{equation}
     {\Delta r(i,\xi)}=\frac{\Phi_{i+1,\xi}^{\rm G}-\Phi_{i,-\xi}^{\rm G}}{\Phi_{i,-\xi}^{\rm G}-\Phi_{i-1,\xi}^{\rm G}}{\Delta \ell(i,\xi)}.
     \label{eq:overdet}
 \end{equation}
For each of the $2(\Omega-2)$ transient states we therefore have a constraint given by Eq.~\eqref{eq:overdet}. However, $\Delta r(i,\xi)$ and $\Delta \ell (i,\xi)$ are controlled by $|S_1|=|S_2|=p$ degrees of freedom (system parameters) where $p \ll 2(\Omega-2)$. Therefore, the system is overdetermined and Eq.~\eqref{eq:overdet} is only generally satisfied across all $(i,\xi)$ for the trivial solution $\Delta r(i,\xi)=\Delta \ell(i,\xi)=0$, i.e. $S_1=S_2$.
Thus, in time-fluctuating environments  the fixation probability $\Phi^{\rm G}(\nu,m)$ is expected to depend on $m$
 and ${\rm G}$. 
 
Interestingly however, Fig.~\ref{fig:weak-bottleneck}(c)
illustrates the almost unnoticeable dependence of 
 $\Phi^{\rm G}(\nu,m)$ on the specific spatial structure. This is due to the overall similar impact of the factor $E_{\rm G}(i)/q_G$ for the various graphs. While differences arise when $m$ varies at fixed $\nu$ due to large variations in the timescales of the competition dynamics, varying spatial structure produces small 
 changes in these timescales, and as such leads to only unnoticeable changes in $\Phi^{\rm G}$. 

\section{Asymmetric dichotomous
Markov noise \& environmental bias}
\label{appendix:bias}
For the sake of simplicity and clarity, in the main text we have focussed on symmetric environmental switching. In this section, we relax this assumption and outline how the results of the paper can be generalised to the case when there is an environmental bias, i.e. when there is a different average time spent in the  states $\xi=\pm1$.

Here, we consider the coloured asymmetric dichotomous
Markov noise (aDMN), also called telegraph process,
$\xi(t)\in \{-1,1\}$ that switches between $\pm 1$  according to 
$\xi \to -\xi$ at rate $\nu_{\pm}$ when $\xi=\pm 1$~\cite{bena2006,HL06,Ridolfi11}. 
It is convenient to write these asymmetric switching rates as $\nu_{\pm}=\nu(1\mp \delta)$, where $\nu\equiv (\nu_{-}+\nu_{+})/2$ is the mean switching rate ,  \(\delta\equiv (\nu_{-}-\nu_{+})/(\nu_{-}+\nu_{+})=(\nu_{-}-\nu_{+})/(2\nu)\) denotes  the switching bias, with $|\delta|\leq 1$ and $\delta>0$ when more time is spent on average in the mild environment~\cite{taitelbaumPopulationDynamicsChanging2020a,taitelbaum2023evolutionary}.  At stationarity, this aDMN has average $\langle \xi(t)\rangle=\delta$ and autocovariance $\langle \xi(t)\xi(t')\rangle-\langle \xi(t)\rangle\langle \xi(t')\rangle=(1-\delta^2)e^{-2\nu|t-t'|}$~\cite{bena2006,HL06,Ridolfi11,wienandEvolutionFluctuatingPopulation2017,wienandEcoevolutionaryDynamicsPopulation2018}. 

The aDMN drives the
{\it time-switching} carrying capacity
 \begin{equation}
  \label{eq:Kasym}
    K(t)=\frac{1}{2}\left[
    K_+ + K_- +\xi(t)\left( K_+ - K_-\right)
    \right],
\end{equation}    
which is the same expression as  Eq.~\eqref{eq:K(t)}, but now driven by the aDMN $\xi(t)$~\cite{
wienandEvolutionFluctuatingPopulation2017,wienandEcoevolutionaryDynamicsPopulation2018,taitelbaumPopulationDynamicsChanging2020a,taitelbaum2023evolutionary,
hernandez-navarroCoupledEnvironmentalDemographic2023,askerCoexistenceCompetingMicrobial2023,LKUM2024}.
The carrying capacity switches back and forth between $K_+$  (mild environment, $\xi=+1$) and $K_-< K_+$  (harsh environment, $\xi=-1$)
at rates $\nu_\pm=\nu(1\mp\delta)$  according to 
\begin{equation*}
    K_+\xrightleftharpoons[\nu_-]{\nu_+}K_-.
\label{Eq:Kswitches}
\end{equation*}
At stationarity, the 
 expected value of the carrying capacity is  $\langle K(t)\rangle=\frac{1}{2}(K_++K_-+\delta(K_+ -K_-))$, and its auto-covariance is $\langle K(t)K(t')\rangle-\langle K(t)\rangle\langle K(t')\rangle=\left(\frac{K_{+}-K_{-}}{2}\right)^2\left(1-\delta^2\right)e^{-2\nu|t-t'|}$~\cite{HL06,bena2006,Ridolfi11}.
\begin{figure}
    \centering
    \includegraphics[width=\linewidth]{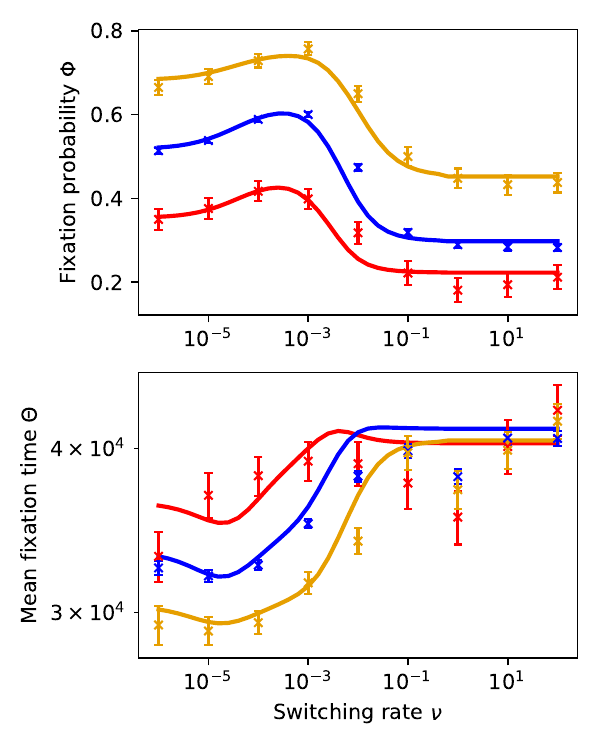}
    \caption{Fixation probability $\Phi^{\rm G}$ and mean fixation time $\Theta^{\rm G}$ against switching rate $\nu$ for different values of $\delta$ for a clique metapopulation. Red, blue, and yellow represent $\delta=-0.5$, $\delta=0.0$, and $\delta=0.5$, respectively.
    Markers show simulation results and lines are predictions of 
    Eq.~\eqref{eq:weak-bottleneck-solution}. 
    Other parameters are $\Omega=16$, $s=0.01$, $m=10^{-4}$, $K_+=200$, and $K_-=20$.
    }
    \label{fig:weak-bottleneck-bias}
\end{figure}
Under asymmetric switching, the 
 stationary population distribution
of a single deme
is well approximated by the stationary density of the piecewise Markov process Eq.~\eqref{PDMP_eq} ($n$-PDMP) \cite{wienandEcoevolutionaryDynamicsPopulation2018}, now driven by the aDMN,
whose joint density is 
\begin{equation}
\label{PDMP_dens_as}
 \begin{aligned}
 \hspace{-7mm}
p_{\xi}(n;\nu,\delta)\propto \begin{cases}
                \frac{1+\delta}{n^2} \left(\frac{K_+ - n}{n}\right)^{\nu(1-\delta)-1} \left(\frac{n-K_-}{n}\right)^{\nu(1+\delta)} \quad \text{if $\xi=+1$,}\\
                \frac{1-\delta}{n^2} \left(\frac{K_+ - n}{n}\right)^{\nu(1-\delta)} \left(\frac{n-K_-}{n}\right)^{\nu(1+\delta)-1}
                \quad \text{if $\xi=-1$},
               \end{cases}
\end{aligned}
\end{equation}
where the proportional factor accounts for the normalisation constants.
 The stationary marginal density of this $n$-PDMP, up to the normalisation constant, is
\begin{equation}
\begin{aligned}
   p(n;\nu,\delta)&=\sum_{\xi}
   \left(\frac{1+\xi\delta}{2}\right) p_{\xi}(n;\nu,\delta),\\
&\propto \frac{1}{n^2} \left(\frac{K_+ - n}{n}\right)^{\nu(1-\delta)-1} \left(\frac{n-K_-}{n}\right)^{\nu(1+\delta)-1},
    \label{eq:nPDMP_dens_as}
\end{aligned}
\end{equation}
where we have again omitted the normalisation constant. 
The $n$-PDMP density captures the mean features of the deme size distribution: 
It is bimodal with peaks at $K_+$ and $K_-$ ($n\approx K_{\pm}$) when  $\nu\ll 1$, and is unimodal and centred around  $n\approx 2K_+K_-/[(1-\delta)K_+ + (1+\delta)K_-]$ when $\nu \gg 1$
~\cite{taitelbaumPopulationDynamicsChanging2020a,hernandez-navarroCoupledEnvironmentalDemographic2023,askerCoexistenceCompetingMicrobial2023,hernandez-navarroEcoevolutionaryDynamicsCooperative2024,LKUM2024}. When \(\nu(1\pm \delta)\lesssim 1\), the  size $n$ of each deme tracks the carrying capacity, and a  bottleneck occurs at an average frequency $\nu_{+}\nu_{-}/(2\nu)=\nu(1-\delta^2)/2$, each time $K$ switches from $K_+$ to $K_-$~\cite{taitelbaumPopulationDynamicsChanging2020a,hernandez-navarroCoupledEnvironmentalDemographic2023,LKUM2024}. 

In the realm of the coarse-grained description discussed in the main text, the regime of weak bottlenecks dominated by the $W/M$ competition can be characterised by the $M$  fixation probability $\Phi^{\rm G}(\nu,\delta,m)$
and unconditional mean fixation time  $\Theta^{\rm G}(\nu,\delta,m)$ by Eq.~\eqref{eq:weak-bottleneck-solution}
obtained by solving the first-step analysis equations Eq.~\eqref{eq:firststepphi} and Eq.~\eqref{eq:firststeptau}
with the transition rates Eq.~\eqref{eq:effectT}
and Eq.~\eqref{eq:rhoMWweak}
obtained using $N_{\xi}(\nu,\delta)$ averaged over Eq.~\eqref{eq:nPDMP_dens_as}, i.e.
$N_{\xi}(\nu,\delta)=\int np_{\xi}(n;\nu,\delta) {\rm d} n$. The results of Fig.~\ref{fig:weak-bottleneck-bias} for a clique metapopulation show that the predictions of the coarse-grained description
based on the PDMP approximation Eq.~\eqref{eq:nPDMP_dens_as} are in good agreement with simulation results. $\Phi^{\rm G}(\nu,\delta,m)$
and  $\Theta^{\rm G}(\nu,\delta,m)$ are again found to exhibit a non-monotonic dependence on $\nu$, with extrema in the range of intermediate $\nu$. The main effect of $\delta$ is to increase the $M$ fixation probability and reduce the mean fixation time when $\delta>0$, which is intuitively clear since this corresponds to a bias towards the mild state favouring the fixation of $M$.

\begin{figure*}
    \centering
    \includegraphics[width=\linewidth]{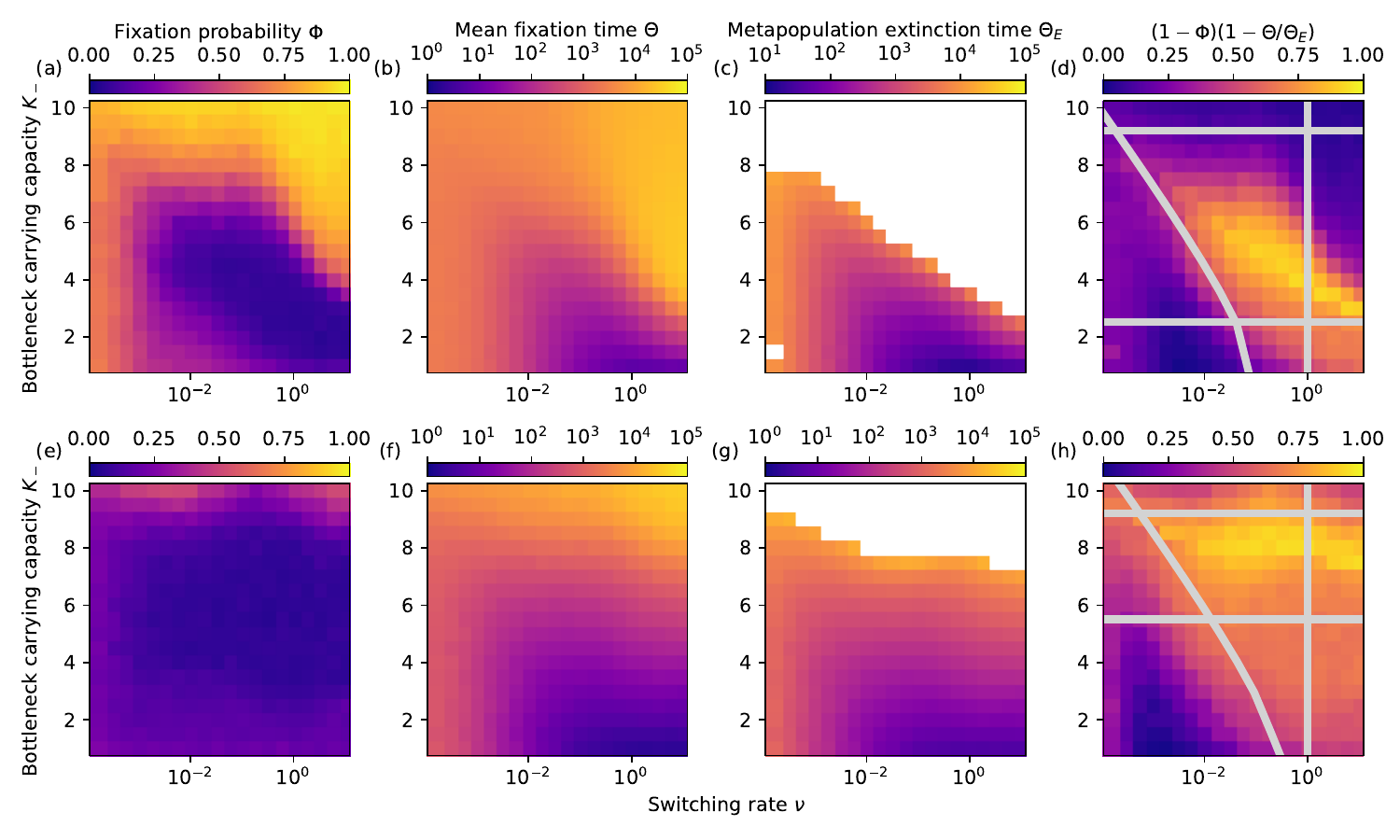}
    \caption{Near-optimal condition for the idealised treatment strategy on the clique metapopulation. $(\nu,K_-)$ heatmaps of 
    $\Phi$, $\Theta$, $\Theta_E$ and $(1-\Phi)(1-\Theta/\Theta_E)$ 
   for the $\delta=0.5$ (a-d) and $\delta=-0.5$ (e-h) metapopulations; see Appendix~\ref{appendix:bias}. 
   Whitespace in panels (c) and (g) indicate the region of the parameter where at least one realisation for those parameters did not reach extinction by $t=10^5$. Grey lines in panels (d) and (h) show the near-optimal conditions for the idealised treatment strategy in the asymmetric environment:  $\psi(m,K_-)<1$ below the top horizontal line,   $mK_+\theta_E \frac{1+\delta}{1-\delta}>1$ above the bottom horizontal line, and $\nu(1+\delta)\theta_E>1$
   above  the curved line, while the vertical line indicates where $\nu<1$ and $\theta_E$ from Eq.~\eqref{eq:thetaE-approx}.
   The near-optimal treatment conditions is the yellowish cloud at the centre of the area enclosed by these lines.
   Other parameters are $\Omega=16$, $m=10^{-4}$, $s=0.1$, and $K_+=200$. In all panels, initially there is a single 
   $M$ deme and $\Omega -1$ demes occupied by $W$.}
    \label{fig:strong-bottleneck-bias}
\end{figure*}

The regime of strong bottlenecks is dominated by the interplay between $M/W$ competition in the mild state ($K=K_+$) and deme extinction in the harsh environmental state ($K=K_-$), occurring in time $\theta_E\equiv\theta_E(K_-, \Omega)$. In this regime, the near-optimal conditions for the removal of the mutant strain can be obtained as under symmetric switching (given by Eq.~\eqref{eq:near-opt}) and read
 \begin{equation}
 \label{eq:near-opt-bias}
\begin{aligned}
 &\psi(m,K_-)<1,~\nu(1\pm \delta)\lesssim 1,\\
&\nu(1+\delta)\theta_E\gtrsim 1, 
~mK_+\theta_E \frac{1+\delta}{1-\delta}\gtrsim1,
\end{aligned}
 \end{equation}
which, as Eq.~\eqref{eq:near-opt}, are conditions depending on $m$ but not on the spatial structure ${\rm G}$.
The main differences from Eq.~\eqref{eq:near-opt} are in the conditions $\theta_E\nu_-=\theta_E\nu(1+\delta)\gtrsim 1$ and $mK_+\theta_E\frac{\nu_-}{\nu_+}=mK_+\theta_E\frac{1+\delta}{1-\delta} \gtrsim 1$. The first of these changes ensures that a switch occurs before the metapopulation mean extinction time in the harsh environment, $\theta_E$. The second ensures there are enough recolonisations in the mild environment to maintain the metapopulation given the minimum switching rate required to prevent extinction in the harsh environment, i.e. rearranging $\theta_E \nu_- \gtrsim 1$ gives $\nu \gtrsim 1/(\theta_E (1+\delta))$ and for sufficient recolonisations we require $mK_+/\nu_+\equiv mK_+/(\nu(1-\delta)) \gtrsim 1$, where we substitute our expression for $\nu$. Since $\theta_E$ is independent of $\delta$, we expect that the conditions Eq.~\eqref{eq:near-opt-bias} define a region in the parameter space that is similar  to that obtained 
 under symmetric switching, shifted towards higher (lower)  values of $\nu$ and $K_-$ when $\delta<0$ ($\delta>0$). This picture is confirmed by the heatmaps of Fig.~\ref{fig:strong-bottleneck-bias}.  
 
\section{Simulation methods \& plots}
\label{appendix:model-and-methods_simulation}
In this section, we explain how the simulation of the individual-based dynamics of the full model has been implemented. We also outline how we have  
 plotted the simulation/numerical data 
that we have obtained 
 to produce the figures discussed in the main text and appendices.

In addition to the  coarse-grained descriptions of the model that provide us with analytical  approximations of 
metapopulation dynamics in different regimes, we have employed Monte Carlo (MC) methods to simulate the full individual-based model and mirror the dynamics encoded in the ME of Eq.~\eqref{eq:ME}. In this section, we outline how we have performed the stochastic simulations that we have notably used to test our analytical predictions. 

While not exact-like other simulation methods (e.g. the Gillespie algorithm~\cite{Gillespie76}) due to time-discretisation, the MC algorithm used here improves on computational efficiency making simulations with larger numbers of cells on the metapopulation that run for long times feasible and straightforward to implement. The questions of computational efficiency and tractability are particularly critical in the context of this work in which we study fixation and extinction of spatially arranged populations, a notoriously computationally demanding 
problem. In this context MC algorithms similar to the one used are useful tools to investigate the properties of spatially extended systems; see, for example, Ref.~\cite{dobramyslStochasticPopulationDynamics2018}.

In the case of symmetric environmental switching, 
the MC algorithm that we have employed can be described as follows: The graph, consisting of the spatially arranged $\Omega$ deme forming the metapopulation, is initialised by randomly picking an initial value for the carrying capacity $K(0)=\{K_-,K_+\}$ with equal probability (in the constant environment case $K(0)=K$ with probability 1), populating a single deme with $K(0)$ $M$-cells, and the remaining $\Omega-1$ of the demes with $K(0)$ $W$-cells.  Time is then discretised in units of Monte Carlo steps (MCS) whereby in each MCS we perform $2N$ birth/death events, where $N$ is measured at the start of the MCS. We choose $2N$ as we typically have $N$ births and $N$ deaths per unit time according to the transition rates Eq.~\eqref{eq:intra_transition_rates} and Eq.~\eqref{eq:migration-transition} when summing over all cells on the metapopulation. Therefore, our units of time are consistent between the theoretical model and the Monte Carlo simulation. \edit{We emphasise that migration events and environmental switches do not contribute to the $2N$ events comprising an MCS. Therefore, $2N$ birth/death events are performed in a single MCS, and the expected number of migrations and environmental switches in the MCS corresponds to the migration and switching events expected in one unit of time.}  Each of the $2N$ events in an MCS occurs sequentially and the rates are updated for the next event in the MCS. The type of each event in an MCS is selected sequentially based on the rates of the events, where a higher rate means a larger probability of that event being selected. Concretely, the following steps for a single MCS occur:
\begin{itemize}
    \item Check if an environmental switch, occurring with rate $\nu$, occurs on the metapopulation. The rate of reaction of birth/death/migration events on the entire metapopulation is $N(1+m+N/K(t))$. Therefore, the probability of an environmental switch is given by $\nu/(N(1+m+N/K(t)) + \nu)$.
    \item Otherwise, a deme $x$ is picked for an event to occur based on the total rate of events on that deme, $n(x)(1+m+n(x)/K(t))$. Therefore, the probability of selecting a deme $x$ for an event is $\frac{n(x)(1+m+n(x)/K(t))}{N(1+m+N/K(t))}$.
    \item A species is picked for an event to occur based on the total rate of events of that species on the selected deme. The propensity of a given species $\alpha$ on deme $x$ is $n_\alpha(x)(1+m+n(x)/K(t))$, and the total rate of events on the deme is as in the previous step. Therefore, that species is selected with probability $\frac{n_\alpha(x)(1+m+n(x)/K(t))}{n(x)(1+m+n(x)/K(t))}=\frac{n_\alpha(x)}{n(x)}$.
    \item The species on the deme can then either undergo a birth, death, or migration event, with the probability of each depending on the rates of these events. Birth, death, or migration is selected with probability $\frac{T_\alpha^{+,-,m}(x)}{n_\alpha(x)(1+m+n(x)/K(t))}$.
    \item The selected event is performed.
    \item \edit{The above steps are repeated until $2N$ birth/death events are performed.}
\end{itemize}
\edit{Since each event is selected based on the current propensities of the system, and those are subsequently updated before the selection of the next event. The probability of events
in our MC algorithm is thus statistically equivalent to that of a  simulation generating the statistically correct sample paths
(e.g. Gillespie \cite{Gillespie76}). The approximation of this algorithm is to assign each MCS to $2N$ birth/death events with $N$ defined at the start of the MCS, unlike a statistically exact simulation algorithm where each event time is drawn individually.}

In all simulations, each realisation is simulated until fixation, tracking when fixation occurs and which species fixates. In simulations for Figs.~\ref{fig:constant-environment}(b), \ref{fig:clique-strong-bottleneck}, \ref{fig:other-strong-bottleneck}, and \ref{fig:strong-bottleneck-bias} the simulation then continues until metapopulation extinction or a large fixed time $T$ (here we set $T=10^5$). For a given set of parameters, if extinction does not occur in any realisation, the extinction time is not recorded. The data is averaged to obtain the fixation probability, the mean fixation time, and the mean time to extinction where applicable. Furthermore, we record the standard error on the mean for each quantity. 
In Figs.~\ref{fig:constant-environment}, \ref{fig:weak-bottleneck}, and \ref{fig:weak-bottleneck-bias}, $10^3$ realisations are ran for each set of parameters, and these data are plotted with the standard error on the mean shows as error bars. In the heatmaps, each data point corresponds to the average value for $10^3$ simulations at that point in parameter space. The standard error on the mean is not plotted for the heatmaps. 

In the case that $K(t)=K=$ constant, we set $\nu=0$ and $K_+=K_-=K$, such that the first step of the above process is effectively skipped. In the case that the switching is asymmetric, the starting state is chosen according to the stationary distribution of $K(t)$, i.e. $K(0)=K_\pm$ with probability $\frac{1\pm\delta}{2}$. The probability for an environmental switch in the first step then depends on the current environmental state, where a switch occurs with probability $\nu_\pm/(N(1+m+N/K(t) + \nu_\pm)$ for $K(t)=K_\pm$.

\bibliography{bibliography}

\end{document}